\documentclass[12pt]{article}
\usepackage{graphicx}
\usepackage{bm}
\usepackage{amsmath}
\usepackage{amssymb}
\usepackage{amsfonts}
\usepackage{mathtools}
\usepackage{hyperref}
\usepackage{braket}
\usepackage{anyfontsize}
\usepackage[normalem]{ulem}
\usepackage{wrapfig}
\usepackage{textpos}
\usepackage[left=1in,right=1in,top=1in,bottom=1in]{geometry}

\makeatletter
\newcommand{\customlabel}[2]{%
   \protected@write \@auxout {}{\string \newlabel {#1}{{#2}{\thepage}{#2}{#1}{}} }%
   \hypertarget{#1}{}
}
\makeatother

\DeclareMathOperator{\tr}{\text{tr}}

\hypersetup{colorlinks=true,urlcolor=[rgb]{0,0,0.5},citecolor=[rgb]{0.5,0,0},linkcolor=[rgb]{0,0,0.4}}

\renewcommand{\title}[1]{\vbox{\center\bf{\Large{#1}}}\vspace{5mm}}
\renewcommand{\author}[1]{\vbox{\center#1}\vspace{5mm}}
\newcommand{\address}[1]{\vbox{\center\em#1}}

\newcommand\emails[1]{\begingroup
	\renewcommand\thefootnote{}\footnote{#1}
	\addtocounter{footnote}{-1}\endgroup}

\def\tr{{\rm tr}}

\def\1den{\hbox{$1\hskip -1.2pt\vrule depth 0pt height 1.53ex width 0.7pt
                  \vrule depth 0pt height 0.3pt width 0.12em$}}

\def\and{\quad {\rm and} \quad}

\begin{document}

\begin{titlepage}

\begin{textblock*}{5cm}(13.7cm,-.7cm)
\makebox{MIT-CTP/4949}
\end{textblock*}
\begin{textblock*}{5cm}(13.7cm,0cm)
\makebox{SU-ITP-17/12}
\end{textblock*}

\begin{center}
\vspace*{2cm}
\title{{\fontsize{0.75cm}{0.75cm}\selectfont Superdensity Operators for \\ Spacetime Quantum Mechanics}}
\author{Jordan Cotler,${}^a$ Chao-Ming Jian,${}^{b,c}$ Xiao-Liang Qi,${}^{a,d}$ and Frank Wilczek\,${}^{e,f,g,h}$}
\address{{\fontsize{0.4cm}{.4cm}\selectfont
${}^a$ Stanford Institute for Theoretical Physics,\\ Stanford University, Stanford, California 94305, USA\\
\vspace*{2mm}
${}^b$ Station Q, Microsoft Research\\ Santa Barbara, California 93106, USA\\
\vspace*{2mm}
${}^c$ Kavli Institute of Theoretical Physics,\\ University of California, Santa Barbara, California 93106, USA\\
\vspace*{2mm}
${}^d$ Institute for Advanced Study, Princeton, New Jersey 08540 USA\\
\vspace*{2mm}
${}^e$ Center for Theoretical Physics, MIT\\ Cambridge, Massachusetts 02139, USA\\
\vspace*{2mm}
${}^f$ The Oskar Klein Centre for Cosmoparticle Physics,\\ Stockholm University, Albanova, 10691 Stockholm, Sweden\\
\vspace*{2mm}
${}^g$ T. D. Lee Institute and Wilczek Quantum Center, \\
Shanghai Jiao Tong University, Shanghai 200240, China \\
\vspace*{2mm}
${}^h$ Department of Physics and Origins Project \\ Arizona State University, Tempe, Arizona 25287, USA
}}

\emails{ \hspace*{-8mm}
\href{mailto:jcotler@stanford.edu}{\tt jcotler@stanford.edu},
\href{mailto: cmjian@kitp.ucsb.edu}{\tt cmjian@kitp.ucsb.edu},
\href{mailto:xlqi@stanford.edu}{\tt xlqi@stanford.edu},
\href{mailto:wilczek@mit.edu}{\tt wilczek@mit.edu}
}

\end{center}

\begin{abstract}
We introduce superdensity operators as a tool for analyzing quantum information in spacetime.  Superdensity operators encode spacetime correlation functions in an operator framework, and support a natural generalization of Hilbert space techniques and Dirac's transformation theory as traditionally applied to standard density operators.  Superdensity operators can be measured experimentally, but accessing their full content requires novel procedures.  We demonstrate these statements on several examples.  The superdensity formalism suggests useful definitions of spacetime entropies and spacetime quantum channels.  For example, we show that the von Neumann entropy of a superdensity operator is related to a quantum generalization of the Kolmogorov-Sinai entropy, and compute this for a many-body system.  We also suggest experimental protocols for measuring spacetime entropies.
\end{abstract}

\end{titlepage}

\tableofcontents
\newpage

\section{Introduction}

Large parts of the existing formalism of quantum mechanics, and its interpretative apparatus, treat space and time on very different footings.  Yet in classical physics it is often advantageous, especially in the analytical theory of dynamical systems, to consider time as an extra dimension on the same footing as spatial dimensions \cite{Arnold1}.  And of course grossly asymmetric treatment of time and space is, from the point of view of relativity, disturbing and unnatural.  In this paper, we propose a formalism for \textit{spacetime} quantum theory which ameliorates the asymmetry.  This formalism suggests new ways to analyze the dynamics of quantum information and entanglement in spacetime, and new experiments to elucidate that dynamics.  

Before attempting to put space and time on similar footing in quantum theory, it is instructive to recall how space is treated in quantum theory -- in other words, the relationship between physical space and the Hilbert space of (single-time) states.  In simple cases, we can decompose Hilbert state space into tensor factors
\vskip-.1cm
\begin{equation}
\mathcal{H} = \bigotimes_{\textbf{x}} \mathcal{H}_{\textbf{x}}
\end{equation}
\vskip-.1cm
with each factor corresponding to a particular point in space $\textbf{x}$.  In this way, the tensor factor decomposition identifies local degrees of freedom.  

Given a quantum state, we can analyze it relative to the tensor factors $\mathcal{H}_\textbf{x}$, using a basis of the Hilbert space which is the tensor product of bases of the individual tensor factors.  We learn about spatial correlations by considering entanglement between partial traces of the quantum state (i.e., spatial entanglement) or, in practice, by measuring correlation functions of operators that act on different tensor factors (i.e., local observables).  Another, more general way to identify local degrees of freedom is to provide a ``net of observables,'' essentially defining which operators on the Hilbert space are local \cite{Haag1}.   Also, notoriously, important subtleties arise in taking infinite products of Hilbert spaces.  In this paper we will prioritize simplicity over maximum generality.  

To upgrade time from its parametric manifestation, it is natural to consider an expanded Hilbert space, containing tensor products for different times, which we will call the history Hilbert space $\mathcal{H}_{\text{hist.}}$.  Supposing a tensor factorization into spacetime points is possible, then the history Hilbert space takes the form
\vskip-.1cm
\begin{equation}
\mathcal{H}_{\text{hist.}} = \bigotimes_{\textbf{x},t} \mathcal{H}_{\textbf{x},t}\,.
\end{equation}
\vskip-.1cm
More compactly, if $\mathcal{H}_t$ is the Hilbert space of states at time $t$, then $\mathcal{H}_{\text{hist}.} = \bigotimes_t \mathcal{H}_t$.  Now we require a spacetime generalization of quantum states.  It is tempting to think that the right notion of spacetime states would simply be states in $\mathcal{H}_{\text{hist.}}$ with the standard inner product, but this turns out not to be useful.  In fact, finding a suitable notion of spacetime states is rather subtle.  

History Hilbert space has appeared before in discussions of quantum theory, perhaps most notably in Griffiths' foundational work on consistent histories \cite{Griffiths1, Griffiths2, Griffiths3}. In the consistent histories framework and its variations the central objects are projection operators on $\mathcal{H}_{\text{hist.}}$ \cite{DF1}--\cite{Isham1}.  In the entangled histories formalism developed by two of the authors of the present paper, the consistent histories formalism was generalized to define quantities which are more akin to spacetime states (in particular, allowing superposition) and some of their characteristic phenomenology was explored \cite{CW1, CW2, CW3, CW4}.     
Other attempts at defining spacetime states include multi-time states \cite{Aharonov1, Aharonov2, Aharonov3} and pseudo-density matrices \cite{Pseudo1}. We should also mention that the multi-time correlation operator, which was developed as a tool to study dynamical entropies in quantum systems \cite{Lindblad1}--\cite{Fannes1}, embodies a special case of our superdensity operator.  We will return to comparative discussion of some of these approaches later; for now, let us only mention that each of them can be expressed within the superdensity formalism, which appears to us more systematic and comprehensive.  We have been inspired by the elegance and power that Dirac's transformation theory achieves for quantum states, and have attempted to achieve something analogous for spacetime analogs of quantum states.  There has also been related work on spacetime quantum circuits and quantum measurements \cite{Hardy1}--\cite{Pollock3} which can be interfaced with our formalism.

In the approach pursued here, superdensity operators play a central role.  Mathematically, superdensity operators are {\it quadratic forms on the space of operators on the history Hilbert space}.  Physically, the superdensity operator of a physical system codifies its response to experimental probes, allowing that those probes may be applied at different times (and places).  More abstractly, just as a standard density operator represents a state and encodes the data of all correlation functions at a fixed time, we propose that the superdensity operator represents a spacetime state, and encodes the data of all \textit{spacetime} correlation functions.  As the name suggests, superdensity operators share many formal properties with density operators. We will demonstrate, in particular, that the information theoretic properties of superdensity operators -- such as their entanglement, entropies, mutual informations, etc. -- are meaningful notions with operational physical significance.  Thus the superdensity operator provides a compelling definition of the spacetime state of a system, which also appears to be fruitful.    

The paper is organized as follows:

\begin{itemize}
\item In Section 2 we will further motivate, define, and exemplify superdensity operators, and discuss their formal properties.  Within this framework, the concepts of spacetime observables and spacetime entanglement arise as naturally as do the corresponding concepts for (single-time) states.  

\item In Section 3 we show that the superdensity operator is in principle observable, and discuss the interesting, novel kinds of measurements its full exploration requires. 

\item In Section 4 we show how the superdensity operator suggests a definition of quantum dynamical entropy which generalizes the classical Kolmogorov-Sinai entropy, and we compute it in some tractable, yet non-trivial models. 

\item In Section 5 we continue the discussion of observability, and suggest realistic protocols for some of the new observables introduced earlier.   

\item In Section 6 we briefly summarize and suggest further implications.

\item In the Appendices, we collect some technical results and also discuss the relationship of the superdensity formalism to preceding approaches.

\end{itemize}

\section{Superdensity formalism}

\subsection{Motivation}

\subsubsection{Coupling to auxiliary apparatus}

Before providing the general definition, which is somewhat abstract, we will first discuss an important, concrete special case.  It is obtained by coupling auxiliary apparatus to an evolving quantum system.  In this setup, the density operator of the auxiliary apparatus will embody the superdensity operator of the evolving quantum system.


Consider a system with initial state $|\psi_0\rangle$ living in some initial time Hilbert space $\mathcal{H}_{t_1}$ of dimension $d$.  Suppose we have a set of operators $\{X_i\}$ on $\mathcal{H}_{t_1}$ which form a complete orthonormal basis for all operators on $\mathcal{H}_{t_1}$.  In particular,
\begin{equation}
\tr(X_i^\dagger X_j) = \delta_{ij}
\end{equation}
for $i,j = 0,...,d^2 - 1$.  To deal with correlations involving $n$ times, we introduce $n$ independent $d^2$--level auxiliary systems, each with orthonormal basis $\{|0\rangle,...,|d^2-1\rangle\}$.  The initial state of the joint system is taken to be
\begin{equation}
|0\rangle_1 \otimes \cdots \otimes |0\rangle_{n} \otimes |\psi_0\rangle\,.
\end{equation}
At the initial time $t_1$, we apply a unitary to $|0\rangle_1$, mapping it to the superposition $\frac{1}{d} \sum_{i=0}^{d^2-1} |i\rangle_1$.  Next, we couple this auxiliary state to the main system via the mapping
\begin{equation}
\left(\frac{1}{d} \sum_{i=0}^{d^2-1} |i\rangle_1 \right) \otimes |0\rangle_2 \cdots \otimes |0\rangle_{n} \otimes |\psi_0\rangle \, \longrightarrow \, \frac{1}{\sqrt{d}}\sum_{i=0}^{d^2 - 1}|i\rangle_1 \otimes |0\rangle_2 \cdots \otimes |0\rangle_{n} \otimes X_i|\psi_0\rangle
\end{equation}
which can be implemented by a unitary transformation (see Section \ref{sec:mapping_time_to_space} for details).  Evolving the main system with the unitary time evolution operators $U(t_{k+1},t_k)$ and similarly coupling each auxiliary system to the main system in turn, we obtain
\begin{equation}
\label{superstate1}
|\Psi\rangle = \frac{1}{d^{n/2}}\sum_{i_1,...,i_n=0}^{d^2 - 1}|i_1,...,i_n\rangle \otimes X_{i_n} U(t_n,t_{n-1}) X_{i_{n-1}} \cdots X_{i_2}U(t_2, t_1)X_{i_1}|\psi_0\rangle
\end{equation}
where we have written $|i_1\rangle_1 \otimes \cdots \otimes |i_n\rangle_n = |i_1,...,i_n\rangle$ for ease of notation.  We call $|\Psi\rangle$ a \textit{superstate}.  It encodes all time-ordered operator insertions on the initial state.  Specifically, since the $X_i$'s form complete operator bases, any time-ordered operator insertions at the times $t_1,...,t_n$ can be written as
\begin{equation}
\mathcal{O} := \sum_{i_1,...,i_n=0}^{d^2-1} c_{i_1,...,i_n} \, X_{i_n} U(t_n,t_{n-1}) X_{i_{n-1}} \cdots X_{i_2} U(t_2,t_1) X_{i_1}
\end{equation}
for some complex coefficients $c_{i_1,...,i_n}$.  Thus, we can extract $\mathcal{O}|\psi_0\rangle$ from the superstate $|\Psi\rangle$ through the projection
\begin{equation}
\left(\sum_{i_1,...,i_n=0}^{d^2-1} c_{i_1,...,i_n} \langle i_1,...,i_n| \otimes \textbf{1}_{d \times d}\right) |\Psi\rangle = \mathcal{O}|\psi_0\rangle\,.
\end{equation}
The central idea here is that the joint state of the auxiliary apparatus and main system encodes a large class of potential operators insertions on the initial state, in superposition.  Crucially, the multiple-time insertions are \textit{entangled} with the auxiliary apparatus, such that the multilinear structure of the entanglement (i.e., each register system is entangled to operator insertions at a particular time) is matched to the multilinear structure of the operator insertions themselves.

More generally, given an initial density operator $\rho_0$, we can couple the system to the auxiliary apparatus in similar fashion to obtain the density operator
\begin{equation}
\frac{1}{d^{n}} \sum_{\substack{i_1,...,i_n = 0 \\ j_1,...,j_n = 0}}^{d^2 - 1} |i_1,...,i_n\rangle \langle j_1,...,j_n| \otimes X_{i_n} U(t_n,t_{n-1}) \cdots U(t_2, t_1) X_{i_1} \, \rho_0 \, X_{j_1}^\dagger \, U^\dagger(t_2,t_1) \cdots U(t_{n},t_{n-1})^\dagger X_{j_n}^\dagger\,.
\end{equation}
Tracing over the state of the main system, we are left with the state of the auxiliary system
\begin{equation}
\label{superdensity1}
\varrho = \frac{1}{d^{n}} \sum_{\substack{i_1,...,i_n = 0 \\ j_1,...,j_n = 0}}^{d^2 - 1} \tr\left(X_{i_n} U(t_n,t_{n-1}) \cdots U(t_2, t_1) X_{i_1} \, \rho_0 \, X_{j_1}^\dagger \, U^\dagger(t_2,t_1) \cdots U(t_{n},t_{n-1})^\dagger X_{j_n}^\dagger\right) \, |i_1,...,i_n\rangle \langle j_1,...,j_n| 
\end{equation}
which encodes \textit{spacetime correlation functions} of the main system.  This is our first example of a \textit{superdensity operator}.  This particular example is also a multitime correlation operator as per \cite{Lindblad1}--\cite{Fannes1}.  

An important feature of the superdensity operator in Eqn.~\eqref{superdensity1} is that we can use a quantum channel (that is, a completely positive trace-preserving map) to map it to the quantum state of the original system.  Details of such a channel are in Appendix \ref{sec:AppA}.  It is useful to keep in mind that $\varrho$ is not just an abstract object -- it can be can obtained by coupling an evolving system to auxiliary apparatus.  We explain a more detailed procedure for measuring superdensity operators in Section 3 below.

Next, we will provide a more abstract motivation for the superdensity operator.

\subsubsection{Spacetime states}

In quantum mechanics, the description of a system at a fixed time is captured by a density operator $\rho$.  Correlations of the system can be extracted by measuring the expectation values of operators $\mathcal{O}$.  These expectation values are computed by taking the trace $\tr(\mathcal{O} \, \rho)$.

More generally, we might compute multiple-time correlation functions like
\begin{equation}
\tr\bigg(\mathcal{O}_i \, U(t_2,t_1) \, \mathcal{O}_j\,\rho_0 \, \mathcal{O}_k^\dagger \, U(t_2,t_1)^\dagger \, \mathcal{O}_\ell^\dagger\bigg)
\end{equation}
where $\rho_0$ is the initial state and $U(t_2,t_1)$ is the unitary time evolution.  Such correlation functions have the added feature that operator probes in the past affect the outcome of future operator probes.  We would like to find a generalization of the density operator $\rho$ which captures the measurable outcome of a sequence probes at \textit{multiple times}, and hence encodes spacetime correlations.  To do so, we first need to find the right mathematical structure.

Before generalizing the density operator, we need to think about it in a slightly different way than usual.  It is convenient to capture the data of a density operator $\rho$ from a dual perspective, namely as a map from operators to numbers:
\begin{equation}
\tr(\,\boldsymbol{\cdot}\,\,\rho) : \, \mathcal{O} \, \longmapsto \, \tr(\mathcal{O} \, \rho)
\end{equation}
where $\tr(\,\boldsymbol{\cdot}\,\,\rho)$ is a ``trace with a slot,'' into which an operator can be inserted to output an expectation value.  In more physical terms, $\tr(\,\boldsymbol{\cdot}\,\,\rho)$ takes as input an operator to be measured, and outputs the measured expectation value with respect to that operator.  To highlight the mathematical similarity between the objects $\rho$ and $\tr(\,\boldsymbol{\cdot}\,\,\rho)$, we consider expanding each in terms of a complete orthonormal basis of operators $\{X_i\}$ of dimension $d$, satisfying $\tr(X_i^\dagger X_j) = \delta_{ij}$\,:
\begin{align}
\label{rhoexpand1}
\rho &= \sum_i \tr(X_i \, \rho) \, X_i \\
\label{dualrhoexpand1}
\tr(\,\boldsymbol{\cdot}\,\,\rho) &= \sum_i \tr(X_i \, \rho) \, \tr(\,\boldsymbol{\cdot}\,\,X_i)\,.
\end{align}
In words, to translate from $\rho$ to $\tr(\,\boldsymbol{\cdot}\,\,\rho)$, we replace each basis element $X_i$ with the dual map $\tr(\,\boldsymbol{\cdot}\,\,X_i)$ which takes $\mathcal{O}\mapsto \tr(\mathcal{O}\, X_i)$.

Although $\rho$ and $\tr(\,\boldsymbol{\cdot}\,\,\rho)$ capture the same data, there is a conceptual difference as to how each object accesses that data.  For example, $\rho$ answers the question, ``What is the description of the system?'' whereas $\tr(\,\boldsymbol{\cdot}\,\,\rho)$ answers the question ``What information can we \textit{extract} from the system via measurement?''  This latter perspective inspires a temporal generalization.

As mentioned above, the outcome of probing a system at different times is a multiple-time expectation value such as $\tr\left(\mathcal{O}_i \, U(t_2,t_1) \, \mathcal{O}_j\,\rho_0 \, \mathcal{O}_k^\dagger \, U(t_2,t_1)^\dagger \, \mathcal{O}_\ell^\dagger\right)$.  More broadly, we can consider
\begin{equation}
\sum_{\substack{i_1,...,i_n \\ j_1,...,j_n}} c_{\substack{i_1,...,i_n \\ j_1,...,j_n}}\tr\bigg(X_{i_n} U(t_n,t_{n-1}) X_{i_{n-1}} \, ... \, X_{i_2} \, U(t_2, t_1)\,  X_{i_1} \, \rho_0 \, X_{j_1}^\dagger \, U(t_1, t_2)^\dagger X_{j_2}^\dagger\,...\, X_{j_{n-1}}^\dagger\, U(t_n,t_{n-1})^\dagger \, X_{j_n}^\dagger\bigg)
\end{equation}
which is an expectation value with operators at $n$ times $t_1,...,t_n$.  Considering these more general types of expectation values, we might ask, ``What information can we extract from the system by probing it at \textit{multiple} times with auxiliary systems and then measuring those auxiliary systems?''  The object which gives us the desired information is
\begin{equation}
\label{tracewithslots2}
\tr\bigg(\,\boldsymbol{\cdot}\, U(t_n,t_{n-1})\,\boldsymbol{\cdot} \, ... \, \boldsymbol{\cdot} \, U(t_2, t_1)\,  \boldsymbol{\cdot} \, \rho_0 \, \boldsymbol{\cdot} \, U(t_1, t_2)^\dagger \boldsymbol{\cdot}\,...\, \boldsymbol{\cdot}\, U(t_n,t_{n-1})^\dagger \boldsymbol{\cdot}\,\bigg)
\end{equation}
which is a ``trace with $2n$ slots.''  This object takes as input a sequence of operators which probe the system at a sequence of times, and outputs the expected value of measurement at the final time.

To more carefully understand this object, it is convenient to define the history Hilbert space
\begin{equation}
\mathcal{H}_{\text{hist.}} := \mathcal{H}_{t_n}\otimes \mathcal{H}_{t_{n-1}} \otimes \cdots \otimes \mathcal{H}_{t_2} \otimes \mathcal{H}_{t_1}\,.
\end{equation}
Next, we consider the space of bounded operators on $\mathcal{H}_{\text{hist}.}$ which we denote by $\mathcal{B}(\mathcal{H}_{\text{hist}.})$.  We can write any operator in $\mathcal{B}(\mathcal{H}_{\text{hist}.})$ as
\begin{equation}
\sum_{i_1,i_2,...,i_n} c_{i_1,i_2,...,i_n} \, X_{i_n} \otimes \cdots \otimes X_{i_2} \otimes X_{i_1}
\end{equation}
for some coefficients $c_{i_1,i_2,...,i_n}$.

Then Eqn.~\eqref{tracewithslots2} can be thought of as a \textit{superdensity operator} $\varrho$, which is a bilinear map
\begin{equation}
\label{superaction1}
\varrho : \mathcal{B}(\mathcal{H}_{\text{hist.}}) \otimes \mathcal{B}^*(\mathcal{H}_{\text{hist.}}) \longrightarrow \mathbb{C}
\end{equation}
that acts by
\begin{align}
\label{superaction2}
&\varrho[W_1 \otimes W_2 \otimes \cdots \otimes W_n\,,\,V_1 \otimes V_2 \otimes \cdots \otimes V_n ] \nonumber \\
=\,\,&\frac{1}{\dim \mathcal{H}_{\text{hist.}}} \, \tr\bigg(W_n \, U(t_n,t_{n-1}) \,W_{n-1} \, ... \, W_2 \, U(t_2, t_1)\,  W_1 \, \rho_0 \, V_1^\dagger \, U(t_1, t_2)^\dagger\, V_2^\dagger\,...\, V_{n-1}^\dagger\, U(t_n,t_{n-1})^\dagger \, V_{n}^\dagger\bigg)
\end{align}
and extends to more general operators in $\mathcal{B}(\mathcal{H}_{\text{hist.}}) \otimes \mathcal{B}^*(\mathcal{H}_{\text{hist.}})$ by multilinearity with respect to each $\mathcal{B}(\mathcal{H}_{t_i})$ and $\mathcal{B}^*(\mathcal{H}_{t_i})$.  A diagrammatic representation of $\varrho$ is given in Figure \ref{TN1} below.
${}$  \\
\begin{center}
\customlabel{TN1}{1}
\includegraphics[scale=.6]{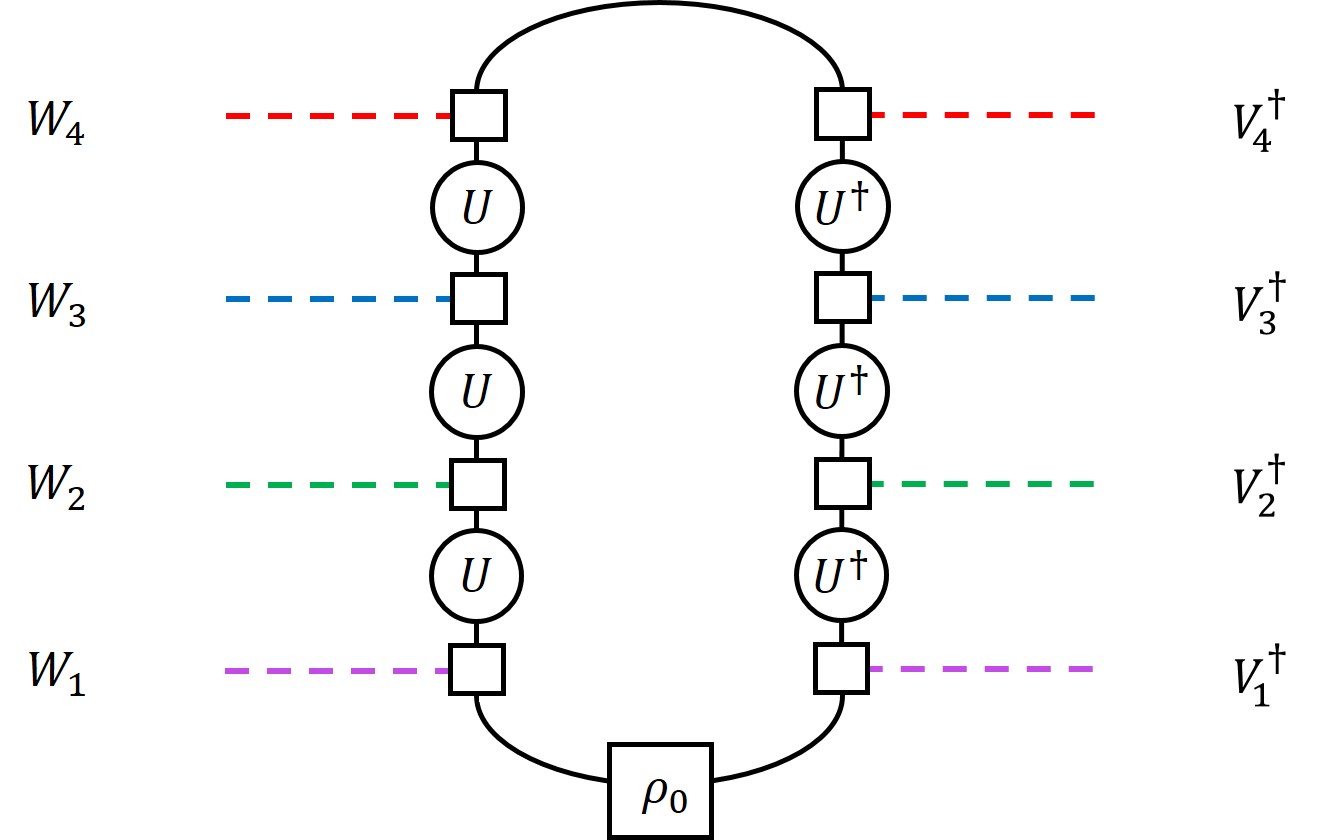}\\
\end{center}
\vskip.7cm
Figure 1: Diagrammatic representation of the superdensity operator $\varrho$.  Operators are contracted with the superdensity operator by inserting them into the dotted lines.  We label where the $W_i$ and $V_j^\dagger$ operators would be contracted.
\\ \\ \\
The prefix ``super'' in superdensity operator denotes that it acts on a vector space of operators.  By contrast, a standard density operator is a map $\rho : \mathcal{H} \otimes \mathcal{H}^* \to \mathbb{C}$.  There are more general types of superdensity operators than the form given in Eqn.~\eqref{superaction1}.   We will examine other types below, after giving a general definition of superdensity operators.

The superdensity operator $\varrho$ has the novel feature that it encodes the time evolution of a system.  Thus, it is a natural temporally extended analog of the usual density operator.  We will show that several naturally defined properties of the superdensity operator reveal physically significant properties of the system it describes, making the superdensity operator an attractive focus for study in quantum information.

\subsection{Operator bra-ket notation}

Since the superdensity formalism makes heavy use of mappings from operators to complex numbers, it is convenient to introduce a version of bra-ket notation for operators, to make expressions more compact.  We will denote an operator $W$ in $\mathcal{B}(\mathcal{H})$ by $|W)$ and an element $\tr(V^\dagger\,\,\boldsymbol{\cdot}\,)$ of the dual space $\mathcal{B}^*(\mathcal{H})$ by $(V|$ so that
\begin{equation}
(V|W) := \tr(V^\dagger \, W)\,.
\end{equation}
The soft bracket is used to distinguish this notation from the usual bra-ket notation for quantum states.  We have the linearity relations
\begin{align}
c_1 \, |A) + c_2 \, |B) &= |c_1 \, A + c_2 \, B) \\ \nonumber \\
c_1 \, (A| + c_2 \, (B| &= (c_1^* \, A + c_2^* \, B|
\end{align}
which can be extended to multilinear relations by taking tensor products like
\begin{align}
&\sum_{i_1,i_2,...,i_n} c_{i_1, i_2,...,i_n }|A_{i_n}) \otimes \cdots \otimes |A_{i_2})\otimes |A_{i_1}) \quad \text{or} \quad\sum_{i_1,i_2,...,i_n} c_{i_1, i_2,...,i_n }(A_{i_n}| \otimes \cdots \otimes (A_{i_2}|\otimes (A_{i_1}|\,.
\end{align}
Hermitian conjugation acts by
\begin{equation}
|W)^\dagger = (W| \quad \text{and} \quad (W|^\dagger = |W)\,.
\end{equation}

If $\{|X_i)\}$ is an orthonormal basis of operators so that $(X_i|X_j) = \tr(X_i^\dagger X_j) = \delta_{ij}$, we can write a resolution of the identity \textit{on the space of operators} as
\begin{equation}
\textbf{1} = \sum_i |X_i)(X_i|\,.
\end{equation}
For example, applying the identity on the space of operators to $|\rho)$ gives
\begin{equation}
|\rho) = \sum_i |X_i)(X_i|\rho) = \sum_i \tr(X_i \, \rho) \, |X_i)
\end{equation}
which is the same as Eqn.~\eqref{rhoexpand1}, and
\begin{equation}
(\rho| = \sum_i \tr(X_i \, \rho) \, (X_i|
\end{equation}
which is just Eqn.~\eqref{dualrhoexpand1}.  In similar fashion, we can write the superdensity operator $\varrho$ defined by Eqn.~\eqref{superaction2} as
\begin{align}
\label{sdexpand1}
\varrho &= \frac{1}{d^{2}}\sum_{\substack{i_1,...,i_n \\ j_1,...,j_n}} \tr\left(X_{i_n} \, U(t_n,t_{n-1}) \cdots U(t_2,t_1)\, X_{i_1}\,\rho_0\, X_{j_1}^\dagger U(t_1,t_2)^\dagger \cdots U(t_n,t_{n-1})^\dagger X_{j_n}^\dagger\right) \nonumber \\
& \qquad \qquad \qquad \qquad \qquad \qquad \qquad \qquad \qquad \qquad \qquad \times |X_{i_1})(X_{j_1}| \otimes \cdots \otimes |X_{i_n})(X_{j_n}|\,.
\end{align}
This is formally the same as Eqn.~\eqref{superdensity1}, but with $|i\rangle \to |X_i)$.

In operator bra-ket notation, traces and partial traces are straightforward.  For example, the trace of a superoperator $\mathcal{S} : \mathcal{B}(\mathcal{H}) \to \mathcal{B}(\mathcal{H})$ is given by
\begin{equation}
\text{tr}(\mathcal{S}) = \sum_{i=1}^{\dim \mathcal{B}(\mathcal{H})} (X_i| \mathcal{S} |X_i) 
\end{equation}
where we note that $\dim \mathcal{B}(\mathcal{H}) = (\dim \mathcal{H})^2$.  Similarly, if $\mathcal{B}(\mathcal{H}) = A \otimes B$, the partial trace with respect to $A$ is given by
\begin{equation}
\text{tr}_A(\mathcal{S}) = \sum_{i=1}^{\dim A} \bigg((Y_i|_A\otimes \textbf{1}_B\bigg) \, \mathcal{S} \,  \bigg(
|Y_i)_A\otimes \textbf{1}_B\bigg)
\end{equation}
where $\{|Y_i)_A\}$ is a complete orthonormal basis for $A$ satisfying $\tr(Y_i^\dagger Y_j) = \delta_{ij}$.

Although a superdensity operator $\varrho$ is formally a bilinear map
$$\varrho : \mathcal{B}(\mathcal{H}_{\text{hist.}}) \otimes \mathcal{B}^*(\mathcal{H}_{\text{hist.}}) \longrightarrow \mathbb{C}$$
we can also treat is as a superoperator
$$\varrho : \mathcal{B}(\mathcal{H}_{\text{hist.}}) \longrightarrow \mathcal{B}(\mathcal{H}_{\text{hist.}})$$
much in the same way an operator (or matrix) can be thought of as either a bilinear form, or as a map from vectors to vectors.  Indeed, the trace and partial trace operations are well-defined for the superdensity operator, and tracing out tensor factors corresponds to losing the ability to probe the system with specified spacetime observables.

\subsection{Superdensity operator definition}
\label{sec:definition}

Having established our motivation and convenient notation, we are ready to give an appropriate, general definition of superdensity operators.  Let $\mathcal{B}$ denote a space of bounded operators (which we typically take to be $\mathcal{B} = \mathcal{B}(\mathcal{H}_{\text{hist.}})$) and let $\mathcal{B}^*$ be its dual space.  Then we have: \\

\noindent \textbf{Definition (superdensity operator):} \textit{A superdensity operator $\varrho$ is a bilinear form}
\begin{equation*}
\varrho : \mathcal{B} \otimes \mathcal{B}^* \longrightarrow \mathbb{C}
\end{equation*}
\textit{satisfying the conditions:} \\ \\
\indent 1. $\varrho^\dagger = \varrho$ \qquad \qquad \qquad \qquad \qquad \qquad \qquad \qquad \quad (\textit{Hermitian}) \\ \\
\indent 2. $\varrho \succeq 0 $\,, \textit{meaning $(W|\varrho|W) \geq 0$ for all} $|W)$ \qquad \!\!\!(\textit{positive semi-definite}) \\ \\
\indent 3. $\tr(\varrho) = 1$ \qquad \qquad \qquad \qquad \qquad \qquad \qquad \quad \,\,\,\,\,(\textit{unit trace}) \\ \\ \\
\indent Note that the definition of the superdensity operator is analogous to that of the standard density operator, and so many of the standard properties of density operators carry over.  For example, the partial trace of a superdensity operator is also a superdensity operator.

While Eqn.~\eqref{sdexpand1} is an example of a superdensity operator, it is not the most general form that a superdensity operator can have.  Nonetheless, let us check that Eqn.~\eqref{sdexpand1} satisfies the properties given in the definition above.  Hermiticity and positive semi-definiteness are manifest, so it remains to check if $\varrho$ has unit trace.  Taking the trace, we find
\begin{align}
\label{sdexpand2}
\tr(\varrho) &= \frac{1}{\dim \mathcal{H}_{\text{hist.}}}\sum_{\substack{i_1,...,i_n \\ j_1,...,j_n}} \tr\left(X_{i_n} \, U(t_n,t_{n-1}) \cdots U(t_2,t_1)\, X_{i_1}\,\rho_0\, X_{j_1}^\dagger U(t_1,t_2)^\dagger \cdots U(t_n,t_{n-1})^\dagger X_{j_n}^\dagger\right) \nonumber \\
& \qquad \qquad \qquad \qquad \qquad \qquad \qquad \qquad \qquad \qquad \qquad \qquad \qquad \qquad \times \delta_{i_1,j_1} \cdots \delta_{i_n, j_n} \nonumber \\ \nonumber \\
&= \frac{1}{\dim \mathcal{H}_{\text{hist.}}}\sum_{\substack{i_1,...,i_n}} \tr\left(X_{i_n} \, U(t_n,t_{n-1}) \cdots U(t_2,t_1)\, X_{i_1}\,\rho_0\, X_{i_1}^\dagger U(t_1,t_2)^\dagger \cdots U(t_n,t_{n-1})^\dagger X_{i_n}^\dagger\right)\,.
\end{align}
Note that $\dim \mathcal{H}_{\text{hist.}} = \dim \mathcal{H}_{t_1} \cdots \dim \mathcal{H}_{t_n}$.  Since $\dim \mathcal{H}_{t_i}$ is the same for all $i$, we can define $d := \dim \mathcal{H}_{t_i}$ so that $\dim \mathcal{H}_{\text{hist.}} = d^n$.  

Recalling that $\{X_i\}$ forms an orthonormal basis (e.g., $\tr(X_i^\dagger X_j) = \delta_{ij}$), we have the completeness relation
$$\frac{1}{d}\sum_i X_i^\dagger X_i = \textbf{1}\,.$$
Using the cyclicity of the trace in the last line of Eqn.~\eqref{sdexpand2}, we can pull around the $X_{i_n}^\dagger$ to get a $X_{i_n}^\dagger X_{i_n}$ term and so:
\newpage
\begin{align}
& \frac{1}{d^{n}}\sum_{\substack{i_1,...,i_n}} \tr\left(X_{i_n} \, U(t_n,t_{n-1}) \cdots U(t_2,t_1)\, X_{i_1}\,\rho_0\, X_{i_1}^\dagger U(t_1,t_2)^\dagger \cdots U(t_n,t_{n-1})^\dagger X_{i_n}^\dagger\right) \nonumber \\
=\, & \frac{1}{d^{n-1}}\sum_{\substack{i_1,...,i_{n-1}}} \tr\left(\frac{1}{d}\sum_{i_n} X_{i_n}^\dagger X_{i_n} \, U(t_n,t_{n-1}) \cdots U(t_2,t_1)\, X_{i_1}\,\rho_0\, X_{i_1}^\dagger U(t_1,t_2)^\dagger \cdots U(t_n,t_{n-1})^\dagger \right) \nonumber \\
=\, & \frac{1}{d^{n-1}}\sum_{\substack{i_1,...,i_{n-1}}} \tr\left(X_{i_{n-1}} \, U(t_{n-1},t_{n-2}) \cdots U(t_2,t_1)\, X_{i_1}\,\rho_0\, X_{i_1}^\dagger U(t_1,t_2)^\dagger \cdots U(t_{n-1},t_{n-2})^\dagger\,X_{i_{n-1}}^\dagger \right)
\end{align} 
where we have again used the cyclicity of the trace to cancel out $U(t_n,t_{n-1})$ and $U(t_n, t_{n-1})^\dagger$ in going from the second to third lines.  Iterating the procedure $n-1$ more times, we are left with
$$\tr(\varrho) = \tr(\rho_0) = 1\,.$$
So indeed, the type of superdensity operator given in Eqn.~\eqref{sdexpand1} satisfies the conditions given in the formal definition.

The superdensity operator given in Eqn.~\eqref{sdexpand1} also has additional structure which is very useful.  Consider the decomposition
\begin{equation}
B(\mathcal{H}_{\text{hist.}}) = \mathcal{B}(\mathcal{H}_{t_n}) \otimes \cdots \otimes \mathcal{B}(\mathcal{H}_{t_2}) \otimes \mathcal{B}(\mathcal{H}_{t_1})\,.
\end{equation}
Furthermore, we decompose each $\mathcal{B}(\mathcal{H}_{t_k})$ into $\mathcal{B}(\mathcal{H}_{t_k}) = A_k \otimes B_k$, where $A_k$ and $B_k$ form unital subalgebras of operators (i.e., they contain an identity with respect to multiplication) acting on $\mathcal{H}_{t_k}$.  Letting $A = A_n \otimes \cdots \otimes A_2 \otimes A_1$ and $B = B_n \otimes \cdots \otimes B_2 \otimes B_1$, we can write
\begin{equation}
\mathcal{B}(\mathcal{H}_{\text{hist.}}) = A \otimes B\,.
\end{equation}
We will show that the superdensity operator $\varrho$ given in Eqn.~\eqref{sdexpand1} has a \textit{restriction property} with respect to the spacetime subalgebras $A$ and $B$, namely that the \textit{restriction to} $A$ given by
\begin{equation}
\label{restriction0}
\big(\textbf{1}_{A} \otimes (\textbf{1}|_{B}\big) \, \varrho \,\big(\textbf{1}_{A} \otimes |\textbf{1})_{B}\big)
\end{equation}
is \textit{also} a superdensity operator, and similarly if we instead restricted to $B$.  The intuition behind Eqn.~\eqref{restriction0} is that inserting $m$ identity operators into an $n$-point function should yield an $(n-m)$--point function.  The restriction in Eqn.~\eqref{restriction0} is filling up the ``$B$ slots'' with identity operators, but does not affect the ``$A$ slots.''  To be clear, note that in the expression $\big(\textbf{1}_{A} \otimes (\textbf{1}|_{B}\big) \, \varrho \,\big(\textbf{1}_{A} \otimes |\textbf{1})_{B}\big)$\,:
\begin{itemize}
\item $\textbf{1}_{A}$ acts on $|Y)_{A}$ for $Y \in A$ by $\textbf{1}_{A} |Y)_{A} = |Y)_{A}$\,;
\item $\textbf{1}_{A}$ acts on $(Y|_{A}$ for $Y \in A$ by $(Y|_{A} \textbf{1}_{A} = (Y|_{A}$\,;
\item  $(\textbf{1}|_{B}$ acts on $|Z)_{B}$ for $Z \in B$ by $(\textbf{1}|_{B} |Z)_{B} = \text{tr}(Z)$\.;
\item  $|\textbf{1})_{B}$ acts on $(Z|_{B}$ for $Z \in B$ by $(Z|_{B} |\textbf{1})_{B} = \text{tr}(Z^\dagger)$\,.
\end{itemize}
Thus, if we write $\varrho = \sum_{i,j,k,\ell} c_{ijk\ell} \, |Y_i)_{A} (Y_j|_{A} \otimes |Z_k)_{B} (Z_\ell|_{B}$, then
\begin{equation}
\big(\textbf{1}_{A} \otimes (\textbf{1}|_{B}\big) \, \varrho \,\big(\textbf{1}_{A} \otimes |\textbf{1})_{B}\big) = \sum_{i,j,k,\ell} c_{ijk\ell} \, |Y_i)_{A} (Y_j|_{A} \cdot \tr(Z_k) \, \tr(Z_\ell^\dagger)\,.
\end{equation}
If $Z_i$\,, $i=0,...,\dim B - 1$ is an orthonormal basis of $B$ such that $Z_0 = \textbf{1}/\sqrt{\tr(\textbf{1})}$, then $\tr(Z_0^\dagger \, Z_i) = \tr(Z_i) = \delta_{i,0}$\,, and so the above equation would simplify to
\begin{equation}
\big(\textbf{1}_{A} \otimes (\textbf{1}|_{B}\big) \, \varrho \,\big(\textbf{1}_{A} \otimes |\textbf{1})_{B}\big) = \sum_{i,j} c_{ij00} \, |Y_i)_{A} (Y_j|_{A} \,.
\end{equation}

Now let us check the restriction property of $\varrho$ given in Eqn.~\eqref{sdexpand1}.  Namely, we will show that $\big(\textbf{1}_{A} \otimes (\textbf{1}|_{B}\big) \, \varrho \,\big(\textbf{1}_{A} \otimes |\textbf{1})_{B}\big)$ is a superdensity operator.  Suppose we have $\mathcal{B}(\mathcal{H}_{\text{hist.}}) = A \otimes B$ as above, and that $\dim \mathcal{H}_{t_1} = \cdots = \dim \mathcal{H}_{t_n} = d$.  It follows that $\dim \mathcal{B}(\mathcal{H}_{\text{hist.}}) = d^{2n}$.  Further suppose that $\dim A_k = d_{A_k}$ and $\dim B_k = d_{B_k}$.  Then there is an orthonormal basis of $\mathcal{B}(\mathcal{H}_{\text{hist.}})$ of the form
\begin{align}
&\left(X_{i_{n}^{A_n}}^{A_n} \otimes X_{i_n^{B_n}}^{B_n} \right) \otimes \cdots \otimes \left(X_{i_2^{A_2}}^{A_2} \otimes X_{i_2^{B_2}}^{B_2} \right) \otimes \left(X_{i_1^{A_1}}^{A_1} \otimes X_{i_1^{B_1}}^{B_1} \right) \nonumber \\ \nonumber \\
=& \left(X_{i_n^{A_n}}^{A_n} \otimes \textbf{1}_{B_n} \right) \otimes \cdots \otimes \left(X_{i_2^{A_2}}^{A_2} \otimes \textbf{1}_{B_2} \right) \otimes \left(X_{i_1^{A_1}}^{A_1} \otimes \textbf{1}_{B_1} \right) \nonumber \\ \nonumber \\
& \qquad \quad \cdot \,\,\left(\textbf{1}_{A_n} \otimes X_{i_n^{B_n}}^{B_n} \right) \otimes \cdots \otimes \left(\textbf{1}_{A_2} \otimes X_{i_2^{B_2}}^{B_2} \right) \otimes \left(\textbf{1}_{A_1} \otimes X_{i_1^{B_1}}^{B_1} \right)
\end{align}
for $i_k^{A_k} = 0,...,d_{A_k} - 1$ and $i_k^{B_k} = 0,...,d_{B_k} - 1$ where $d_{A_k} \cdot d_{B_k} = d^2$ for all $k$.  Indeed, we have
\begin{align*}
\left(X_{i_n^{A_n}}^{A_n} \otimes \textbf{1}_{B_n} \right) \otimes \cdots \otimes \left(X_{i_2^{A_2}}^{A_2} \otimes \textbf{1}_{B_2} \right) \otimes \left(X_{i_1^{A_1}}^{A_1} \otimes \textbf{1}_{B_1} \right) \,\, &\in A \\ \\
\left(\textbf{1}_{A_n} \otimes X_{i_n^{B_n}}^{B} \right) \otimes \cdots \otimes \left(\textbf{1}_{A_2} \otimes X_{i_2^{B_2}}^{B_2} \right) \otimes \left(\textbf{1}_{A_1} \otimes X_{i_1^{B_1}}^{B_1} \right) \,\, & \in B\,.
\end{align*}
Then we have
\begin{align}
\label{restriction1}
&\big(\textbf{1}_{A} \otimes (\textbf{1}|_{B}\big) \, \varrho \,\big(\textbf{1}_{A} \otimes |\textbf{1})_{B}\big) \nonumber \\
&= \frac{1}{\sqrt{d_{A_1} \cdots d_{A_n}}}\sum_{\substack{i_1^{A_1},...,i_n^{A_n} \\  j_1^{A_1},...,j_n^{A_n}}} \tr\left(X_{i_n^{A_n}}^{A_n} \, U(t_n,t_{n-1}) \cdots U(t_2,t_1)\, X_{i_1^{A_1}}^{A_1}\,\rho_0\, X_{j_1^{A_1}}^{A_1\,\dagger} U(t_1,t_2)^\dagger \cdots U(t_n,t_{n-1})^\dagger X_{j_n^{A_n}}^{A_n\,\dagger}\right) \nonumber \\
& \qquad \qquad \qquad \qquad \qquad \qquad \qquad \qquad \qquad \qquad \qquad \times |X_{i_1^{A_1}}^{A_1})(X_{j_1^{A_1}}^{A_1}| \otimes \cdots \otimes |X_{i_n^{A_n}}^{A_n})(X_{j_n^{A_n}}^{A_n}|\,.
\end{align}
It is clear that $\big(\textbf{1}_{A} \otimes (\textbf{1}|_{B}\big) \, \varrho \,\big(\textbf{1}_{A} \otimes |\textbf{1})_{B}\big)$ is Hermitian and positive definite, so let us show that it has unit trace.  We find
\begin{align}
& \tr \big(\textbf{1}_{A} \otimes (\textbf{1}|_{B}\big) \, \varrho \,\big(\textbf{1}_{A} \otimes |\textbf{1})_{B}\big) \nonumber \\
=&\, \frac{1}{\sqrt{d_{A_1} \cdots d_{A_n}}}\sum_{\substack{i_1^{A_1},...,i_n^{A_n}}} \tr\left(X_{i_n^{A_n}}^{A_n} \, U(t_n,t_{n-1}) \cdots U(t_2,t_1)\, X_{i_1^{A_1}}^{A_1}\,\rho_0\, X_{i_1^{A_1}}^{A_1\,\dagger} U(t_1,t_2)^\dagger \cdots U(t_n,t_{n-1})^\dagger X_{i_n^{A_n}}^{A_n\,\dagger}\right) \nonumber \\
=&\, \tr\left(\tr_{A_n}\left[ U(t_n,t_{n-1}) \cdots \left[\tr_{A_2}\left[U(t_2,t_1)\, \left[\tr_{A_1}(\rho_0) \otimes \frac{\textbf{1}_{A_1}}{d_{A_1}} \right] U(t_1,t_2)^\dagger\right] \otimes \frac{\textbf{1}_{A_2}}{d_{A_2}}\right] \cdots U(t_n,t_{n-1})^\dagger \right] \otimes \frac{\textbf{1}_{A_n}}{d_{A_n}}\right) \nonumber \\
=& \, 1\,.
\end{align}
Therefore $\varrho$ satisfies the restriction property with respect to $A$.  Indeed, the proof that $\varrho$ satisfies the restriction property with respect to $B$ is identical, with the roles of $A$ and $B$ switched.

There are many other types of superdensity operators beyond the form given in Eqn.~\eqref{sdexpand1}. We will now explore several examples, in the following section.

\subsection{Examples of superdensity operators}

Here we collect several examples of superdensity operators to demonstrate their range of utility. \\

\noindent \textbf{Example 1:} We repeat for completeness the form of superdensity operator given by Eqn.~\eqref{sdexpand1} above:
\begin{align*}
\varrho &= \frac{1}{\dim \mathcal{H}_{\text{hist.}}}\sum_{\substack{i_1,...,i_n \\ j_1,...,j_n}} \tr\left(X_{i_n} \, U(t_n,t_{n-1}) \cdots U(t_2,t_1)\, X_{i_1}\,\rho_0\, X_{j_1}^\dagger U(t_1,t_2)^\dagger \cdots U(t_n,t_{n-1})^\dagger X_{j_n}^\dagger\right) \nonumber \\
& \qquad \qquad \qquad \qquad \qquad \qquad \qquad \qquad \qquad \qquad \qquad \times |X_{i_1})(X_{j_1}| \otimes \cdots \otimes |X_{i_n})(X_{j_n}|\,.
\end{align*}
A diagrammatic representation can be seen in Figure \ref{TN1} above.   \\ \\
\noindent \textbf{Example 2:}  Suppose that $\mathcal{C}_i : \mathcal{B}(\mathcal{H}_{t_i}) \to \mathcal{B}(\mathcal{H}_{t_{i+1}})$ is a quantum channel.  (Recall that a quantum channel is a completely positive trace preserving (CPTP) map.)  This means that $\mathcal{C}_i \otimes \textbf{1}_{n\times n}$ maps positive operators to positive operators for all $n$, and $\mathcal{C}_i$ preserves the trace of operators.  Then the following is a superdensity operator:
\begin{align}
\label{channelnesting1}
\varrho &=  \frac{1}{\dim \mathcal{H}_{\text{hist.}}}\sum_{\substack{i_1,...,i_n \\ j_1,...,j_n}} \tr\left(X_{i_n} \, \mathcal{C}_{n-1}[ \cdots \mathcal{C}_1[\, X_{i_1}\,\rho_0\, X_{j_1}^\dagger ] \cdots ] X_{j_n}^\dagger\right)\, |X_{i_1})(X_{j_1}| \otimes \cdots \otimes |X_{i_n})(X_{j_n}|\,.
\end{align}
In words, we can replace the unitary evolution in Example 1 by quantum channel evolution, which is more general.  A diagram of this superdensity operator is shown below in Figure \ref{TN2}\,:
\vskip.8cm
\begin{center}
\customlabel{TN2}{2}
\includegraphics[scale=.5]{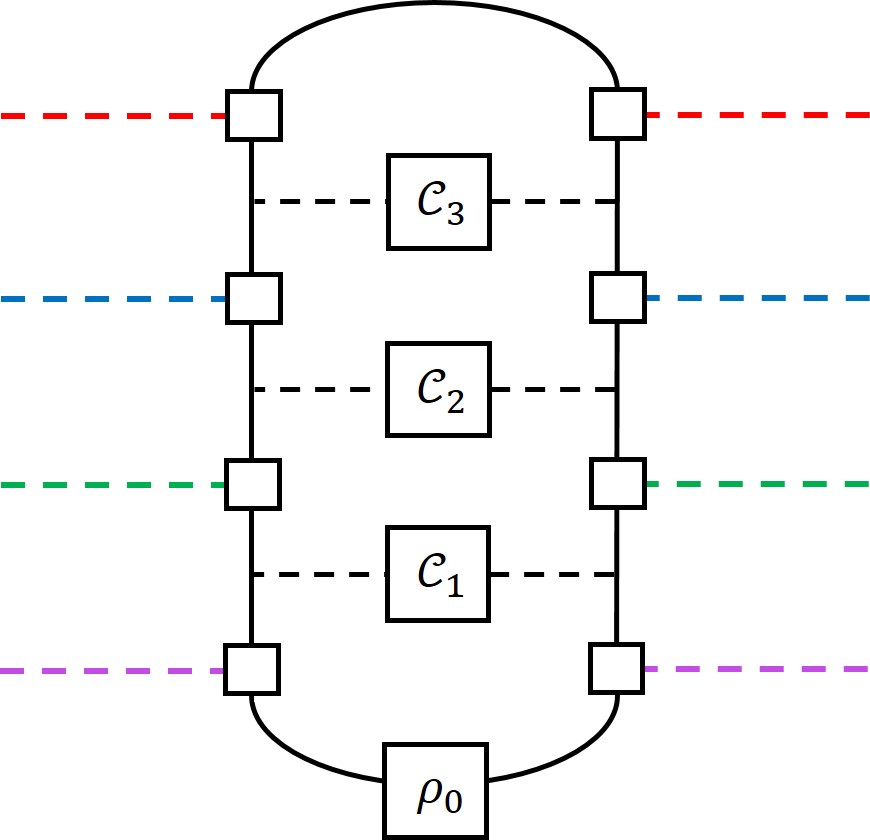}\\
\vskip1cm
Figure 2: Diagram for a superdensity operator with evolution by quantum channels.
\end{center}
\vskip.4cm

\indent Instead of constructing sequentially nested quantum channels as per Eqn.~\eqref{channelnesting1}, we can consider \textit{spacetime} quantum channels that act at \textit{more} than one time, and hence are non-local in time.  Nonetheless, we still get a valid superdensity operator -- a diagram is shown in Figure \ref{TN3}.
\vskip.7cm
\begin{center}\customlabel{TN3}{3}
\includegraphics[scale=.5]{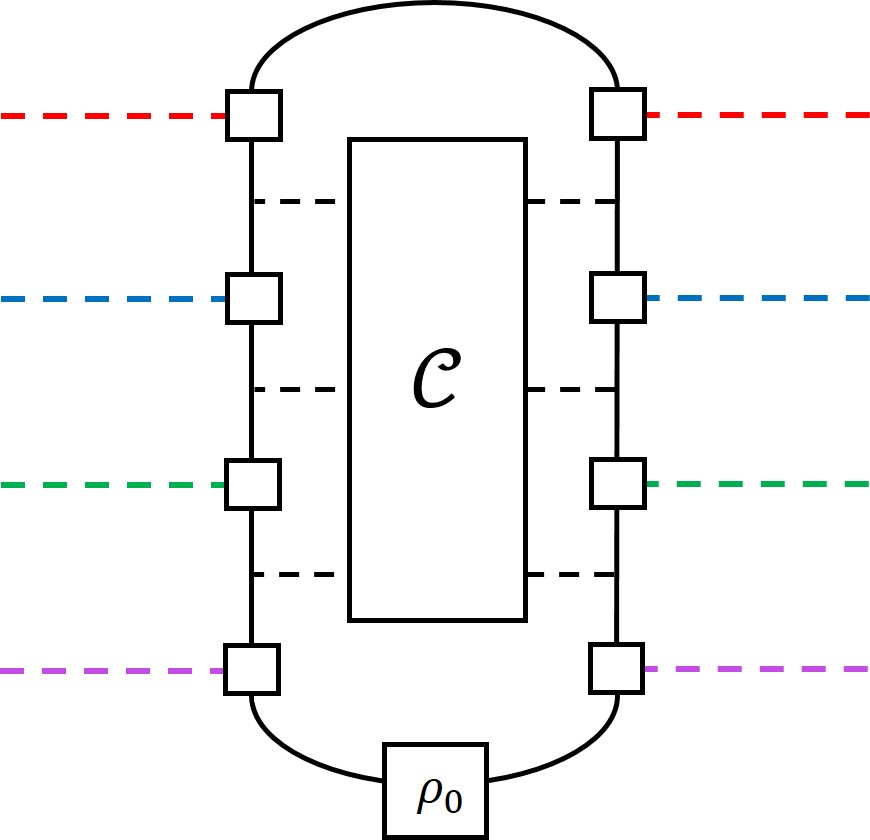}\\
\vskip1.1cm
Figure 3: Superdensity operator with a spacetime quantum channel.
\end{center}
\vskip.4cm
To construct this object algebraically, it is convenient to start with the superdensity operator:
\newpage
\begin{align}
\varrho &=  \frac{1}{\dim \mathcal{H}_{\text{hist.}}}\sum_{\substack{i_1,...,i_{2n-1} \\ j_1,...,j_{2n-1}}} \tr\left(X_{i_{2n-1}} \, X_{i_{2n-2}} \cdots\, X_{i_2}\,X_{i_1}\,\rho_0\, X_{j_1}^\dagger X_{j_2}^\dagger \cdots X_{j_{2n-2}}^\dagger \, X_{j_{2n-1}}^\dagger\right) \nonumber \\
& \qquad \qquad \qquad \qquad \qquad \qquad \qquad \qquad \qquad \qquad \qquad \times |X_{i_1})(X_{j_1}| \otimes \cdots \otimes |X_{i_{2n-1}})(X_{j_{2n-1}}|\,.
\end{align}
\vskip-.2cm
Note that this superdensity operator has no unitary time evolution -- it is instead comprised of $4n-2$ empty slots.  This superdensity operator is a bilinear form which maps $\mathcal{B}(\mathcal{H}) \otimes \mathcal{B}^*(\mathcal{H}) \to \mathbb{C}$, where $\mathcal{H} = \mathcal{H}_{2n-1} \otimes \mathcal{H}_{2n-2} \otimes \cdots \otimes \mathcal{H}_2 \otimes \mathcal{H}_1 \simeq \mathcal{H}_{\text{even}} \otimes \mathcal{H}_{\text{odd}}$.  Here we have used the notation $\mathcal{H}_{\text{even}} := \mathcal{H}_{2n-2} \otimes \mathcal{H}_{2n-4} \otimes \cdots \otimes \mathcal{H}_{4} \otimes \mathcal{H}_2$ and $\mathcal{H}_{\text{odd}} := \mathcal{H}_{2n-1} \otimes \mathcal{H}_{2n-3} \otimes \cdots \otimes \mathcal{H}_{3} \otimes \mathcal{H}_1$.

Now suppose we have a quantum channel $\mathcal{C} : \mathcal{B}(\mathcal{H}_{\text{even}}) \to \mathcal{B}(\mathcal{H}_{\text{even}})$.  We let $\mathcal{C}$ act as the identity on $\mathcal{B}(\mathcal{H}_{\text{odd}})$.  In operator bra-ket notation, $\mathcal{C}$ can be written as
\vskip-.1cm
\begin{equation}
\mathcal{C} = \sum_i \textbf{1}_{\text{odd}} \otimes |M_i)_{\text{even}}(M_i|_{\text{even}}\,.
\end{equation}
\vskip-.1cm
\noindent where $\{M_i\}$ are the Kraus operators of $\mathcal{C}$, satisfying $\sum_i M_i^\dagger M_i = \textbf{1}$.  Finally, we can contract $\mathcal{C}$ with $\varrho$ as
\begin{equation}
\sum_i \bigg(\textbf{1}_{\text{odd}} \otimes (M_i|_{\text{even}}\bigg) \, \varrho \, \bigg(\textbf{1}_{\text{odd}} \otimes |M_i)_{\text{even}}\bigg)  =: \widetilde{\varrho}
\end{equation}
which gives us a new superdensity operator $\widetilde{\varrho}$ that has the form expressed diagrammatically in Figure \ref{TN3} above.
\\ \\
\noindent \textbf{Example 3:} We can act on the superdensity operator with a superchannel $\mathcal{SC}$, which is a quantum channel mapping superdensity operators to superdensity operators.  A superchannel $\mathcal{SC}$ acting on a superdensity operator to produce another superdensity operator is shown in Figure \ref{superchannel1} below. 
\vskip.5cm
\begin{center}\customlabel{superchannel1}{4}
\includegraphics[scale=.4]{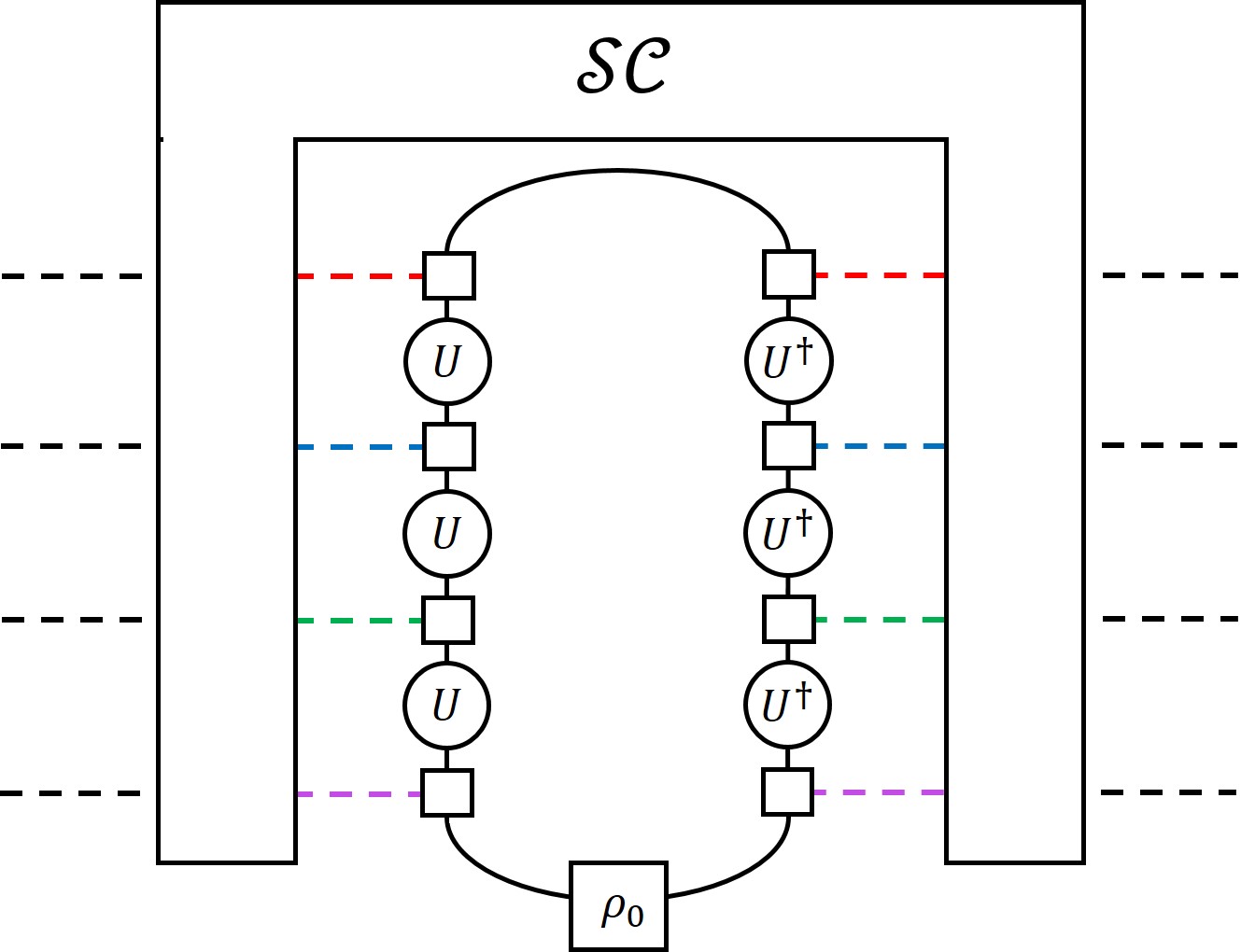}\\
\end{center}
Figure 4: Diagram of a superchannel $\mathcal{SC}$ applied to a superdensity operator, outputting another superdensity operator.
\newpage
If $\mathcal{B} = A \otimes B$ and $\varrho$ has the restriction property with respect to $A$, then one may desire that $\mathcal{SC}$ be compatible with the restriction.  For instance, if
\begin{equation}
\big(\textbf{1}_{A} \otimes (\textbf{1}|_{B}\big) \,\mathcal{SC}[\varrho] \,\big(\textbf{1}_{A} \otimes |\textbf{1})_{B}\big)\,= \mathcal{SC}\big[\big(\textbf{1}_{A} \otimes (\textbf{1}|_{B}\big) \, \varrho \,\big(\textbf{1}_{A} \otimes |\textbf{1})_{B}\big)\big]\,,
\end{equation}
then $\mathcal{SC}$ maps superdensity operators that have the restriction property with respect to $A$ to other superdensity operators that also have the restriction property with respect to $A$. \\

\noindent \textbf{Example 4:} We can construct superdensity operators for out-of-time order evolution by choosing some of the unitaries in Example 1 to evolve backwards in time.  A diagram of an out-of-time-order superdensity operator is shown below in Figure \ref{TN4}.
\vskip.5cm
\begin{center}\customlabel{TN4}{5}
\includegraphics[scale=.45]{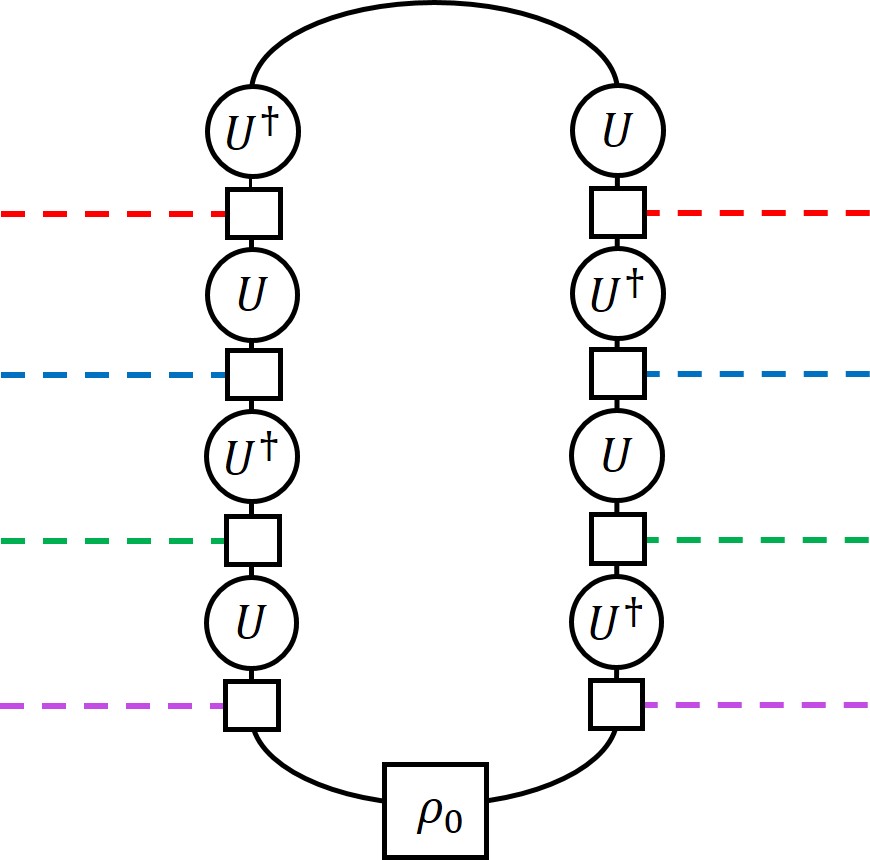}
\vskip.3cm
Figure 5: Diagram for an out-of-time-order superdensity operator.
\end{center}
\vskip.2cm
This object is a map from operators to out-of-time-order correlators, and contains information about the spectrum of Lyapunov exponents associated with such correlators \cite{Stanford1, Stanford2}. \\

\noindent \textbf{Example 5:} Superdensity operators can also have spacetime structure.  For example, contracted tensor networks with dangling legs provide a wealth of examples of superdensity operators, such as the one pictured in Figure \ref{TN5} below.
\vskip.5cm
\begin{center}\customlabel{TN5}{6}
\includegraphics[scale=.45]{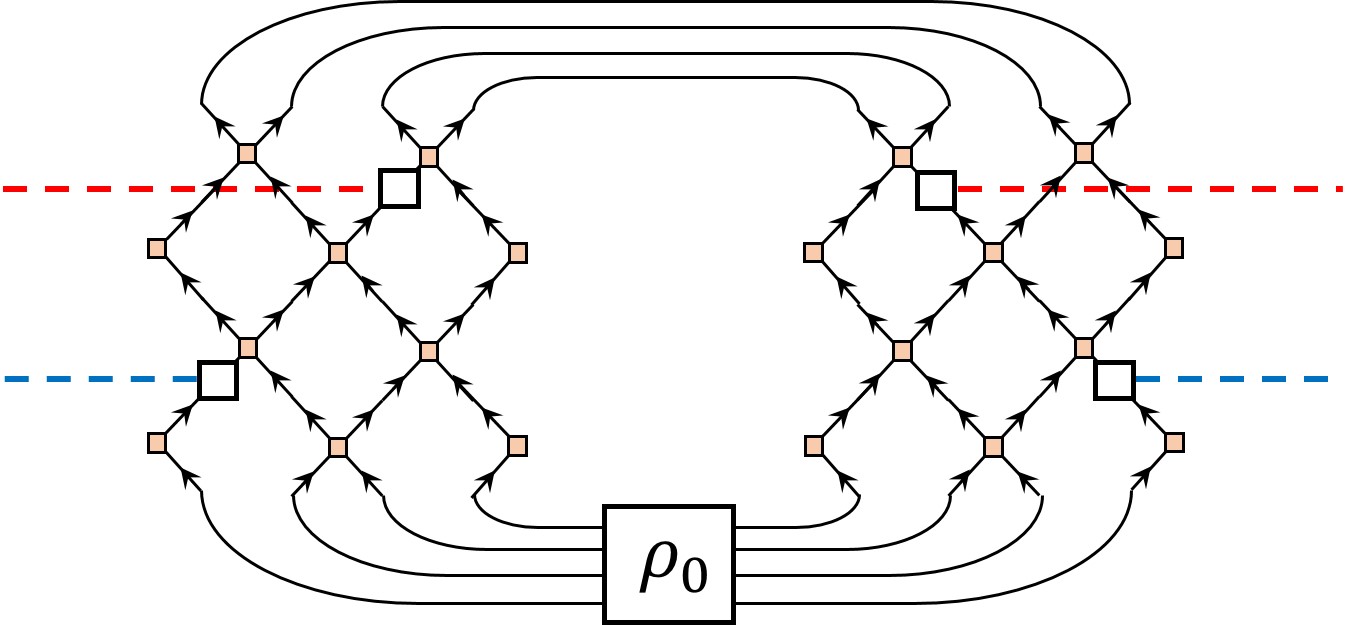}
\end{center}
\vskip.3cm
Figure 6: A superdensity operator given by a contracted tensor network with dangling legs. 

\section{Measuring superdensity operators}

In this section, we elaborate on how to prepare a superdensity operator by coupling an evolving system to auxiliary apparatus, so that the state of the auxiliary apparatus accurately mirrors the superdensity operator.  Then we explain how if one is given many copies of a superdensity operator, one can perform measurements to characterize the superdensity operator completely.  For this purpose we adapt a procedure called quantum tomography, which we shall briefly review.   In this Section, our discussion is primarily mathematical.  In Section \ref{sec:ExperimentalProposals}, we will outline experimental protocols.

\subsection{Mapping time to space}
\label{sec:mapping_time_to_space}
Suppose we have a system in an initial state $|\psi_0\rangle$ in the Hilbert space $\mathcal{H}_{t_1}$ which has dimension $d$.  We will call this the ``main system.''  Further suppose we have a single auxiliary $d^2$--level system with orthonormal basis $\{|0\rangle,...,|d^2 - 1\rangle\}$.  We will prepare the auxiliary system in the initial state $\frac{1}{d} \sum_{i=0}^{d^2 - 1}|i\rangle$, which is a uniform superpostion over all states in the auxiliary Hilbert space.  We can couple the main and auxiliary systems with a controlled unitary gate on the joint system, namely
\begin{equation}
U = \sum_{i,j=0}^{d^2 - 1} = |i\rangle \langle i |\otimes U_i
\end{equation}
where $\{U_i\}$ for $i=0,...,d^2-1$ is a set of unitaries.  Note that $U$ acts by
\begin{equation}
\label{CUnitary1}
U \big(|i\rangle \otimes |\psi\rangle\big) = |i\rangle \otimes U_i |\psi\rangle
\end{equation}
and hence
\begin{equation}
U\bigg(\frac{1}{d} \sum_{i=0}^{d^2 - 1}|i\rangle \otimes |\psi_0\rangle \bigg) = \frac{1}{d} \sum_{i=0}^{d^2 - 1}|i\rangle \otimes U_i|\psi_0\rangle\,.
\end{equation}
In order to construct a superdensity operator, we would like to create states of the form $\frac{1}{\sqrt{d}} \sum_{i=0}^{d^2 - 1}|i\rangle \otimes X_i|\psi_0\rangle$ where $\{X_i\}$ is an orthonormal basis of operators on $\mathcal{H}_{t_1}$.  However, the controlled unitary operation we utilized in Eqn.~\eqref{CUnitary1} does not generalize to a mapping like $|i\rangle \otimes |\psi\rangle \to |i\rangle \otimes X_i |\psi\rangle$, since that is generally a \textit{non}-unitary operation.

Fortunately, this problem can be remedied.  Suppose we pick the $U_i$'s in Eqn.~\eqref{CUnitary1} such that they form a complete orthogonal basis for operators on $\mathcal{H}_{t_1}$.  By orthogonal, we mean $\tr(U_i^\dagger U_j) = d \, \delta_{ij}$.  An example of such a set of unitaries is the set of $d$-dimensional Pauli operators, namely the set of $d^2$ unitaries
\begin{equation}
\Sigma_1^m \Sigma_3^n\,\,\,\,\,\text{for }m,n=0,...,d-1
\end{equation}
where
\begin{align}
\Sigma_1 |j\rangle &:= |j+1\rangle \\
\Sigma_3 |j\rangle &:= \omega^j |j\rangle
\end{align}
where the $j$ indices are treated modulo $d$ and $\omega = e^{2\pi i/d}$ is a primitive $d$th root of unity.  Since the set $\{U_i\}$ forms a complete basis for operators on $\mathcal{H}_{t_1}$, any operator $X_i$ in $\mathcal{B}(\mathcal{H}_{t_1})$ can be written as
\begin{equation}
X_i = \frac{1}{d}\sum_{j=0}^{d^2 - 1} \tr(X_i U_j^\dagger) \, U_j
\end{equation}
where the factor of $1/d$ is due to the orthogonality relation $\tr(U_i^\dagger U_j) = d \, \delta_{ij}$.  Now to transform $\frac{1}{d} \sum_{i=0}^{d^2 - 1}|i\rangle \otimes U_i|\psi_0\rangle$ into our desired state $\frac{1}{\sqrt{d}} \sum_{i=0}^{d^2 - 1}|i\rangle \otimes X_i|\psi_0\rangle$, the trick is to act with a unitary on the auxiliary system that will rearrange the $U_i$'s in exactly the right linear combination to produce the $X_i$'s.  Consider the operator on the auxiliary system
\begin{equation}
V = \frac{1}{\sqrt{d}}\sum_{i,j=0}^{d^2 - 1} \text{tr}(X_i U_j^\dagger)\,|i\rangle \langle j| \otimes \textbf{1}_{d \times d}\,.
\end{equation}
The operator $V$ is unitary since
\begin{align}
V^\dagger V &= \frac{1}{d} \sum_{i,j,k = 0}^{d^2 - 1} \tr(X_i U_j^\dagger)\,  \tr(U_j X_k^\dagger) \, |i\rangle \langle \ell| =  \sum_{i,k = 0}^{d^2 - 1} \tr(X_i X_k^\dagger) \, |i\rangle \langle j| = \textbf{1}_{d^2 \times d^2} \\
V V^\dagger &= \frac{1}{d} \sum_{i,j,k = 0}^{d^2 - 1} \tr(X_i^\dagger U_j\,  \tr(U_j^\dagger X_k) \, |i\rangle \langle \ell| =  \sum_{i,k = 0}^{d^2 - 1} \tr(X_i^\dagger X_k) \, |i\rangle \langle j| = \textbf{1}_{d^2 \times d^2}
\end{align}
where we have used the completeness and orthogonality of the $U_i$'s, as well as the completeness and orthonormality of the $X_i$'s.  Then we have
\begin{align}
V \left(\frac{1}{d}\sum_{i=0}^{d^2 - 1}|i\rangle \otimes U_i |\psi_0\rangle \right) &= \frac{1}{\sqrt{d}}\sum_{m=0}^{d^2 - 1}|m\rangle \otimes \left(\frac{1}{d}\sum_{n=0}^{d^2 - 1} \tr(X_m U_n^\dagger)\,U_n \right) \, |\psi_0\rangle \\
&= \frac{1}{\sqrt{d}}\sum_{m=0}^{d^2 - 1}|m\rangle \otimes X_m |\psi_0\rangle
\end{align}
and so more compactly,
\begin{equation}
VU\left(\frac{1}{d} \sum_{i=0}^{d^2 - 1} |i\rangle \otimes |\psi_0\rangle\right) = \frac{1}{\sqrt{d}} \sum_{i=0}^{d^2 - 1}|i\rangle \otimes X_i |\psi_0\rangle\,.
\end{equation}
We emphasize that this procedure works for \textit{any} orthonormal basis of operators $\{X_i\}$ acting on $\mathcal{H}_{t_1}$.

Letting $|\phi\rangle = \frac{1}{d} \sum_{i=0}^{d^2 - 1} |i\rangle$, suppose we have $n$ auxiliary $d^2$-level systems in the initial state $|\phi\rangle_1 \cdots |\phi\rangle_n = |\phi \cdots \phi\rangle$.  Starting with the joint state $|\phi\cdots \phi\rangle \otimes |\psi_0\rangle$, if we apply $V U$ between the $i$th auxiliary system and the main system at time $t_i$\,, and in between evolve the main system with $U(t_{i+1},t_i)$, then we have
\begin{align}
& V_n U_n \big(1_{\text{aux.}}\otimes U(t_{n}, t_{n-1})\big) V_{n-1} U_{n-1} \cdots V_2 U_2 \big(1_{\text{aux.}}\otimes U(t_2, t_1)\big)V_1 U_1 \big( |\phi\rangle_1 \otimes \cdots \otimes |\phi\rangle_n \otimes |\psi_0\rangle \big) \nonumber \\
=&\, \frac{1}{d^{n/2}}\sum_{i_1,...,i_n = 0}^{d^2 - 1} |i_1,..., i_n \rangle \otimes X_{i_n} U(t_n,t_{n-1}) X_{i_{n-1}} \cdots X_{i_2} U(t_2, t_1) X_{i_1} |\psi_0\rangle
\end{align}
which is a \textit{superstate} $|\Psi\rangle$, as defined in Section 2 above.  If we only make measurements on the auxiliary system, then this corresponds to computing the reduced density operator $\tr_{\text{main}}(|\Psi\rangle \langle \Psi|)$, which is the usual superdensity operator
\begin{align}
\label{SuperdensityMeasured1}
&\frac{1}{d^{n}}\sum_{\substack{i_1,...,i_n = 0 \\ j_1,...,j_n = 0}}^{d^2 - 1} \tr( X_{i_n} U(t_n,t_{n-1}) X_{i_{n-1}} \cdots X_{i_2} U(t_2, t_1) X_{i_1} \, \rho_0 \, X_{j_1}^\dagger U(t_2, t_1)^\dagger X_{j_2}^\dagger \cdots X_{j_{n-1}}^\dagger U(t_n,t_{n-1})^\dagger X_{j_n}^\dagger) \nonumber \\
& \qquad \qquad \qquad \qquad \qquad \qquad \qquad \qquad \qquad \qquad \qquad \qquad \qquad \qquad \qquad \times |i_1,..., i_n \rangle \langle j_1,...,j_n| \nonumber \\
\end{align}
where $\rho_0 = |\psi_0\rangle \langle \psi_0|$.  The same form holds even if the initial state was mixed, i.e., if $\rho_0$ is not a pure state.

In summary, we have mapped the superdensity operator to a conventional density operator, by coupling the main system to an auxiliary system.  The density operator describing the auxiliary system functions as the superdensity operator of the main system.

We have constructed superdensity operators which at each time probes the entire main system with a set of operators $\{X_i\}$.  Instead, we can construct a superdensity operator which at each time $t_j$ probes the main system with respect to the operator \textit{subalgebra} $A_{j}$ where $\mathcal{B}(\mathcal{H}_{t_j}) = A_j \otimes B_j$\,, by choosing operators $\{X_i^{A_j}\}$ supported on $A_j$.  Running through the same procedure, we obtain
\begin{align}
\label{SuperdensityMeasured2}
&\frac{1}{\sqrt{d_{A_1}\cdots d_{A_n}}}\sum_{\substack{i_1,...,i_n\\ j_1,...,j_n}}\tr( X_{i_n}^{A_n} U(t_n,t_{n-1}) \cdots U(t_2, t_1) X_{i_1}^{A_1} \, \rho_0 \, X_{j_1}^{A_1\,\dagger} U(t_2, t_1)^\dagger \cdots U(t_n,t_{n-1})^\dagger X_{j_n}^{A_n\,\dagger}) \nonumber \\
& \qquad \qquad \qquad \qquad \qquad \qquad \qquad \qquad \qquad \qquad \qquad \qquad \qquad \qquad \qquad \times |i_1,..., i_n \rangle \langle j_1,...,j_n| \nonumber \\
\end{align}
where $d_{A_j}$ is the dimension of $A_j$.

\subsection{Spacetime quantum tomography}

Suppose we are given an unknown state $\rho$.  How do we determine what $\rho$ is? Generally a measurement on $\rho$ will extract some information, but act irreversibly on $\rho$.  What we actually need is many identical copies of $\rho$, namely $\rho^{\otimes N}$, so that we can perform many measurements.  We would like a measurement procedure on multiple copies of $\rho$ that determines $\rho$ to some specified precision, which will determine the number of identical copies $N$ that we require.  Such a procedure is called a \textit{quantum tomography}, and is an essential tool for characterizing quantum states \cite{Tomog1, Tomog2}.

In experiment, quantum tomography is used to characterize unknown states and unknown sources.  For example, if one wants to design a source that produces entangled Bell pairs, one would need to perform a quantum tomography on that source to check that it indeed produces Bell pairs.  From a theoretical standpoint, quantum tomography is interesting since the trade-off between the precision to which we can know $\rho$, and the number of copies of $\rho$ which we require to achieve that precision, places fundamental bounds on our ability to learn about a system by measuring it.
  
Superdensity operators characterize spacetime correlations of an evolving quantum system.  Specifically, knowing an $n$-time superdensity operator is equivalent to knowing all $n$-time correlation functions with respect to a specified class of spacetime observables.  Thus, it is interesting to consider quantum tomographies on superdensity operators, including partial quantum tomographies, in which we learn features of the superdensity operator but not the exact state.  Analyzing such tomographies quantifies our ability to reveal spacetime correlations through measurement.  

Quantum tomography of the superdensity operator is \textit{spacetime quantum tomography}.  Since the superdensity operator can be embodied as the density operator of an auxiliary apparatus, as previously explained, we can gain insight by utilizing known quantum tomography results, which we partially review below.  On the other hand, the superdensity operator has additional structure, which can be leveraged for more efficient spacetime quantum tomography, as we will see below.

In the most general setup, we would have an $n$-time superdensity operator in which we have access to the subalgebra of observables $A_i$ at time $t_i$, for $i=1,...,n$.  Then the dimensions of the superdensity operator would be $\prod_{i=1}^n \dim A_i$ by $\prod_{i=1}^n \dim A_i$.  Let us analyze a simpler example, which captures all of the essential features of the general case.  Suppose we have the superdensity operator $\varrho : \mathcal{B}(\mathbb{C}^{2^n}) \otimes \mathcal{B}^*(\mathbb{C}^{2^n}) \to \mathbb{C}$ of a single qubit at $n$ times, so that the dimensions of the superdensity operator are $4^n \times 4^n$.  Furthermore, suppose we have $N$ copies of $\varrho$, namely $\varrho^{\otimes N}$.

The simplest method of performing quantum tomography is to choose a complete basis of $4^n \times 4^n$ operators, and measure the expectation value of $\varrho$ with respect to these operators.  We will partially follow the exposition of \cite{ NielsenChuang}.  For example, letting our complete basis of operators be the Pauli strings $\sigma_{i_1}\otimes \cdots \otimes \sigma_{i_{2n}}$ for $i_1,...,i_{2n} \in \{0,1,2,3\}$, we can reconstruct $\varrho$ by performing measurements to estimate the expectation values \newline $\tr(\sigma_{i_1}\otimes \cdots \otimes \sigma_{i_{2n}} \, \varrho)$, and then using our estimates $M_{i_1,...,i_{2n}}$ to compute our approximation to $\varrho$, namely
\begin{equation}
\widehat{\varrho} = \frac{1}{4^n} \sum_{i_1,...,i_{2n}=0}^3 M_{i_1,...,i_{2n}} \, \sigma_{i_1}\otimes \cdots \otimes \sigma_{i_{2n}}\,.
\end{equation}
Concretely, suppose we want to find an approximation $\widehat{\varrho}$ to $\varrho$ that satisfies
$$\| \widehat{\varrho} - \varrho \|_1 < \epsilon$$
for some specified $\epsilon$.  Then how many measurements do we need to perform, and how many copies $N$ of the superdensity operator do we require?

In our simple procedure outlined above, we note that each observable is a Pauli string which only has eigenvalues $\{+1,-1\}$.  Then every time we perform a single measurement of a single Pauli string $\sigma_{i_1} \otimes \cdots \otimes \sigma_{i_{2n}}$, we output either $+1$ or $-1$.  Repeating the measurement $m$ times and averaging the outputs, if $m$ is sufficiently large then we expect the average to be a Gaussian random variable with mean $\tr(\sigma_{i_1} \otimes \cdots \otimes \sigma_{i_{2n}} \, \varrho)$ and standard deviation $\Delta_{i_1,...,i_{2n}}/\sqrt{m}$, where $\Delta_{i_1,...,i_{2n}}$ is the standard deviation for a single measurement.  Since $\Delta_{i_1,...,i_{2n}}$ is at most one, the standard deviation of our $m$ measurements is at most $1/\sqrt{m}$.  So if we perform a large number $m$ measurements for each Pauli string, we require $m \cdot 4^{2n}$ total measurements requiring $N = m \cdot 4^{2n}$ copies of $\varrho$.  In this case, we can treat our approximation $\widehat{\varrho}$ to $\varrho$ as a function of i.i.d.\! Gaussian independent variables $R_{i_1,...,i_{2n}}$, namely
\begin{equation}
\widehat{\varrho}(\{R_{i_1,...,i_{2n}}\}) = \frac{1}{4^n}\sum_{i_1,...,i_{2n} = 0}^3 \bigg(\tr(\sigma_{i_1} \otimes \cdots \otimes \sigma_{i_{2n}} \, \varrho) + R_{i_1,...,i_{2n}}\bigg) \, \sigma_{i_1} \otimes \cdots \otimes \sigma_{i_{2n}}
\end{equation} 
where each $R_{i_1,...,i_{2n}}$ has zero mean and standard deviation upper bounded by $1/\sqrt{m}$.  We would like to find $m$ (and accordingly, $N = m \cdot 4^{2n}$) such that
\begin{equation}
\label{expectEstimate1}
\mathbb{E}\,\|\widehat{\varrho}(\{R_{i_1,...,i_{2n}}\}) - \varrho\|_1 < \epsilon\,.
\end{equation}
Now we derive a simple bound to build intuition.  We have the relations
\begin{align}
\mathbb{E}\,\|\widehat{\varrho}(\{R_{i_1,...,i_{2n}}\}) - \varrho\|_1 &\geq \mathbb{E}\,\|\widehat{\varrho}(\{R_{i_1,...,i_{2n}}\}) - \varrho\|_2 \\ \nonumber \\
&= \mathbb{E}\, \sqrt{\tr\left(\frac{1}{4^n} \sum_{i_1,...,i_{2n} = 0}^3 R_{i_1,...,i_{2n}} \, \sigma_{i_1} \otimes \cdots \otimes \sigma_{i_{2n}} \right)^2} \\ \nonumber \\
&= \mathbb{E}\, \sqrt{\frac{1}{4^n}\sum_{i_1,...,i_{2n} = 0}^3 R_{i_1,...,i_{2n}}^2 } \\ \nonumber \\
&= \frac{\sqrt{2}}{2^{n}\sqrt{m}} \frac{\Gamma((4^{2n}+1)/2)}{\Gamma(4^{2n}/2)} \sim \frac{2^{n}}{\sqrt{m}} 
\end{align}
where we have used the orthogonality of the Pauli strings in the Hilbert-Schmidt inner product.  Then comparing with Eqn.~\eqref{expectEstimate1}, we find $\epsilon > \mathcal{O}(2^{n}/\sqrt{m}) = \mathcal{O}((4^n)^{3/2}/\sqrt{N})$ and thus $N \geq \mathcal{O}((4^n)^{3}/\epsilon^2)$.  More generally, if our superdensity operator is a $D \times D$ matrix, we would require at least
\begin{equation}
\label{bound0}
N \geq \mathcal{O}(D^{3}/\epsilon^2)
\end{equation}
copies of $\varrho$ so that $\| \widehat{\varrho} - \varrho \|_1 < \epsilon$.  A lower bound on $N$ must scale at least as $(D^2 - 1)$ since we are measuring this many parameters, and indeed our lower bound has a larger scaling of $D^3$.  In fact, the bound in Eqn.~\eqref{bound0} is tight: in the context of standard quantum tomography, it was recently shown \cite{Harrow1} that for \textit{any} quantum tomography scheme in which $N$ copies of a $D \times D$ density operator are measured independently, we have\footnote{In \cite{Harrow1}, the authors use the convention $\frac{1}{2}\| \widehat{\varrho} - \varrho \|_1 < \epsilon$, so the $\epsilon$ in their bounds differs by a factor of $2$ from our $\epsilon$.}
\begin{equation}
N \geq \Omega(D^3/4\epsilon^2)
\end{equation}
where the big--$\Omega$ notation means that $N \geq D^3/4\epsilon^2$ for large $D$ with no additional $\mathcal{O}(1)$ prefactors.

There are more sophisticated methods of quantum tomography, which involve making joint measurements on all $N$ copies at once.  However, this requires having access to all $N$ copies at once, instead of preparing and measuring them sequentially.  The references \cite{Harrow1, Wright1} show that for a quantum tomography scheme in which all $N$ copies are accessible at once, it is necessary that $N \geq \Omega(D^2/4\epsilon^2)$ and sufficient that $N \leq \mathcal{O}(D^2/\epsilon^2)$.  These bounds are better by a factor of $D$ from the case of independent measurements.  The new feature of joint measurements which allows for such an improvement is that the joint measurements of the $N$ copies can be entangled.

So far, we have discussed how to apply standard quantum tomographic techniques to measuring superdensity operators.  We have only been using the fact that the superdensity operator is a density operator of the auxiliary apparatus, and then applying known results.  However, superdensity operators have additional structure since they are not \textit{generic} density operators of the auxiliary apparatus, and capture the data of spacetime correlation functions in particular.  There has been much recent work on developing quantum tomographic techniques optimally tailored to restricted classes of density operators, and no doubt similar techniques can be applied to superdensity operators.

The simplest feature of a superdensity operator which we can exploit is its rank.  Of course, the superdensity operator has much more structure than its rank, but we will not pursue a study of more specially tailored quantum spacetime tomography protocols in this paper.  As an example of how to compute the rank of a superdensity operator, consider the two-time superdensity operator
\begin{align}
\varrho = \frac{1}{d^2} \sum_{i_1,i_2,j_1,j_2=0}^{d^2 - 1} \tr\big(X_{i_2} \, U \, X_{i_1} \, \rho_0 \, X_{j_1}^\dagger \, U^\dagger \, X_{j_2}^\dagger \big) \, |X_{i_1})(X_{j_1}| \otimes |X_{i_2})(X_{j_2}|\,,
\end{align}
where the initial state $\rho_0$ is a state on $d$-dimensional Hilbert space, and has rank $r$.  Suppose that $U = e^{-i H t}$, and that $H$ has the orthonormal eigenbasis $|\lambda_i\rangle$ with eigenvalues $\lambda_i$.  Choosing the basis of operators $\{X_i\}_{i=0}^{d^2-1} = \{\Lambda_{mn} = |\lambda_m\rangle \langle \lambda_n|\}_{m,n=0}^{d-1}$, we have
\begin{align}
\varrho &= \frac{1}{d^2} \sum_{\substack{m_1,n_1,p_1,q_1 = 0 \\ m_2,n_2,p_2,q_2 = 0}}^{d-1} \tr\bigg(|\lambda_{p_1}\rangle \langle \lambda_{q_1}| U|\lambda_{m_1}\rangle \langle \lambda_{n_1}|\rho_0 |\lambda_{n_2}\rangle \langle \lambda_{m_2}| U^\dagger |\lambda_{q_2}\rangle \langle \lambda_{p_2}|\bigg) \, |\Lambda_{m_1 n_1})(\Lambda_{m_2,n_2}|\otimes|\Lambda_{p_1 q_1})(\Lambda_{p_2,q_2}| \nonumber \\ \nonumber \\
&= \frac{1}{d^2} \sum_{\substack{m_1,n_1,p_1 = 0 \\ m_2,n_2 = 0}}^{d-1} e^{-i(\lambda_{m_1} - \lambda_{m_2})t}\langle\lambda_{n_1}|\rho_0 |\lambda_{n_2}\rangle \, |\Lambda_{m_1 n_1})(\Lambda_{m_2,n_2}|\otimes|\Lambda_{p_1 m_1})(\Lambda_{p_1,m_2}|\,.
\end{align}
Relabeling the basis of the superdensity operator, we can write $|\Lambda_{m n}) = |\lambda_m\rangle |\lambda_n\rangle$, which gives us
\begin{align}
\varrho &= \frac{1}{d^2} \sum_{\substack{m_1,n_1,p_1 = 0 \\ m_2,n_2 = 0}}^{d-1} e^{-i(\lambda_{m_1} - \lambda_{m_2})t}\langle\lambda_{n_1}|\rho_0 |\lambda_{n_2}\rangle \,|\lambda_{n_1}\rangle \langle \lambda_{n_2}|\otimes |\lambda_{m_1}\rangle |\lambda_{m_1}\rangle \langle \lambda_{m_2} | \langle \lambda_{m_2}| \otimes |\lambda_{p_1}\rangle \langle \lambda_{p_1}| \nonumber \\ \nonumber \\
&= \rho_0 \otimes \left(\frac{1}{\sqrt{d}}\sum_{m_1=0}^{d-1} |\lambda_{m_1}\rangle |\lambda_{m_1}\rangle \right)\left(\frac{1}{\sqrt{d}}\sum_{m_2=0}^{d-1} \langle \lambda_{m_2}| \langle\lambda_{m_2}| \right) \otimes \frac{1}{d} \, \textbf{1}_{d \times d}\,.
\end{align}
We see that $\varrho$ can be expressed as a tensor product of three density operators with ranks $r$, $1$ and $d$.  Hence the total rank of $\varrho$ is their product, namely $r \cdot d$.

More generally, the rank of a superdensity operator involves an interplay between the rank of the initial state, the time evolution, and the algebra of spacetime operators used as probes.  If the rank of a superdensity operator is at most $r_{\text{super}}$, then using \cite{Harrow1}, we need at least
\begin{equation}
N \geq \Omega(D\,r_{\text{super}}^2/4\epsilon^2 \,\log(1/2\epsilon))
\end{equation}
copies of $\varrho$ if we measure each $\varrho$ independently.  If we measure all $\varrho$ jointly, also using \cite{Harrow1, Wright1} we require at least $N \geq \Omega(r_{\text{super}}\,D/4\epsilon^2)/\log(D/2 r_{\text{super}}\,\epsilon)$ copies and at most $N \leq \mathcal{O}(r_{\text{super}}\, D/4\epsilon^2 )$ copies.

\section{Spacetime entropies}
\subsection{Features of spacetime entropies}

Since superdensity operators are also density operators, we can utilize the standard definitions of quantum entropy and all of their properties.  However, simply carrying over the definitions to the spacetime setting does not guarantee that the spacetime generalizations are natural or useful.  Furthermore, it is not immediately obvious that spacetime entropies will reduce to standard ``spatial'' entropies if the superdensity operator is taken to be at a single time.  We will show that spacetime entropies are natural and useful, and in particular prove that the spacetime mutual information constrains correlations between two algebras of \textit{spacetime} operators.  Furthermore, we show that spacetime quantum entropies reproduce standard spatial quantum entropies for single-time superdensity operators.

First, we point out a subtlety in applying standard definitions of quantum entropy to reduced \textit{superdensity} operators.  Suppose we have a superdensity operator
$$\varrho : \mathcal{B}(\mathcal{H}_{\text{hist.}}) \otimes \mathcal{B}^*(\mathcal{H}_{\text{hist.}}) \longrightarrow \mathbb{C}\,.$$  Writing $\mathcal{B}(\mathcal{H}_{\text{hist.}}) = A \otimes B$, suppose we want to compute an entropy of $\varrho$ with respect to the subalgebra of spacetime operators $A$, and that $\varrho$ has the restriction property with respect to $A$ (see the discussion of restriction in Section \ref{sec:definition} above).  We have two options:
\begin{enumerate}
\item We can compute the entropy with respect to the \textit{reduced} superdensity operator $\text{tr}_B(\varrho)$\,; or
\item We can compute the entropy with respect to the \textit{restricted} superdensity operator $\varrho_A := \big(\textbf{1}_A \otimes (\textbf{1}|_B\big) \,\varrho\, (\textbf{1}_A \otimes |\textbf{1})_B\big)$, which is itself a superdensity operator since $\varrho$ has the restriction property with respect to $A$. 
\end{enumerate}
The two options have different interpretations.  For the first option, the spacetime entropy characterizes the amount of information in measuring all spacetime observables in $A$, having \textit{forgotten} the measurement results of all spacetime observables in $B$.  As an example, if we measure the system at late times and throw away knowledge of the system in the past, then we may lose knowledge of the initial state of the system and so can only probe features of the time evolution unitary which is the only information that remains.  For the second option, the spacetime entropy characterizes the amount of information in measuring all spacetime observables in $A$, having \textit{never measured} spacetime observables in $B$.  As a similar example, suppose we only measure the system at late times.  Then we still have access to information contained in the time-evolved initial state since we have not thrown away any information about the past.

Suppose we have an entropic quantity which depends on $k$ subalgebras $A_1,...,A_k$, where $\mathcal{B}(\mathcal{H}_{\text{hist.}}) = A_1 \otimes \cdots \otimes A_k \otimes B$, and that $\varrho$ has the restriction property with respect to $A_1 \otimes \cdots \otimes A_k$.  As an example, we might want to compute the mutual information between $A_1$ and $A_2$, in which case $k = 2$.  In such settings, it is most natural to use a combination of the two options above: Suppose we \textit{only} measure spacetime observables which are in $A_1 \otimes \cdots \otimes A_k$.  Then an entropic quantity depending on $A_1,...,A_k$ characterizes the relationship between the information encoded in measurements with respect to different subsets of the $A_i$'s.  Thus, we should first \textit{restrict} $\varrho$ to $\varrho_{A_1,...,A_k}$, and then study the relationship between the various reduced density operators of $\varrho_{A_1,...,A_k}$ using our entropic quantity.

By example, suppose we decompose $\mathcal{B}(\mathcal{H}_{\text{hist.}}) = A_1 \otimes A_2 \otimes B$, and that we wish to compute the \textit{spacetime mutual information} of $\varrho$ between $A_1$ and $A_2$, denoted by $I_{\varrho}(A_1 : A_2)$.  First, we restrict $\varrho$ to operator insertions with respect to $A_1$ and $A_2$.  This restriction is defined by
\begin{equation}
\label{restrict0}
\varrho_{A_1 A_2} := \big(\textbf{1}_{A_1 A_2} \otimes (\textbf{1}|_{B}\big) \, \varrho \, \big(\textbf{1}_{A_1 A_2} \otimes |\textbf{1})_{B}\big)
\end{equation}
which is itself a superdensity operator since $\varrho$ has the restriction property with respect to $A_1 \otimes A_2$.  Then the spacetime quantum mutual information is simply
\begin{equation}
\label{spacetimeMI0}
I_{\varrho}(A_1:A_2) := S[\tr_{A_2}(\varrho_{A_1 A_2})] + S[\tr_{A_1}(\varrho_{A_1 A_2})] - S[\varrho_{A_1 A_2}]\,.
\end{equation}
Since $\varrho_{A_1 A_2}$ is a superdensity operator, and since superdensity operators have all of the properties of density operators, the spacetime quantum mutual information inherits all of the properties of the standard quantum mutual information. We will further explore properties of the spacetime mutual information below.

We would like to show that spacetime quantum entropies reproduce standard ``spatial'' entropies for superdensity operators at a single time.  This provides evidence that spacetime entropies are natural generalizations of spatial entropies.  Specifically, let us show that given an initial state $\rho_0 \in \mathcal{S}(\mathcal{H}_1 \otimes \mathcal{H}_2)$ with $\dim \mathcal{H}_1 = d_1$ and $\dim \mathcal{H}_2 = d_2$, the single-time superdensity operator
\begin{equation}
\varrho = \frac{1}{d_1} \sum_{i,j=0}^{d_1^2 - 1}\tr\big(X_i^1 \otimes \textbf{1}_{2\times 2} \,\rho_0 \, X_j^{1\,\dagger} \otimes \textbf{1}_{2 \times 2} \big) \, |X_i^1)(X_j^1|
\end{equation}
which probes the $\mathcal{H}_1$ tensor factor has von Neumann entropy
\begin{equation}
S[\varrho] = S[\tr_{\mathcal{H}_2}(\rho_0)] + \log(d_1)\,.
\end{equation}
First, we can write
\begin{equation}
\frac{1}{d_1} \sum_{i,j=0}^{d_1^2 - 1}\tr\big(X_i^1 \otimes \textbf{1}_{\mathcal{H}_2} \, \rho_0 \, X_j^{1\,\dagger} \otimes \textbf{1}_{\mathcal{H}_2} \big) \, |X_i^1)(X_j^1| = \frac{1}{d_1} \sum_{i,j=0}^{d_1^2 - 1}\tr\big(X_i^1 \, \tr_{\mathcal{H}_2}(\rho_0)\, X_j^{1\,\dagger} \big) \, |X_i^1)(X_j^1|\,.
\end{equation}

Let $\{|\lambda_i\rangle\}$ be an eigenbasis of $\tr_{\mathcal{H}_2}(\rho_0)$, where $|\lambda_i\rangle$ has eigenvalue $\lambda_i$.  Further letting $\Lambda_{mn} = |\lambda_m\rangle \langle \lambda_n|$ for $m,n=0,...,d-1$, we see that $\{\Lambda_{mn}\}$ forms a complete orthonormal basis for operators on $\mathcal{H}_1$ satisfying $\tr(\Lambda_{mn} \, \Lambda_{pq}) = \delta_{mq}\, \delta_{np}$.  Expressing the superdensity operator in this basis, we find
\begin{align}
\varrho = \frac{1}{d_1} \sum_{i,j,k,\ell=0}^{d_1 - 1}\tr\big(\Lambda_{ij} \, \tr_{\mathcal{H}_2}(\rho_0)\, \Lambda_{k\ell} \big) \, |\Lambda_{ij})(\Lambda_{\ell k}| = \sum_{i,j=0}^{d_1 - 1} \frac{\lambda_j}{d_1} \, |\Lambda_{ij})(\Lambda_{ij}|\,.
\end{align}
which is diagonal.  We see that the spectrum of $\varrho$ is $d_1$ copies of the spectrum of $\tr_{\mathcal{H}_2}(\rho_0)/d_1$.  It follows that
\begin{equation}
S[\varrho] = d_1 \, \left(- \sum_{i=1}^{d_1} \frac{\lambda_i}{d_A} \log(\lambda_i/d_1) \right) = S[\tr_{\mathcal{H}_2}(\rho_0)] + \log(d_1)
\end{equation}
as claimed. If we consider the R\'{e}nyi $\alpha$-entropy, defined by $S_\alpha[\rho] := \frac{1}{1-\alpha} \log \tr (\rho^\alpha)$, then we similarly have
\begin{equation}
S_\alpha[\varrho] = S_\alpha[\tr_{\mathcal{H}_2}(\rho_0)] + \log(d_1)\,.
\end{equation}

\subsection{Spacetime quantum mutual information}

Here we discuss the \textit{spacetime quantum mutual information} defined by Eqn.'s~\eqref{restrict0} and~\eqref{spacetimeMI0} above.  To recap, suppose we have a superdensity operator $\varrho$ with respect the history Hilbert space $\mathcal{H}_{\text{hist.}}$.  As explained above, we decompose $\mathcal{B}(\mathcal{H}_{\text{hist.}})$ by $\mathcal{B}(\mathcal{H}_{\text{hist.}}) = A_1 \otimes A_2 \otimes B$ and suppose that $\varrho$ has the restriction property with respect to $A_1 \otimes A_2$.  Considering the restriction $\varrho_{A_1 A_2} := \big(\textbf{1}_{A_1 A_2} \otimes (\textbf{1}|_{B}\big) \, \varrho \, \big(\textbf{1}_{B} \otimes |\textbf{1})_{A_1 A_2}\big)$, the spacetime quantum mutual information between $A_1$ and $A_2$ is simply
\begin{equation}
\label{spacetimeMI}
I_{\varrho}(A_1 : A_2) := S[\tr_{A_2}(\varrho_{A_1 A_2})] + S[\tr_{A_1}(\varrho_{A_1 A_2})] - S[\varrho_{A_1 A_2}]\,.
\end{equation}
As stated above, the spacetime quantum mutual information inherits all of the properties of the standard quantum mutual information since $\varrho_{A_1 A_2}$ has all of the properties of a density operator.

To make the definition concrete, let us consider two examples.  Suppose that the initial time Hilbert space $H_{t_1}$ decomposes as $\mathcal{H}_{t_1} = \mathcal{H}_1 \otimes \mathcal{H}_2 \otimes \mathcal{H}_3$, and let $A_1 = \mathcal{B}(\mathcal{H}_1)$ and $A_2 = \mathcal{B}(\mathcal{H}_2)$.  Then we have
\begin{equation}
\varrho_{A_1 A_2} = \frac{1}{d_{1} \, d_{2}}\sum_{i,j=0}^{d_{1}^2 - 1} \sum_{k,\ell = 0}^{d_{2}^2 - 1} \tr(X_k^{2} X_i^{1} \, \rho_0 \, X_j^{1\,\dagger} X_{\ell}^{2\,\dagger}) \, |X_i^1)(X_j^1|\otimes |X_k^2)(X_\ell^2|\,.
\end{equation}
Since
\begin{align}
S[\tr_{\mathcal{B}(\mathcal{H}_{2})}(\varrho_{A_1 A_2})] &= S[\tr_{23}(\rho_0)] + \log(d_1) \\ \nonumber \\
S[\tr_{\mathcal{B}(\mathcal{H}_{1})}(\varrho_{A_1 A_2})] &= S[\tr_{12}(\rho_0)] + \log(d_2) \\ \nonumber \\
S[\tr_{\mathcal{B}(\mathcal{H}_{1})\otimes \mathcal{B}(\mathcal{H}_{2})}(\varrho_{A_1 A_2})] &= S[\tr_{3}(\rho_0)] + \log(d_1 \, d_2)
\end{align}
it follows that
\begin{equation}
I_{\varrho}(A_1 : A_2) = S[\tr_{23}(\rho_0)] + S[\tr_{12}(\rho_0)]  - S[\tr_{3}(\rho_0)]
\end{equation}
and thus $I_{\varrho}(A_1 : A_2) = I_{\rho_0}(1 : 2)$.  Therefore, at a single time, we reduce to the usual definition of quantum mutual information.

For a more interesting example, suppose now that $\mathcal{H}_{t_1} = \mathcal{H}_{1} \otimes \mathcal{H}_{2}$, and $\mathcal{H}_{t_2} = \mathcal{H}_{3} \otimes \mathcal{H}_{4}$.  Letting $A_1 = \mathcal{B}(\mathcal{H}_{1})$ and $A_3 = \mathcal{B}(\mathcal{H}_{3})$, restricting the superdensity operator to $A_1$ and $A_3$ we obtain 
\begin{equation}
\varrho_{A_1 A_3} = \frac{1}{d_1 \, d_3}\sum_{i,j=0}^{d_1^2 - 1} \sum_{k,\ell = 0}^{d_3^2 - 1} \tr(X_k^{A_3} \, U \, X_i^{A_1} \, \rho_0 \, X_{j}^{A_1\,\dagger} \, U^\dagger \, X_{\ell}^{A_3\,\dagger}) \, |X_{i}^{A_1})(X_{j}^{A_1}|\otimes |X_{k}^{A_3})(X_{\ell}^{A_3}|\,.
\end{equation}
The reduced superdensity operators $\tr_{\mathcal{B}(\mathcal{H}_{A_3})}(\varrho_{A_1 A_3})$ and $\tr_{\mathcal{B}(\mathcal{H}_{A_1})}(\varrho_{A_1 A_3})$ can be written as
\begin{align}
\tr_{\mathcal{B}(\mathcal{H}_{3})}(\varrho_{A_1 A_3}) &= \frac{1}{d_1} \sum_{i,j=0}^{d_1^2 - 1} \tr(X_i^{1} \, \rho_0 \, X_j^{{1}\,\dagger}) \, |X_i^{1})(X_j^{1}| \\ \nonumber \\
\tr_{\mathcal{B}(\mathcal{H}_{1})}(\varrho_{A_1 A_3}) &= \frac{1}{d_3} \sum_{i,j=0}^{d_3^2 - 1} \tr\big(X_i^3 \,U\, \big(\tr_{\mathcal{H}_1}(\rho_0) \otimes \textbf{1}_{\mathcal{H}_1}/d_1\big)\,U^\dagger \, X_j^{3\,\dagger}\big) \, |X_i^3)(X_j^3|
\end{align}
so it follows that
\begin{align}
\label{spacetimeTrace1}
S[\tr_{\mathcal{B}(\mathcal{H}_{3})}(\varrho_{A_1 A_3})] &= S[\tr_{\mathcal{H}_2}(\rho_0)] + \log(d_1) \\ \nonumber \\
\label{spacetimeTrace2}
S[\tr_{\mathcal{B}(\mathcal{H}_{1})}(\varrho_{A_1 A_3})] &= S\big[\tr_{\mathcal{H}_4}\big( U \big(\tr_{\mathcal{H}_1}(\rho_0) \otimes \textbf{1}_{\mathcal{H}_1}/d_1\big) U^\dagger \big)\big] + \log(d_3) \,.
\end{align}
The entropy $S[\varrho_{A_1 A_3}]$ does not have a similar simplifying form.  Let us unpack Eqn.'s~\eqref{spacetimeTrace1} and~\eqref{spacetimeTrace2}.  In Eqn.~\eqref{spacetimeTrace1}, we see that $S[\tr_{\mathcal{B}(\mathcal{H}_{3})}(\varrho_{A_1 A_3})]$, which traces out \textit{future} operator insertions on $\mathcal{H}_3$, is up to an additive constant the von Neumann entropy of the reduced density operator $\tr_{\mathcal{H}_2}(\rho_0)$ of the initial time state $\rho_0$. In Eqn.~\eqref{spacetimeTrace2}, $S[\tr_{\mathcal{B}(\mathcal{H}_{1})}(\varrho_{A_1 A_3})]$ which traces out \textit{past} operator insertions on $\mathcal{H}_1$ is up to an additive constant the von Neumann entropy of the reduced initial system (i.e., without $\mathcal{H}_1$) which is then evolved and traced down to $\mathcal{H}_3$.  In other words, it is like the information that we can measure in $\mathcal{H}_3$ at time $t_2$ given that we do not know anything about the part of the state in $\mathcal{H}_1$ at time $t_1$.

An interesting feature of the spacetime mutual information is that it bounds \textit{spacetime correlations.}  We follow the argument of \cite{Verstraete1} for bounding the standard \textit{spatial} mutual information.  Since we can write $I_{\varrho}(A_1 : A_2) = S(\varrho_{A_1 A_2}\|\varrho_{A_1} \otimes \varrho_{A_2}) \geq \frac{1}{2}\|\varrho_{A_1 A_2} - \varrho_{A_1} \otimes \varrho_{A_2}\|_1^2$, using the inequality $\|X\|_1 \geq \tr(XY)/\|Y\|$ we find
\begin{equation}
I_{\varrho}(A_1:A_2) \geq \frac{\big|\tr\left((M_{A_1} \otimes M_{A_2}) \, \varrho_{A_1 A_2}\right) - \tr(M_{A_1} \, \varrho_{A_1})\,\tr(M_{A_2} \, \varrho_{A_2}) \big|^2}{2 \,\|M_{A_1}\|^2 \|M_{A_2}\|^2}
\end{equation}
where $M_{A_1}$ and $M_{A_2}$ are operators on the \textit{space of operators} $A_1$ and $A_2$, respectively.  Ordinary correlation functions in Keldysh contour order are a special example of correlators of $M_{A_1},~M_{A_2}$. For example, suppose $a_1 \in A_1$, $a_1' \in A_1^*$, $a_2 \in A_2$, and $a_2' \in A_2^*$.  Letting $M_{A_1} = |a_1')(a_1|$ and $M_{A_2} = |a_2')(a_2|$, we have
\begin{equation}
I_{\varrho}(A_1:A_2) \geq \frac{\big|\varrho_{A_1 A_2}[a_1 \otimes a_2, a_1' \otimes a_2'] - \varrho_{A_1 A_2}[a_1 \otimes \textbf{1}, a_1' \otimes \textbf{1}] \,\varrho_{A_1 A_2}[\textbf{1} \otimes a_2, \textbf{1} \otimes a_2'] \big|^2}{2 \,\|a_1\|_2^2 \, \|a_1'\|_2^2 \, \|a_2\|_2^2 \, \|a_1'\|_2^2}
\end{equation}
since $\| M_{A_1} \|^2 = \tr(a_1^\dagger a_1) \, \tr(a_1'^\dagger a_1') = \|a_1\|_2^2 \, \|a_1'\|_2^2$ and similarly for $\|M_{A_2}\|^2$.  As a consequence, if $I_{\varrho}(A_1:A_3) = 0$, then
\begin{equation}
\label{spacetimeMutualBound1}
\varrho_{A_1 A_2}[a_1 \otimes a_2, a_1' \otimes a_2'] = \varrho_{A_1 A_2}[a_1 \otimes \textbf{1}, a_1' \otimes \textbf{1}] \,\varrho_{A_1 A_2}[\textbf{1} \otimes a_2, \textbf{1} \otimes a_2']\,,
\end{equation}
i.e., spacetime correlations between $A_1$ and $A_2$ factorize.  This is a nontrivial statement, since operators on $A_1$ and $A_2$ need not be spacelike separated, and can even be interlaced in spacetime.

As an example, suppose that $A_1$ and $A_2$ correspond to two spacetime points, $x$ and $y$, and that we have an initial state is $\rho_0$.  Letting $a_1 = \mathcal{O}_1$, $a_2 = \mathcal{O}_2$, $a_1' = \textbf{1}_x$ and $a_2' = \textbf{1}_y$, Eqn.~\eqref{spacetimeMutualBound1} implies
\begin{equation}
I_\varrho(x\, : \, y) \geq \frac{\big| \langle \mathcal{O}_1(x) \, \mathcal{O}_2(y) \rangle_{\rho_0} - \langle \mathcal{O}_1(x) \rangle_{\rho_0} \, \langle \mathcal{O}_2(y) \rangle_{\rho_0} \big|^2}{2\, \dim \mathcal{H}_x\, \dim \mathcal{H}_y \, \|\mathcal{O}_1\|_2^2\, \|\mathcal{O}_2\|_2^2 \, }
\end{equation}
where we have written correlation functions in the Heisenberg picture.  Thus, if $I_\varrho(x\, : \, y) = 0$, spacetime correlation functions between $x$ and $y$ factorize.

\subsection{Relation to the Kolmogorov-Sinai entropy}

The growth rate of the spacetime entropy as we increase the number of measurements characterizes how fast a quantum system entangles with the apparatus via successive measurements as the quantum system evolves.  We begin by discussing a classical limit of the growth rate of spacetime entropy that reduces to the Kolmogorov-Sinai entropy as defined in classical dynamical systems as per \cite{Lindblad1}--\cite{Fannes1}. 

Suppose we initialize our quantum system in the maximally mixed state $\rho_0 = \textbf{1}/d$ to study the growth rate of the spacetime entropy. The evaluation of $\tr(\,\boldsymbol{\cdot}\,\rho_0)$ sums over every basis state of the physical Hilbert space with equal weight.  Say that our quantum system of interest has a well-defined (semi-)classical limit.  In this classical limit, $\tr(\,\boldsymbol{\cdot}\,\rho_0)$ can be replaced by an integration $ \int_\mathcal{M} dx $ over the classical phase space $\mathcal{M}$ which is a continuous topological space equipped with a measure. The measure is normalized such that 
\begin{align}
\int_\mathcal{M} dx ~ 1 = 1,
\end{align}
where $x$ represents points in the phase space $\mathcal{M}$. The classical limit of an operator is given by an integrable function $f$ in the phase space $\mathcal{M}$ whose expectation value is given by $\int_\mathcal{M} dx f(x) $. The completeness relation for an operator basis $\{f_i\}$ to comprise a classical channel which measures the system is
\begin{align}
\sum_i f_i^* (x) f_i (x) = 1 
\end{align}
for any point $x$ in the phase space $\mathcal{M}$.  (This is directly analogous to the completeness relation of the Kraus operators in the Kraus decomposition of a quantum channel.) If the operator basis is orthonormal, we have
\begin{align}
\label{functionComplete1}
\int_\mathcal{M} f_i^*(x) f_{j}(x) \, dx  = \delta_{ij}.
\end{align}
Following Eq.~\eqref{sdexpand1}, the classical limit of a superdensity operator with $n$ measurements is given by
\begin{align}
\varrho_n &=  \sum_{\substack{i_1,...,i_n \\ j_1,...,j_n}} 
\int_\mathcal{M} d x~ f_{i_n} \big(x(t_n)\big) f_{i_{n-1}}\big(x(t_{n-1})\big) \cdots f_{i_{1}}\big(x(t_{1})\big) f_{j_{1}}^*\big(x(t_{1})\big) \cdots f_{j_{n-1}}^*\big(x(t_{n-1})\big) f_{j_{n}}^*\big(x(t_{n})\big) 
 \nonumber \\
& \qquad \qquad \qquad \qquad \qquad \qquad \qquad \qquad \qquad \qquad \qquad \times |f_{i_1})(f_{j_1}| \otimes \cdots \otimes |f_{i_n})(f_{j_n}|\,.
\end{align}
Here $x(t)$, as a function of $t$, represents the classical phase space trajectory. The growth rate of the spacetime entropy in the limit of large number of measurements $n$ is defined by
\begin{align}
h[\{f_i\}] = \lim_{n\rightarrow \infty} \frac{1}{n} S[\varrho_n].
\end{align}
This growth rate $h[\{f_i\}] $ clearly depends on the choice of operator basis. Interestingly, \cite{Alicki1} showed that the supremum of $h[\{f_i\}] $ over all choices of operator basis $\{f_i\}$ satisfying Eqn.~\eqref{functionComplete1} is equivalent to the Kolmogorov-Sinai (KS) entropy of the classical system:
\begin{align}
h_{KS} = \sup_{\text{all choices of }\{f_i\}} h[\{f_i\}].
\end{align}
In a classical dynamical system, the KS entropy characterizes the long-time limit of the entropy production rate due to chaotic dynamics.  According to Pesin's theorem, the KS entropy provides a lower bound on the sum of all the positive Lyapunov exponents of a classical dynamical system \cite{Pesin1, Pesin2, Pesin3}.  In fact, with some extra assumptions that often hold for physical systems (technically, the dynamics must be closed $C^2$ Anosov systems), the KS entropy is \textit{exactly equal to} the sum of the positive Lyapunov exponents.

We have explained that the classical limit of the quantum spacetime entropy naturally accommodates the definition of the classical KS entropy. In other words, we can view the growth rate of the quantum spacetime entropy as a quantum generalization of the KS entropy.  However, in quantum systems, the meaning of the KS entropy is more subtle.  Unlike in the classical case, a \textit{non-zero} production rate of the spacetime entropy is not necessarily tied to quantum chaos.  Nonetheless, the production rate itself and its temperature dependence may be tied to quantum chaos.

We will now explicitly analyze spacetime entanglement in two model systems -- free fermions, and the SYK model -- and calculate their (non-zero) entropy production.

\subsection{Example: Free fermion model}
\label{sec:free_fermion}
In this section, we calculate the spacetime entropy for a free fermion system.  Consider a lattice system with a Majorana fermion mode on each site and a non-interacting free fermion Hamiltonian $H$:
\begin{align}
H = i\,\sum_{\alpha,\beta} J_{\alpha \beta} \, \chi_\alpha \chi_\beta\,,
\end{align}
where $\alpha,\beta$ labels the sites on the lattice and $J_{\alpha \beta}$ is the coupling between the Majorana fermion operators $\chi_\alpha$ and $\chi_\beta$ on the $\alpha$th and the $\beta$th sites, respectively. The Majorana fermion operators satisfy the commutation relations
\begin{align}
\{\chi_\alpha, \chi_\beta\} = 2\delta_{\alpha \beta}\,.
\end{align}
We consider an initial density operator $\rho_0$ with the general form $\rho_0 \propto \exp \big( \sum_{\alpha \beta} i M_{\alpha \beta} \chi_\alpha \chi_\beta \big )$, where $M_{jk}$ is a real antisymmetric matrix. Density operators of this form will be referred to as Gaussian density operators.  Gaussian density operators have the feature that their multi-point correlation functions satisfy Wick's theorem, namely a multi-point correlation functions of Majorana fermion operators is given by summing over different Wick contractions.  Also, under the time evolution generated by the free fermion Hamiltonian $H$, the evolved state $e^{-iHt} \rho_0 e^{iHt}$ is also a Gaussian density operator but with a different $M_{\alpha \beta}$.

Now we would like to choose an orthonormal basis of operators which we can use to probe the system at some fixed time. Consider the basis operators
\begin{align}
X_0 := \frac{1}{\sqrt{2}} \, \textbf{1}\,, ~~~ X_1 = \frac{1}{\sqrt{2}} \, \chi_\alpha\,.
\label{Op_Basis_SingleMajorana}
\end{align}
Here we have chosen $X_1$ to be a Majorana fermion operator on a specific site (the $\alpha$th site). Consider $n$ times $t_1,...,t_n$, with $t_k = t_0 + k \Delta t$ ($k=1,2,...,n$). We can define the superdensity operator $\varrho$ following Eq. ~\eqref{sdexpand1}. For each time $t_k$, the basis of the history Hilbert space is given by $|X_0)$ and $|X_1)$.  Now, we introduce a complex fermion operator $f_{k}$ in the history Hilbert space for each time $t_k$ and identify $|X_0)$ and $|X_1)$ as states with fermion occupation $f_{k}^\dag f_{k} = 0$ and $f_{k}^\dag f_{k} =1$, namely in the Hilbert space $\mathcal{H}_{t_k}$\,:
\begin{align}
& f^\dag_{k} f_{k} |X_0) = 0\,,~~~~ f^\dag_{k} f_{k} |X_1) = |X_1).
\end{align}
One should bare in mind that the fermion operators $f_k$ and $f_k^\dag$ at different time instants anti-commute with each other. We can show that the superdensity operator $\varrho$ is a Gaussian density operator when expressed in terms of the fermion operators $f_k$ and $f_k^\dag$ (see Appendix \ref{app:SuperDensity_Free_Fermion} for details). All of the information of the Gaussian superdensity operator $\varrho$ can be captured by the two-point functions $G_{k\ell} \equiv \tr(f_k^\dag f_\ell \varrho)$, $\widetilde{G}_{k\ell} \equiv \tr(f_k f_\ell^\dag \varrho) = \delta_{k\ell} - G_{\ell k}$ and $\Delta_{k\ell} \equiv \tr(f_k^\dag f_\ell^\dag \varrho)$. These two-point functions of fermions operators in the history Hilbert space can be expressed in terms of the two point functions of the Majorana operators in the physical space:
\begin{align}
G_{k\ell} 
& =
(\mathcal{K}^{-1})_{2n+k,n+\ell}
\,,~~~~~~
 \Delta_{k\ell} =
-(\mathcal{K}^{-1})_{2n+k,2n+\ell}\,,
\end{align}
where $\mathcal{K}$ is the $4n \times 4n$ matrix:
\begin{align}
\mathcal{K} = \left(
\begin{array}{cccc}
-Q            &  0 & 1 & -P^T \\
 0            &  0 & 1 & 1 \\
-1            & -1 & 0 & 0 \\
P  & -1 & 0 & -\overline{Q}
\end{array}
\right).
\label{eq:CalK_Def}
\end{align}
Here, the matrix $\mathcal{K}$ is written in block form with each entry representing a $n\times n$ block. In the matrix $\mathcal{K}$, each $1$ stands for a $n\times n$ identity matrix. The matrices $P$, $Q$ and $\overline{Q}$ are given by the physical two-point functions of the Majorana fermions:
\begin{align}
\label{mind_your_Ps_and_Qs}
& P_{k\ell} = \tr\left( U(t_n,t_{\ell})\, \chi_\alpha \,  U(t_\ell,t_1) \,\rho_0\, U(t_k,t_1)^\dagger \, \chi_\alpha  \, U(t_n,t_{k})^\dagger \right)
\nonumber \\ \nonumber \\
& Q_{k\ell} = 
\begin{cases} 
\tr\left( U(t_n,t_{k}) \, \chi_\alpha \,  U(t_k,t_\ell)\, \chi_\alpha \,  U(t_\ell,t_1) \,\rho_0\, U(t_n,t_{1})^\dagger \right) &  \quad k > \ell \\
- \tr\left( U(t_n,t_{\ell}) \, \chi_\alpha  \, U(t_\ell,t_k)\, \chi_\alpha \, U(t_k,t_1) \,\rho_0\, U(t_n,t_{1})^\dagger \right) & \quad k < \ell \\
      0 & \quad k=\ell 
\end{cases}
\\ \nonumber \\
& \overline{Q}_{k\ell} = 
\begin{cases} 
-\tr\left( U(t_n,t_{1}) \, \rho_0\, U(t_\ell,t_{1})^\dagger \,\chi_\alpha  \, U(t_k,t_{\ell})^\dagger \, \chi_\alpha \,  U(t_n,t_{k})^\dagger \right) & \,  k > \ell 
\\
\tr\left( U(t_n,t_{1}) \, \rho_0\, U(t_k,t_{1})^\dagger \, \chi_\alpha  \, U(t_\ell,t_{k})^\dagger \, \chi_\alpha \,  U(t_n,t_{\ell})^\dagger \right) & \, k < \ell
\\
0 & \, k= \ell
\end{cases}
\nonumber 
\end{align}
Having obtained the two-point functions $G_{k\ell}$, $\widetilde{G}_{k\ell}$ and $\Delta_{k\ell}$, they can be organized into a $2n \times 2n$ Hermitian matrix $C$, which we call the correlation matrix:
\begin{align}
C = \left( 
\begin{array}{cc}
G & \Delta \\
\Delta^\dag & \widetilde{G}
\end{array}
\right) = \left( 
\begin{array}{cc}
G & \Delta \\
\Delta^\dag & 1- G^T
\end{array}
\right).
\label{eq:Correlation_Matrix_Def}
\end{align}
According to \cite{Peschel}, the entropy associated to a Gaussian density operator $\varrho$ can be expressed in terms of the eigenvalues of its correlation matrix $\mathcal{C}$:
\begin{align}
S[\varrho] = \sum_{\text{eigenvalues }\lambda\text{ of }C} - \lambda \log(\lambda).
\end{align}

The analysis above concentrated on the operator basis choice in Eq.~\eqref{Op_Basis_SingleMajorana}, which is generated by a single Majorana fermion operator $\chi_\alpha$. We can straightforwardly generalize the same analysis to the operator basis generated by a collection of Majorana fermion operators $\{ \chi_\alpha \,|\, \alpha \in A \}$, where $A$ is a set of sites in the lattice. The operator basis is given by
\begin{align}
X_{\{n_\alpha\}} \equiv \frac{1}{2^{|A|/2}} \prod_{\alpha\in A} \chi_\alpha^{n_\alpha}\,,
\label{eq:multi_fermion_basis}
\end{align}
where each configuration of $n_\alpha \in \{0,1\}$ corresponds to an operator $X_{\{n_\alpha\}}$\,, and $|A|$ denotes the number of elements in the set $A$. At a single time, the state $|X_{\{n_\alpha\}})$ can be identified as a state in a Hilbert space with $|A|$ fermion modes with the fermion occupation directly given by $\{n_\alpha\}$. Similarly, we can show that the superdensity operator is a Gaussian density operator when expressed in terms of the fermion operators in the history Hilbert space.

Now we study the specific case of a 1D Majorana chain.  We consider a 1D lattice of Majorana fermion modes with nearest-neighbor coupling. The Hamiltonian of the system is given by
\begin{align}
H  = i \,\frac{J_0}{4} \sum_\alpha \chi_\alpha \chi_{\alpha+1}\,
\end{align}
and we choose the initial state to be the ground state of the Hamiltonian. In the history Hilbert space, we again choose the basis operators to be those in Eq.~\eqref{Op_Basis_SingleMajorana}. The correlation functions of this model are exactly known:
\begin{align}
\langle \chi_\alpha(t_k)  \chi_\alpha(t_\ell)\rangle = \frac{1}{\pi} \int_0^\pi dq ~ e^{-i(t_k - t_\ell) \sin q} = J_0(t_k-t_\ell)  -iH_0(t_k-t_\ell),
\end{align} 
where $J_0(t)$ is the zeroth Bessel function of the first kind and $H_0 (t)$ is the zeroth Struve function. This correlation function directly determines the matrices in Eq.~\eqref{eq:PQQ_matrix}. We can then numerically evaluate the correlation matrix $C$ and the corresponding spacetime entropy. In Figure \ref{fig:Free_Fermion} above, we plot the spacetime entropy as a function of the number of measurements $n$ for different values of $J_0 \Delta t = 1,2,3,4,5,10,20,30$ (with colors from dark to light). First of all, we notice that the spacetime entropy grows linearly as a function of the number of measurements $n$. Here, we have taken the thermodynamic limit of the 1D chain, which means there will be no upper bound on the spacetime entropy.  Therefore, we expect this linear growth of the spacetime entropy to persist for larger values of $n$. Another observation is that, as we increase $J_0 \Delta t$, the spacetime entropy saturates to the straight line $n \log2$ (see dashed line in Figure \ref{fig:Free_Fermion}), meaning that one bit of quantum entropy is produced per each measurement. 

\begin{figure}[t]
\begin{center}
\includegraphics[width=5in]{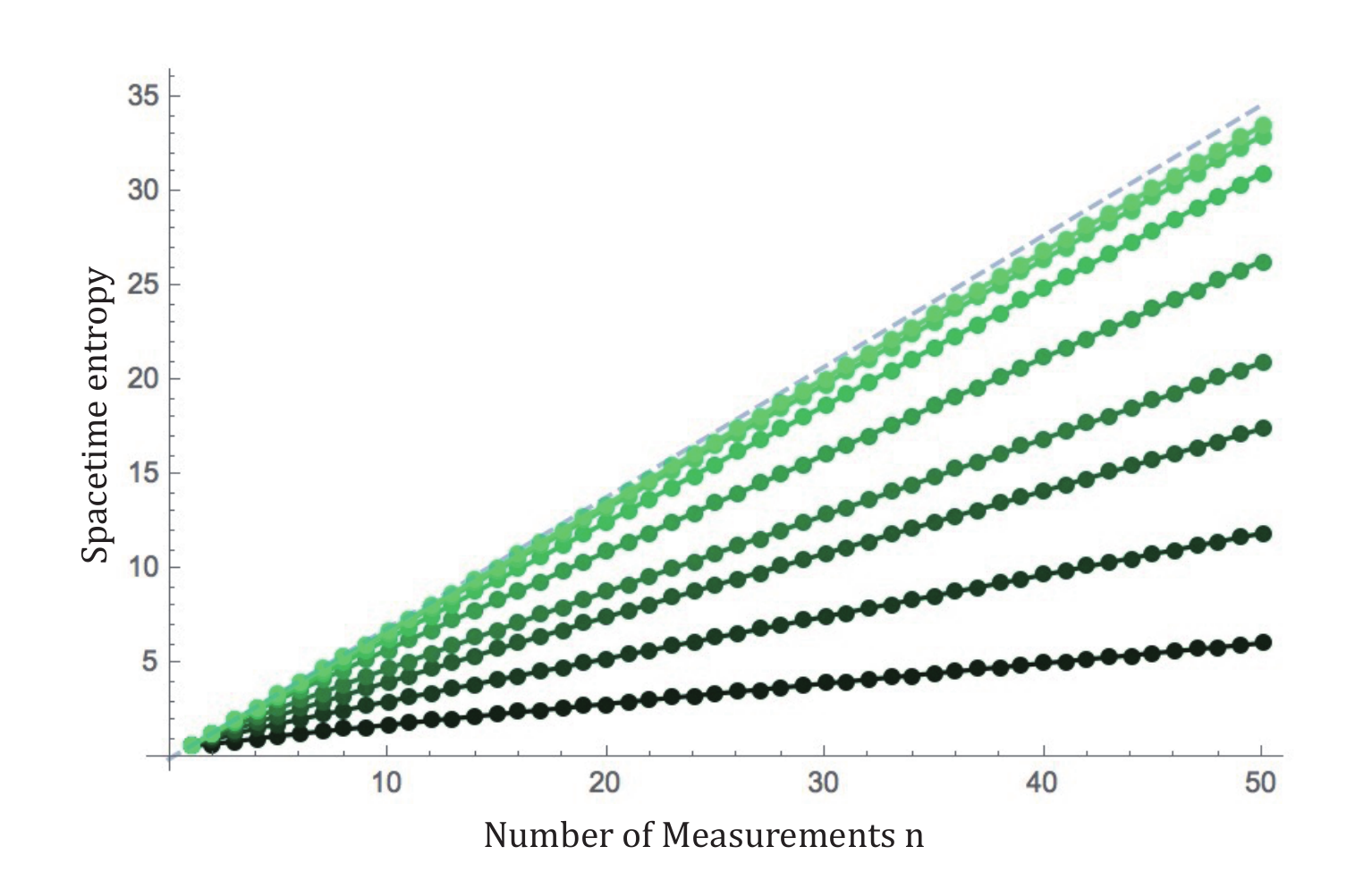}
\customlabel{fig:Free_Fermion}{7}
\end{center}
Figure 7: The spacetime entropy of the 1D Majorana chain with its ground state as the initial state is plotted as a function of the number of measurements $n$. The points of different color (from dark to light) correspond to different values of $J_0 \Delta t = 1,2,3,4,5,10,20,30$. As we increase $J_0 \Delta t$, the spacetime entropy saturates to the dashed line $n\log 2$.
\end{figure}

\indent The saturation at large $J_0 \Delta t$ can be understood analytically. At large values of $J_0 \Delta t$, the correlation function satisfies
\begin{align}
\langle \chi_\alpha(t_k)  \chi_\alpha(t_\ell)\rangle = J_0(t_k-t_\ell)  -iH_0(t_k-t_\ell) = \delta_{k\ell} + \mathcal{O}((J_0 \Delta t)^{-1}).
\end{align} 
Consequently, to leading order in $(J_0 \Delta t)^{-1}$, we have $P_{k\ell} = \delta_{k\ell}$ and $Q_{k\ell} = \overline{Q}_{k\ell} = 0$. By Eqn.'s~\eqref{eq:CorrelMatrix_G} and~\eqref{eq:CorrelMatrix_D}, the resulting correlation matrix $C$ becomes a $2n\times 2n$ identity matrix with the corresponding spacetime entropy exactly given by $n\log 2$.

\subsection{Example: SYK model}

As another example, we numerically study the spacetime entropy in a simple model of many-body chaotic systems, the Sachdev-Ye-Kitaev model \cite{SYK1, SYK2, SYK3}. The model describes $N$ Majorana fermions with the following Hamiltonian:
\begin{equation}
{H}=  \sum_{1\le j<k<\ell<m\le N} J_{jk\ell m} \, {\chi}_j {\chi}_k {\chi}_\ell {\chi}_m\,, \quad \{ {\chi}_j,{\chi}_k \} = \delta_{jk}
\end{equation}
where $\{J_{jk\ell m}\}$  are independent random couplings with zero mean 
$\overline{J_{jk\ell m}}=0$, and the variance of the individual couplings is given by $\frac{1}{3!} N^3\, \overline{J^2_{jk\ell m}} = \mathcal{J}^2$. Consider the operators $X_i$ for $i=0,1,2,...,M$ defined by
\begin{eqnarray}
X_0=\sqrt{1-p}\, \textbf{1},\quad ~X_i=\sqrt{\frac{p}{M}}\,\chi_i
\end{eqnarray}
with $0<p<1$ and $1\leq M\leq N$. The operators $\{X_i\}$ satisfy $\sum_i X_i^\dagger X_i= \textbf{1}$, so that the linear map
\begin{eqnarray}
\rho\longrightarrow \sum_i X_i \,\rho\, X_i^\dagger\nonumber
\end{eqnarray}
is a quantum channel. Considering $n$ moments in time $t_n=t_0+n\Delta t$, we can define the superdensity operator $\varrho$ in the standard way, as per Eqn.~\eqref{sdexpand1}. Physically, inserting a sequence of $X_i$'s in the time evolution of the SYK model is a consequence of introducing a coupling of the SYK model with auxiliary Majorana fermion systems. For a given initial density operator $\rho_0$, we can study the spacetime entropy, i.e. the entropy of $\varrho$. The calculation can be further simplified when the initial state is a pure state, in which case the entropy $S[\varrho]$ is the entanglement entropy of the auxiliary systems with the original SYK model, so that it is equal to the entropy of the SYK model after applying positive maps. Denoting the density operator of SYK model after $n$ steps of couplings by $\rho_n$, we have
\begin{eqnarray}
\rho_n=\sum_i X_i\, U(\Delta t)\,\rho_{n-1}\,U(\Delta t)^\dagger X_i^\dagger\,.
\end{eqnarray}
The entropy of the first $n$ auxiliary systems is thus $S[\rho_n] = S(t_n)$. 

We compute $S(t_n)$ numerically for the simple case of $M=1$ in which only one Majorana fermion is accessed by the auxiliary systems. The result is shown in Figure \ref{fig:syk}. If there is trivial time evolution, the superdensity operator describes repeated measurements of the same qubit. For concreteness, we can organize the Hilbert space of $N$ Majorana fermions into that of $N/2$ qubits.  ($N$ is always even for the Hilbert space to be well-defined.) Defining complex fermion operators
\begin{eqnarray}
f_k :=\frac12\left(\chi_{2k-1}+i\chi_{2k}\right)\,,\quad~k=1,2,...,N/2\,,
\end{eqnarray}
the $N/2$ qubits can be labeled by each $f_k^\dagger f_k$ being $0$ or $1$, which corresponds to each $i\,\chi_{2k-1}\chi_{2k}$ being $+1$ or $-1$, respectively. The coupling of external system to $\chi_1$ will thus only access the first qubit labeled by $f_1^\dagger f_1$. This is why the spacetime entropy saturates to $\log 2$ if the measurement is done very frequently ($\Delta t\rightarrow 0$). It is a consequence of the ``quantum Zeno effect'': time evolution of a quantum system freezes if it is constantly being measured. 

When $\Delta t$ is not too short, the system evolves before the first qubit is measured again. Due to the chaotic time evolution, eventually all qubits can be accessed, and the spacetime entropy can reach as high as the maximal value $N\log\sqrt{2}$. This is consistent with our numerical result. Before reaching its maximum value, the entropy $S(t_n)$ grows linearly as a function of time, and the growth rate is almost independent from $\Delta t$, as is shown in Figure \ref{fig:syk}(a) by the convergence of curves with different $\Delta t$. Our calculation is done for a system with $16$ Majorana fermions, with an initial state
\begin{eqnarray}
\ket{\Psi}=Z^{-1/2}e^{-\beta H/2}\ket{0}\,.
\end{eqnarray}
Here $\ket{0}$ is the state with $f_k^\dagger f_k$ being $0$ for all $k$, i.e.\! the vacuum of the $f_k$ fermion. This is a high energy state, and the imaginary time evolution reduces the energy. The parameter $\beta$ plays the role of an inverse temperature. 

\begin{figure}[t]
\customlabel{fig:syk}{8}
\begin{center}
\includegraphics[width=7in]{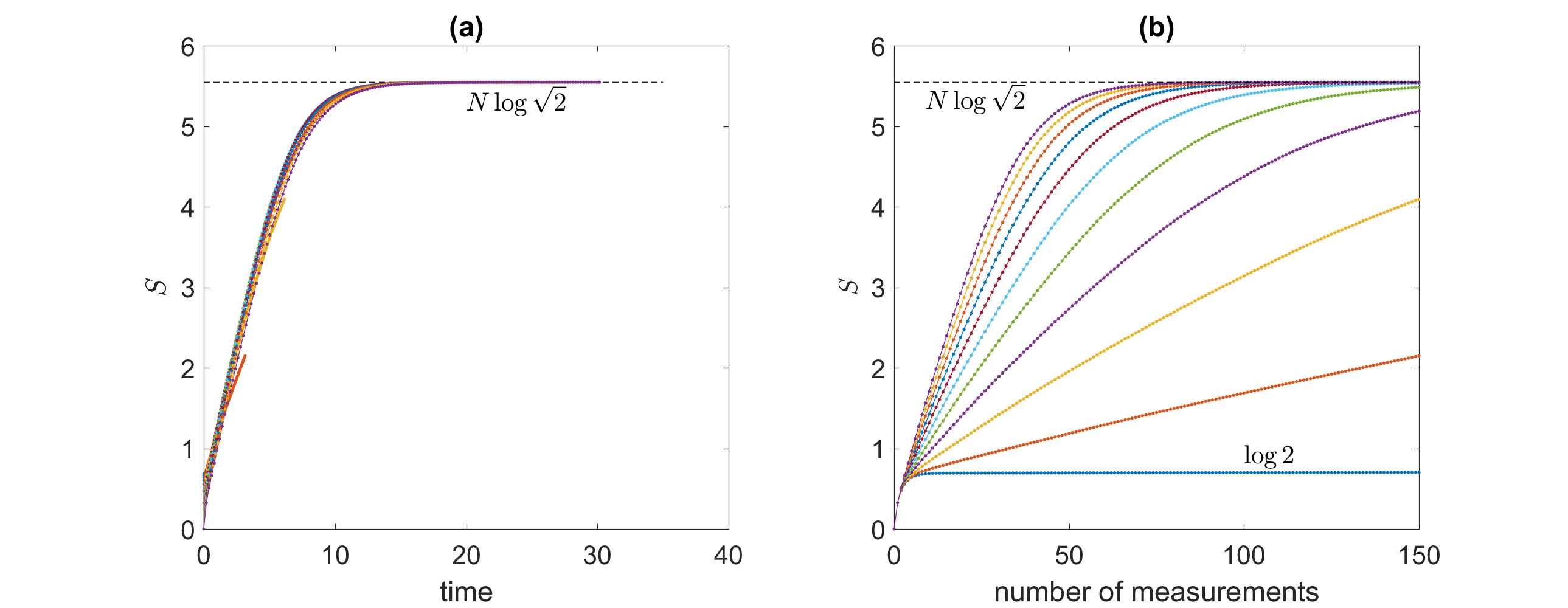}
\end{center}
Figure 8: The spacetime entropy of SYK model with a pure initial state (see text), as a function of (a) $t_n=t_0+n\Delta t$, and (b) number of measurements $n$.
\end{figure}

\subsection{Long--time limit}

We would like to make some further comments on the long--time saturation of the spacetime entropy.  Suppose we have a superdensity operator with an initial-time Hilbert space of dimension $d$, and at $N$ equally spaced times we measure the system in an identical manner.  This is equivalent to applying $N$ identical quantum channels to the main system at equally spaced times, as shown below in Figure \ref{transfer}(a) with the quantum channel $\mathcal{C}$ defined by $\mathcal{C}[\rho] = \sum_i X_i \rho X_i^\dagger$ where $\sum_i X_i^\dagger X_i = \textbf{1}$.  Note that there is unitary evolution $U$ between each application of the quantum channel.

Instead of viewing this procedure as applying quantum channels and time evolution to the initial density operator $N$ times, one can instead view the procedure as applying a ``transfer matrix'' $N$ times to an initial state in a doubled system.  For example, if the main system has a pure initial state $\rho_0 = |\Psi_0\rangle\langle \Psi_0|$, then correspondingly in the doubled system we have an initial state $\ket{\Psi_0}\ket{\Psi_0^*}$ to which the transfer matrix
\begin{eqnarray}
T=\sum_i UX_i\otimes U^* X_i^*
\end{eqnarray}
\begin{figure}[t]
\customlabel{transfer}{9}
\begin{center}
\includegraphics[width=4.3in]{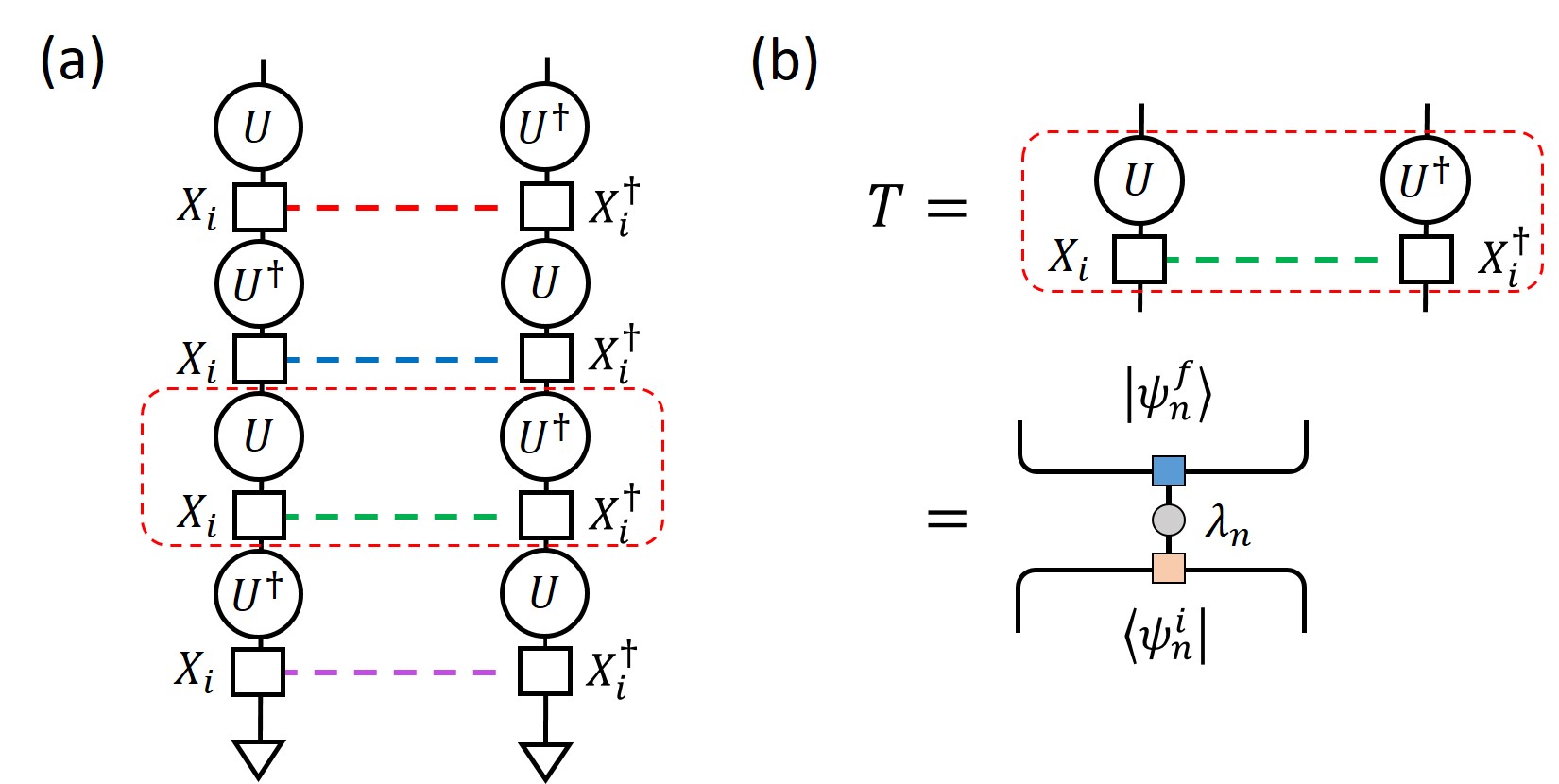}
\end{center}
Figure 9: (a) Tracing over the same apparatus at $N$ times is equivalent to repeatedly applying the transfer matrix to a doubled system.  The transfer matrix is defined by the operators in the red dashed box. (b) Diagonalization of the transfer matrix. \label{fig:transferM1}
\end{figure}
is applied $N$ times, where the stars denote complex conjugation.  Generically, the transfer matrix can be diagonalized (since its corresponding quantum channel $\mathcal{C}$ is generically diagonalizable) which leads to the expansion
\begin{eqnarray}
T=\sum_n\lambda_n\ket{\psi^f_n}\bra{\psi^i_n}
\end{eqnarray}
where the superscripts ``$i$'' and ``$f$'' stand for ``initial'' and ``final.''  This decomposition is depicted diagrammatically in Figure \ref{transfer}(b).  Here the eigenvectors $\ket{\psi_n^i}$ and $\ket{\psi^f_n}$ live in the doubled Hilbert space, and satisfy the orthogonality condition $\langle{\psi^i_n}|{\psi^f_m}\rangle=\delta_{nm}$.  We take the order of eigenvalues $\lambda_n$ such that $|\lambda_n|\geq |\lambda_{n+1}|$.  In fact, $T$ has an eigenvector with eigenvalue $1$, since the condition $\sum_i X_i^\dagger X_i= \textbf{1}$ is equivalent to
\begin{eqnarray}
\frac{1}{\sqrt{d}}\sum_i \langle i | \langle i | \,\cdot T = \frac{1}{\sqrt{d}}\sum_i \langle i | \langle i |\,,
\end{eqnarray}
where $|\Phi\rangle := \frac{1}{\sqrt{d}}\sum_i |i\rangle |i\rangle$ is the maximally entangled state between the two subsystems of the doubled system.  In fact, all of the eigenvalues of $T$ have norm less than or equal to one.  (This follows from the fact that the norms of the eigenvalues of $\mathcal{C}$ are less than or equal to $1$ \cite{Wolf1, Havel1}.)

In general, there are multiple eigenvalues with maximal norm $|\lambda_n|=1$. For example, if the $X_i$'s are all restricted to a subsystem with dimension $d_A$, and the complement system has Hilbert space dimension $d_{\overline{A}}$, all operators acting on the complement system are preserved by the transfer matrix, leading to an eigenvalue degeneracy of at least $d_{\overline{A}}^2$.  Another example is if each $X_i$ is unitary (up to a normalization constant), in which case all eigenvalues $\{\lambda_n\}$ have norm equal to $1$. 

After applying $T$ to the doubled system many times, the only nontrivial effect comes from the leading eigenvalues, since 
\begin{eqnarray}
T^N\simeq \sum_{\left|\lambda_n\right|=1}\lambda_n^N\ket{\psi^f_n}\bra{\psi^i_n}\,.
\end{eqnarray}
\begin{figure}[t]
\begin{center}
\customlabel{fig:transferM2}{10}
\includegraphics[width=6in]{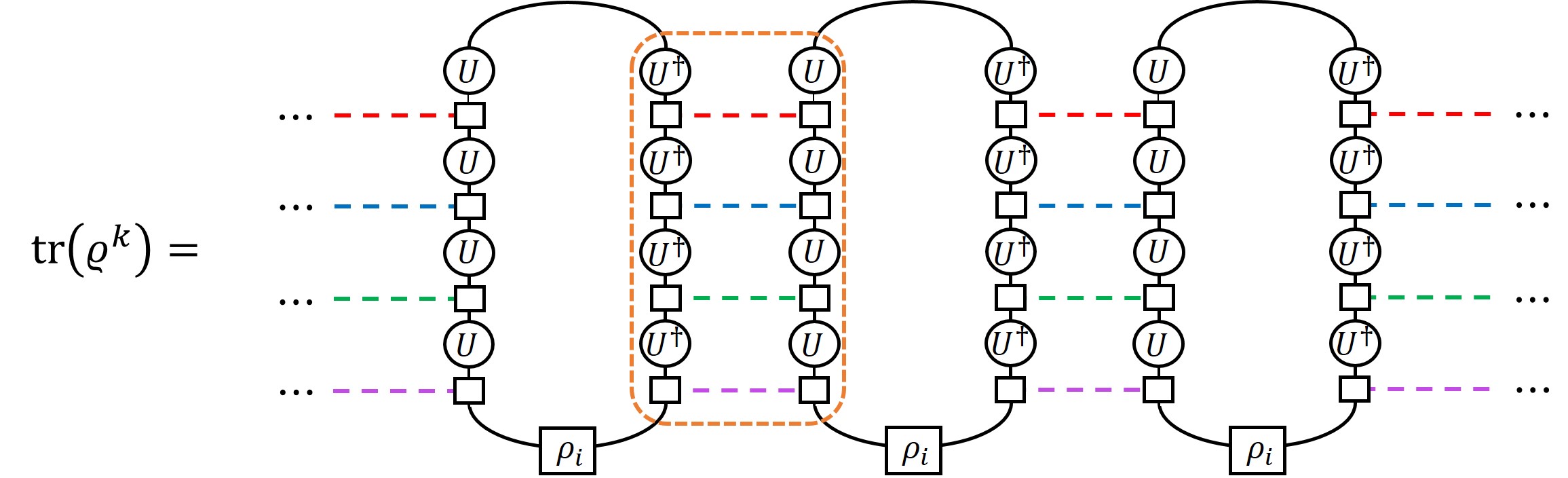}
\end{center}
Figure 10: A diagrammatic representation of $\tr(\varrho^k)$, where different copies of the superdensity operator are coupled together by blocks of transfer matrices.  A block $T^N$ of transfer matrices is outlined with dotted lines in orange.
\end{figure}
\noindent The eigenvectors corresponding to the leading eigenvalues determine the long-time saturation value of the spacetime entropy, and the subleading eigenvalue determines the asymptotic long--time behavior.  Consider Figure \ref{fig:transferM2}, which shows $\tr(\varrho^k)$ with $N$ probes per superdensity operator (shown in the diagram is $N=4$).  We see in Figure \ref{fig:transferM2} that each adjacent pair of superdensity operators is connected by a block $T^N$, where one such block is outlined with dotted lines in orange.  Note that the $T$ operators are flipped horizontally relative to Figure \ref{transfer}(b), since we are applying the $T$'s \textit{between} adjacent superdensity operators and not on a single superdensity operator.  By examining the diagram in Figure \ref{fig:transferM2}, it follows that if the second largest eigenvalue of $T$ is $|\lambda_m|=e^{-\mu}$ for some $\mu > 0$, then the spacetime R\'{e}nyi $k$-entropy will approach its long--time value exponentially
\begin{equation}
S_k[\varrho] = \frac{1}{1-k} \log \tr(\varrho^k) = S_k(\infty) - g(k) \times e^{- \mu N}
\end{equation} 
in the large $N$ limit.  Note that $g(k)$ is some function of $k$.  Similarly, taking the limit $k \to 1$ of the above equation, the von Neumann entropy is 
\begin{equation}
S(N)\simeq S(\infty)- \text{const.}\times e^{-\mu N}\,,
\end{equation}
where $g(k) \to \text{const.}$ as $k \to 1$.

A particularly simple case is when the transfer matrix has only one leading eigenvalue, in which case it must be $\lambda_0=1$, and the corresponding left eigenvector must be $\bra{\Phi} = \sum_i \langle i| \langle i|$. The right eigenvector is generically a different state, which we denote as $\ket{\psi_0^f}$. Therefore,
\begin{eqnarray}
T^N\simeq \ket{\psi^f_0}\bra{\Phi}\,.
\end{eqnarray}
The doubled state $\ket{\psi^f_0} = \sum_{i,j} c_{ij} |i\rangle |j^*\rangle$ corresponds to a final state $\rho_f = \sum_{i,j} c_{ij} \, |i\rangle \langle j|$ of the original system.  Independent from the initial state, the system always becomes the state $\rho_f$ after the positive map and unitary evolution many times in succession. In this case, if we consider a generic initial density operator $\rho_i$, the long--time value of the spacetime R\'{e}nyi $k$-entropy is 
\begin{eqnarray}
\label{LateRenyi1}
S_k(\infty)=S_k[\rho_i]+S_k[\rho_f]
\end{eqnarray}
\begin{figure}[t]
\begin{center}
\customlabel{fig:transferM3}{11}
\includegraphics[width=5.9in]{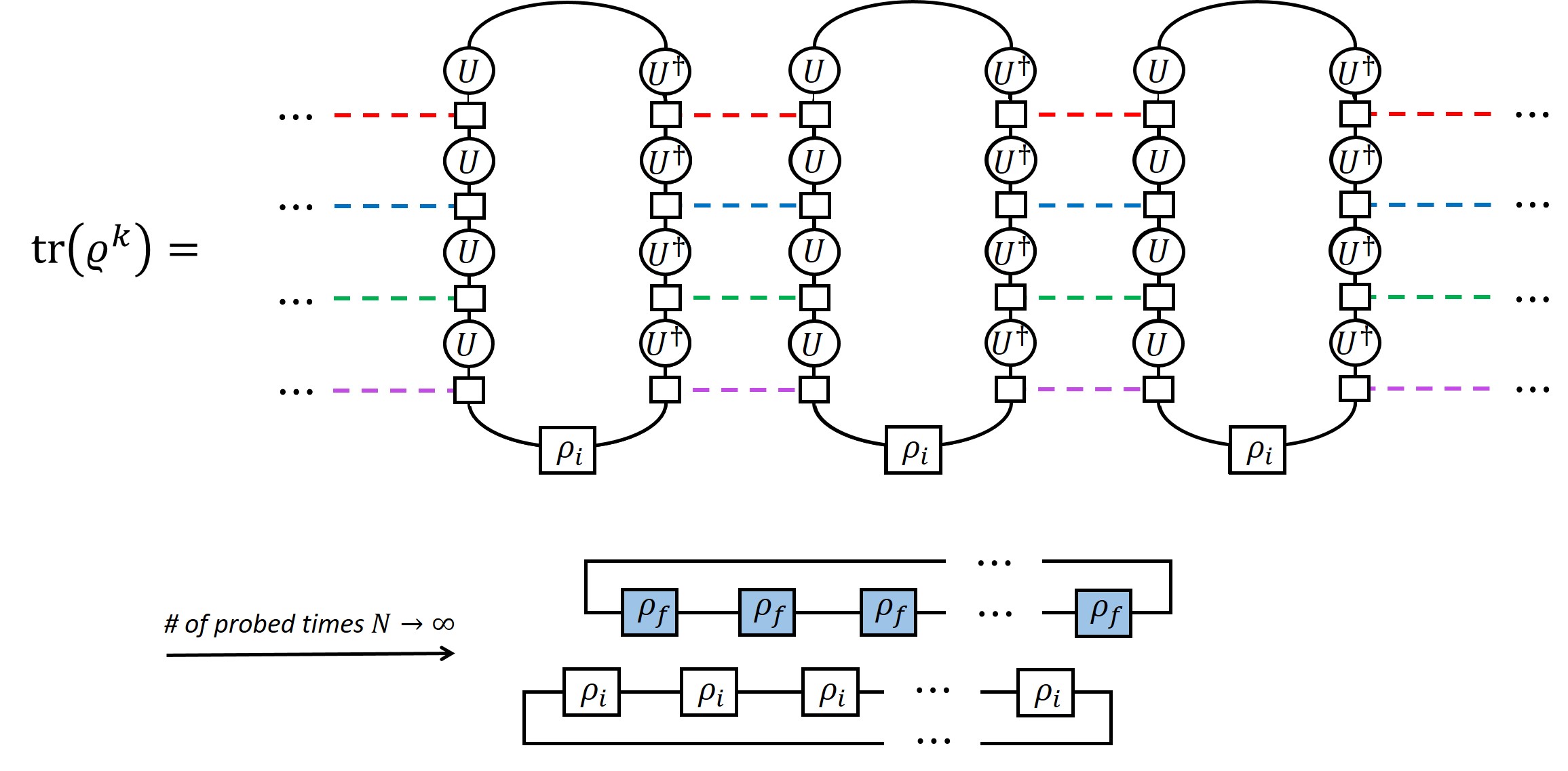}
\end{center}
Figure 11: The spacetime entropy in the long--time limit $N\rightarrow \infty$ if the transfer matrix has only one leading eigenvalue (see text).
\end{figure}
and similarly the long--time value of the spacetime von Neumann entropy is
\begin{eqnarray}
\label{LateVN1}
S(\infty)=S[\rho_i]+S[\rho_f]\,.
\end{eqnarray}
This equation can be proven by studying $\frac{1}{1-k} \log \tr(\varrho^k)$.  As illustrated in Figure \ref{fig:transferM3}, when the transfer matrix $T$ has only one leading eigenvalue, the trace of the superdensity operator $\varrho^k$ in the long--time limit $N\rightarrow \infty$ is equivalent to a  separate trace over $k$ copies of the initial state $\rho_i$ and $k$ copies of the final state $\rho_f$.  Thus $\tr(\varrho^k) \simeq \tr(\rho_i^k)\, \tr(\rho_f^k)$, and Eqn.'s~\eqref{LateRenyi1} and~\eqref{LateVN1} follow.

In the case of the SYK  model we discussed in previous subsection, it is clear from the numerics that $\rho_f=\textbf{1}/d$ is the maximally mixed state, which corresponds to the maximum possible spacetime entropy. In general, if the positive maps $X_i$ also satisfy $\sum_iX_iX_i^\dagger= \textbf{1}$ (as is the case for the SYK model), then the transfer matrix satisfies
\begin{eqnarray}
T\ket{\Phi}=\ket{\Phi}
\end{eqnarray}
which implies that $\ket{\psi^f_0}=\ket{\Phi}$ if there is only one leading eigenvalue.

\section{Experimental protocols} \label{sec:ExperimentalProposals}

There has been much recent experimental work on developing quantum simulators, namely quantum systems with precisely tunable Hamiltonians \cite{Monz1}--\cite{Bernien1}.  These systems will provide an ideal testing ground for the superdensity operator and related quantities like spacetime entropies.

In this section, we present a general protocol for coupling a single qubit degree of freedom to an apparatus degree of freedom comprised of two qubits. Specifically, we construct Hamiltonians which generate time evolution that creates the desired coupling between the single qubit and the auxiliary apparatus.  We show how to apply this procedure sequentially over time to extract the superdensity operator from an evolving system, where we couple to one qubit of main system at each time.  As an interesting application, we discuss how to experimentally measure the spacetime R\'{e}nyi 2-entropy of two qubits, each considered at a different time.  

\subsection{Coupling to auxiliary apparatus via Hamiltonian evolution}

Suppose that the main system comprises of $N$ qubits, and starts in the state $|\psi_{0,N}\rangle$.  Further suppose we want to construct a superdensity operator which probes the main system at $n$ different times, and only probes one qubit at each time.  To construct such a superdensity operator, we desire a superstate like
\begin{equation}
\label{superstate2}
|\Psi_{n,N}\rangle = \frac{1}{2^{n}}\sum_{i_1,...,i_n=0}^{3}|i_1,...,i_n\rangle \otimes \widetilde{\sigma}_{i_n}^{(a_n)} U(t_n,t_{n-1})\, \sigma_{i_{n-1}} \cdots \widetilde{\sigma}_{i_2}^{(a_2)}\,U(t_2, t_1)\,\widetilde{\sigma}_{i_1}^{(a_1)}|\psi_{0,N}\rangle
\end{equation}
which is the same as Eqn.~\eqref{superstate1} above, but with $d = 2$ and $X_i = \widetilde{\sigma}_i^{(a_j)}/\sqrt{2}$ acting on the $a_j$th qubit of the main system, and
\begin{equation}
\widetilde{\sigma}_0^{(a_j)} := \sigma_0^{(a_j)} = \textbf{1}^{(a_j)}\,, \quad \widetilde{\sigma}_1^{(a_j)} := \sigma_1^{(a_j)}\,, \quad \widetilde{\sigma}_2^{(a_j)} := \sigma_2^{(a_j)}\,, \quad \widetilde{\sigma}_3^{(a_j)} := -i \sigma_3^{(a_j)}\,.
\end{equation}
Note that $\widetilde{\sigma}_3^{(a_j)}$ is defined with an extra factor of $-i$ relative to $\sigma_3^{(a_j)}$.  We will see shortly why this factor of $-i$ is useful.

If we trace out the main system from $|\Psi_{n,N}\rangle$, then we will be left with the desired $n$--time superdensity operator.  Let us further simplify the problem by first considering a single-time superstate whose main system comprises of a single qubit (i.e., $N = 1$):
\begin{equation}
\label{superstate3}
|\Psi_{1,1}\rangle = \frac{1}{2}\sum_{j=0}^{3}|j\rangle \otimes \widetilde{\sigma}_{j}^{(1)}|\psi_{0,1}\rangle \,.
\end{equation}
We can treat the auxiliary system as two qubits, since it is a $4$--level system, and rewrite Eqn.~\eqref{superstate3} as
\begin{equation}
\label{superstate4}
|\Psi_{1,1}\rangle = \frac{1}{2}\bigg(|0\rangle |0\rangle \otimes \widetilde{\sigma}_0^{(1)} \, |\psi_{0,1}\rangle + |1\rangle |0\rangle \otimes \widetilde{\sigma}_1^{(1)} \, |\psi_{0,1}\rangle + |0\rangle |1\rangle \otimes \widetilde{\sigma}_2^{(1)} \, |\psi_{0,1}\rangle + |1\rangle |1\rangle \otimes \widetilde{\sigma}_3^{(1)} \, |\psi_{0,1}\rangle \bigg)\,.
\end{equation}

So how do we make this state?  Suppose we start with the initial state
\begin{equation}
|0\rangle |0\rangle \otimes |\psi_{0,1}\rangle\,.
\end{equation}
Applying a Hadamard gate $\text{Had} = \frac{1}{\sqrt{2}}\begin{bmatrix}
1 & 1 \\ 1 & -1
\end{bmatrix} = \frac{1}{\sqrt{2}}(\sigma_1 + \sigma_3)$
to each qubit of the auxiliary system, we obtain
\begin{equation}
\label{intermediateState1}
\text{Had}^{\otimes 2} (|0\rangle |0\rangle) \otimes |\psi_{0,1}\rangle = \frac{1}{2}\bigg(|0\rangle |0\rangle \otimes |\psi_{0,1}\rangle + |1\rangle |0\rangle \otimes |\psi_{0,1}\rangle + |0\rangle |1\rangle \otimes |\psi_{0,1}\rangle + |1\rangle |1\rangle \otimes |\psi_{0,1}\rangle \bigg)\,.
\end{equation}
Next, applying the controlled--$\sigma_1^{(1)}$ unitary $U_1$ between the first auxiliary qubit and the main system, given by
\begin{equation}
U_{1} = |0\rangle \langle 0| \otimes \textbf{1} \otimes \textbf{1} + |1\rangle \langle 1| \otimes \textbf{1} \otimes  \sigma_1^{(1)}\,,
\end{equation}
and then applying the controlled--$\sigma_2^{(1)}$ unitary $U_2$ between the second auxiliary qubit and the main system, given by
\begin{equation}
U_{2} = \textbf{1} \otimes |0\rangle \langle 0| \otimes \textbf{1} + \textbf{1} \otimes |1\rangle \langle 1| \otimes \sigma_2^{(1)}\,,
\end{equation}
we indeed recover
\begin{align}
& U_2 \,U_1\,\big( \text{Had}^{\otimes 2} (|0\rangle |0\rangle) \otimes |\psi_{0,1}\rangle \big) \nonumber \\ \nonumber \\
& \qquad = \frac{1}{2}\bigg(|0\rangle |0\rangle \otimes \widetilde{\sigma}_0^{(1)} \, |\psi_{0,1}\rangle + |1\rangle |0\rangle \otimes \widetilde{\sigma}_1^{(1)} \, |\psi_{0,1}\rangle + |0\rangle |1\rangle \otimes \widetilde{\sigma}_2^{(1)} \, |\psi_{0,1}\rangle + |1\rangle |1\rangle \otimes \widetilde{\sigma}_3^{(1)} \, |\psi_{0,1}\rangle \bigg)
\end{align}
which is the same as Eqn.~\eqref{superstate4} above.  The factor of $-i$ in the definition of $\widetilde{\sigma}_3^{(a_j)}$ allows us to achieve Eqn.~\eqref{superstate4} by the two sequential controlled--$\sigma_1^{(1)}$ and controlled--$\sigma_2^{(1)}$ unitaries, since $\sigma_2^{(1)}\sigma_1^{(1)} = - i \, \sigma_3^{(3)} = \widetilde{\sigma}_3^{(1)}$.  This trick was suggested by Mikhail Lukin.  More compactly, writing $\text{Had} \otimes \textbf{1} = U_{\text{Had}1}$ and $\textbf{1} \otimes \text{Had} = U_{\text{Had}2}$, we have
\begin{equation}
U_2 \, U_1 \, U_{\text{Had}2} \, U_{\text{Had}1} \,\big(|0\rangle|0\rangle \otimes |\psi_{0,1}\rangle \big) = |\Psi_{1,1}\rangle\,.
\end{equation}
Note that even if we had $U_2 \,U_1\, U_{\text{Had}2} \, U_{\text{Had}1}\,\big(|0\rangle|0\rangle \otimes |\psi_{0,1}\rangle \big) = e^{i \phi} |\Psi_{1,1}\rangle$ where $\phi$ is some phase, this would be fine as well since global phases do not affect any of our analysis.

In physical systems such as quantum simulators, unitaries are implemented by Hamiltonian evolution.  Thus, we should work out how to write $U_{\text{Had}1}$, $U_{\text{Had}2}$, $U_1$ and $U_2$ as exponentiated Hamiltonians.  We have
\begin{equation}
\widetilde{U}_{\text{Had}1} = e^{- i \, \frac{\pi}{2} \, H_{\text{Had}1}}\,, \qquad \widetilde{U}_{\text{Had}2} = e^{- i \, \frac{\pi}{2} H_{\text{Had}2}} \,, \qquad \widetilde{U}_{1} = e^{- i \frac{\pi}{4} \,H_{1}}\,, \qquad  \widetilde{U}_{2} = e^{- i \frac{\pi}{4} \,H_{2}}
\end{equation}
where the Hamiltonians $H_{\text{Had}1}$, $H_{\text{Had}2}$, $H_{1}$, $H_{2}$ are given by
\begin{align}
H_{\text{Had}1} &= \vec{m} \cdot \vec{\sigma}^{(1),\text{aux}}\,, \qquad  H_{\text{Had}2} = \vec{m} \cdot \vec{\sigma}^{(2),\text{aux}} \,, \qquad \text{where} \quad \vec{m} = \frac{1}{\sqrt{2}}\,(1,0,1) \\ \nonumber \\
H_{1} &= \sigma_3^{(1),\text{aux}} + \sigma_1^{(1)} -  \sigma_3^{(1),\text{aux}} \sigma_1^{(1)} \\ \nonumber \\
H_{2} &= \sigma_3^{(2),\text{aux}} + \sigma_2^{(1)} -  \sigma_3^{(2),\text{aux}} \sigma_2^{(1)}
\end{align}
where $\sigma_k^{(j),\text{aux}}$ acts on the first or second auxiliary qubits for $j = 1$ or $2$, respectively.  We have also used the standard notation $\vec{\sigma} = (\sigma_1, \sigma_2, \sigma_3)$  Note that $\widetilde{U}_{\text{Had}1} = i \, U_{\text{Had}1}$, $\widetilde{U}_{\text{Had}1} = i \, U_{\text{Had}2}$, $\widetilde{U}_{1} = e^{i \frac{\pi}{4}} U_{1}$, $\widetilde{U}_{2} = e^{i \frac{\pi}{4}} U_{2}$, and so we have $\widetilde{U}_{\text{couple}} \, \widetilde{U}_{\text{Had}} \,\big(|0\rangle|0\rangle \otimes |\psi_{0,1}\rangle \big) = e^{-i \frac{\pi}{2}} \,|\Psi_{1,1}\rangle$ which has an overall global phase which does not matter to us.  In summary, the procedure for obtaining $|\Psi_{1,1}\rangle$ up to a global phase is as follows: \\ \\
\textbf{Procedure to obtain $|\Psi_{1,1}\rangle$ up to a global phase:}
\begin{enumerate}
\item Start with the initial state $|0\rangle|0\rangle \otimes |\psi_{0,1}\rangle$.
\item Apply $e^{- i \, \frac{\pi}{2} \, H_{\text{Had}1}}$ and then $e^{- i \, \frac{\pi}{2} \, H_{\text{Had}2}}$ to the auxiliary system.
\item Apply $e^{- i \frac{\pi}{4} \,H_{1}}$ and $e^{- i \frac{\pi}{4} \,H_{2}}$ then  to the whole system.
\item Output the state $|\Psi_{1,1}\rangle$ up to a global phase.
\end{enumerate}

We can devise a very similar to procedure to obtain $|\Psi_n\rangle$.  First, we fix some notation: suppose we have $n$ auxiliary systems comprising of $2$ qubits each, so that we start with the initial state
\begin{equation}
\big(|0\rangle_1 |0\rangle_1\big)\big(|0\rangle_2 |0\rangle_2\big) \cdots \big(|0\rangle_n |0\rangle_n \big) \otimes |\psi_{0,N}\rangle\,.
\end{equation}
Furthermore, suppose that $H_{\text{Had}1,j}$ and $H_{\text{Had}2,j}$ act on the $j$th pair of auxiliary qubits, and that $H_{1,j,a_j}$ and $H_{2,j,a_j}$ couple the $j$th pair of auxiliary qubits to the $a_j$th qubit of the main system.  Finally, let $U(t_{k+1}, t_{k}) = e^{- i (t_{k+1} - t_{k}) H}$ be the time evolution of the main system, where $H$ is the Hamiltonian of the main system.  Then we have the following procedure to obtain $|\Psi_{n,N}\rangle$ up to a global phase:
\\ \\
\textbf{Procedure to obtain $|\Psi_{n,N}\rangle$ up to a global phase:}
\begin{enumerate}
\item Start with the initial state $\big(|0\rangle_1 |0\rangle_1\big)\big(|0\rangle_2 |0\rangle_2\big) \cdots \big(|0\rangle_n |0\rangle_n \big) \otimes |\psi_0\rangle$.
\item For $j=1,...,n$, apply the unitaries $e^{- i \, \frac{\pi}{2} \, H_{\text{Had}1,j}}$ and $e^{- i \, \frac{\pi}{2} \, H_{\text{Had}2,j}}$ to the $j$th pair of auxiliary qubits.
\item For $j=1,...,n-1$\,:
\subitem (a) Couple the $j$th pair of auxiliary qubits to the $a_j$th qubit of the main system by \subitem applying the unitary $e^{- i \frac{\pi}{4} \,H_{1,j,a_j}}$ followed by the unitary $e^{- i \frac{\pi}{4} \,H_{2,j,a_j}}$.
\subitem (b) Evolve the main system by the unitary $e^{- i (t_{j+1} - t_{j}) H}$.
\item Couple the $n$th pair of auxiliary qubits to the $a_n$th qubit of the main system by applying the unitary $e^{- i \frac{\pi}{4} \,H_{1,j,a_j}}$ followed by the unitary $e^{- i \frac{\pi}{4} \,H_{2,j,a_j}}$.
\item Output the state $|\Psi_{n,N}\rangle$ up to a global phase.
\end{enumerate}
For sufficiently small $n$ (but for $n \geq 2$ so that the superdensity operator is nontrivial), the operations necessary to generate such $|\Psi_{n,N}\rangle$'s are not far beyond existing technology.

\subsection{Measuring spacetime R\'{e}nyi 2-entropies}
Suppose we prepare the superstate
\begin{equation}
|\Psi_{2,N}\rangle = \frac{1}{4} \sum_{i,j=0}^3 |i,j\rangle \otimes \widetilde{\sigma}_j^{(b)} \, U(t_2,t_1)\, \widetilde{\sigma}_i^{(a)} \, |\psi_0\rangle
\end{equation}
using the methods explained above.  Tracing out the main system will leave us with the superdensity operator
\begin{equation}
\varrho = \frac{1}{16} \sum_{i,j,k,\ell=0}^3 \tr\left(\widetilde{\sigma}_j^{(b)} \, U(t_2,t_1)\, \widetilde{\sigma}_i^{(a)} \, |\psi_0\rangle\langle \psi_0| \widetilde{\sigma}_k^{(a)}\,U(t_2,t_1)^\dagger\, \widetilde{\sigma}_\ell^{(b)}\right) \, |i,j\rangle \langle k, \ell|
\end{equation}
which is a state of four qubits, since $|i,j\rangle = |i\rangle|j\rangle$ can be thought of as two systems of two qubits each, as explained above.  We would like to measure quantities like the spacetime purity $\text{tr}(\varrho^2)$ or the spacetime R\'{e}nyi 2-entropy $S_2[\varrho] = - \log \tr(\varrho^2)$.  These quantities characterize the entanglement between the $a$th spin at the initial time, and the $b$th spin at the final time.

There have been many experimental proposals and demonstrations measuring the purity and R\'{e}nyi 2-entropy of a small number of qubits \cite{Abanin1}--\cite{Gray1}.  In particular, by generating two copies of a state $\rho$ of a small number of qubits, one can experimentally measure the purity $\text{tr}(\rho^2)$, and then take minus the logarithm to obtain the R\'{e}nyi 2-entropy $- \log \tr(\rho^2)$.  In our proposed setting, we would need to generate two copies of the \textit{superdensity operator} $\varrho$, and then apply the experimental procedures already developed to extract $\text{tr}(\varrho^2)$ and determine $- \log \tr(\varrho^2)$.  A nice feature of our proposal is that each $\varrho$ is only a state of $4$ qubits, which is small enough to apply the known experimental procedures.

For completeness, we briefly outline the underlying mathematical structure of the known schemes for experimentally measuring the purity and R\'{e}nyi 2-entropy.  Suppose one is given two states $\rho_1$ and $\rho_2$, so that their joint state is $\rho_1 \otimes \rho_2$.  Then the unitary $\text{SWAP}$ operator acts on the joint system by interchanging the two subsystems:
\begin{equation}
\text{SWAP}^\dagger \, \big(\rho_1 \otimes \rho_2\big) \, \text{SWAP} = \rho_2 \otimes \rho_1\,.
\end{equation}
Note that $\text{SWAP} = \text{SWAP}^\dagger$, since $\text{SWAP}$ is its own inverse.  Now the key trick is that if we take the trace of $\rho_1 \otimes \rho_2$ against a \textit{single} copy of the swap operator, we have
\begin{equation}
\tr(\rho_1 \otimes \rho_2 \, \text{SWAP}) = \tr(\rho_1 \, \rho_2)\,.
\end{equation}
Hence, if we have two copies of our superdensity operator $\varrho$, then $\tr(\varrho \otimes \varrho \, \text{SWAP}) = \tr(\varrho^2)$ is the spacetime purity of $\varrho$.  But then we ask, how do we measure an expectation value of the form $\tr(\varrho \otimes \varrho \, \text{SWAP})$?

To do so, we introduce an additional qubit, which we call the ``switch qubit.''  Different experimental proposals have implemented the switch qubit in different ways -- the switch qubit can be a physical qubit, two outgoing spatial modes from a beamsplitter, etc \cite{Abanin1}--\cite{Gray1}.  In any case, the underlying mathematics is the same: we implement a unitary $U$ which acts on both the switch qubit and the two superdensity operators:
\begin{equation}
U = |0\rangle \langle 0| \otimes \textbf{1} + |1\rangle \langle 1| \otimes \text{SWAP}\,.
\end{equation}
This unitary leaves the superdensity operator copies untouched if the switch qubit is in the state $|0\rangle$, and applies the swap gate to the superdensity operator copies if the switch qubit is in the state $|1\rangle$.  Defining $|+\rangle = (|0\rangle + |1\rangle)/\sqrt{2}$ and $|-\rangle = (|0\rangle - |1\rangle)/\sqrt{2}$, suppose we start with the initial state
\begin{equation}
|+\rangle \langle +| \otimes \varrho \otimes \varrho = \frac{1}{2}\left(|0\rangle \langle 0| + |0\rangle \langle 1| + |1\rangle \langle 0| + |1\rangle \langle 1| \right)\otimes \varrho \otimes \varrho\,.
\end{equation}
Then applying $U$ we have
\begin{align}
U^\dagger\left(|+\rangle \langle +| \otimes \varrho \otimes \varrho\right)U &= \frac{1}{2}\bigg(|0\rangle \langle 0|\otimes \varrho \otimes \varrho + |0\rangle \langle 1|\otimes (\varrho \otimes \varrho)\text{SWAP} \nonumber \\
& \qquad \quad  + |1\rangle \langle 0| \otimes \text{SWAP}(\varrho \otimes \varrho) + |1\rangle \langle 1|\otimes \text{SWAP} (\varrho \otimes \varrho) \text{SWAP} \bigg)
\end{align}
which contains desired terms like $\text{SWAP}(\varrho \otimes \varrho)$ and $(\varrho \otimes \varrho)\text{SWAP}$.  To extract the traces we want, we simply measure the switch qubit in the $\{|+\rangle, |-\rangle\}$ basis.  The corresponding probabilities are
\begin{align}
\text{Prob}(+) &= \text{tr}\left(\langle +| \otimes \textbf{1}\otimes \textbf{1} \big(U^\dagger\left(|+\rangle \langle +| \otimes \varrho \otimes \varrho\right)U\big) |+\rangle \otimes \textbf{1} \otimes \textbf{1}\right) = \frac{1}{2}\left(1 + \tr(\varrho^2)\right) \\ 
\text{Prob}(-) &= \text{tr}\left(\langle -| \otimes \textbf{1}\otimes \textbf{1} \big(U^\dagger\left(|+\rangle \langle +| \otimes \varrho \otimes \varrho\right)U\big) |-\rangle \otimes \textbf{1} \otimes \textbf{1}\right) = \frac{1}{2}\left(1 - \tr(\varrho^2)\right)
\end{align}
and so
\begin{equation}
\text{Prob}(+) - \text{Prob}(-) = \tr(\varrho^2)\,. 
\end{equation}
By collecting measurement statistics, one can determine the spacetime purity $\tr(\varrho^2)$ to progressively higher precision, and then use the result to compute the R\'{e}nyi 2-entropy $- \log \tr(\varrho^2)$.

\section{Summary and conclusion}

We have developed  tools for treating quantum information in spacetime.  In particular, the superdensity operator is the spacetime analog of the standard density operator.  Since the superdensity operator is itself a density operator on a larger Hilbert space, the standard tools of quantum information techniques can be upgraded to the spacetime setting.  Furthermore, the superdensity operator is observable, and can be treated as the density operator of auxiliary apparatus which couples to a system of interest at multiple sequential times.  The observability of the superdensity operator suggests new and novel experiments, some of which we have outlined.

We have demonstrated a few applications of the superdensity formalism -- notably, computing spacetime entropies for many-body systems, and studying spacetime mutual information, which bounds spacetime correlation functions.  Several other applications are under active development.  For example, we are exploring the encoding of time and causality in quantum systems, and quantifying the non-local encoding of causal influence in the presence of entanglement \cite{Causality}.  Also, there may be a wide range of new tools for temporal analogs of the renormalization group, as well as phenomena involving temporal phase transitions such as in \cite{Ma1}. 

In the context of quantum information, the superdensity formalism can be used to understand spacetime extensions of quantum protocols and algorithms, for example quantum key distribution and quantum error correction.  For quantum gravity, it would be interesting to apply the superdensity formalism to AdS-CFT \cite{AdS1, AdS2, AdS3}, in which it is argued that entanglement is responsible for the emergence of space \cite{SpaceEnt1}--\cite{SpaceEnt9}.  It seems that a more general form of entanglement is responsible for the emergence of space\textit{time}.  More broadly, we anticipate our formalism will have utility in studying the quantum information of \textit{dynamics}, going beyond the static properties of a quantum state at a fixed time.

\section*{Acknowledgements}
We would like to thank Ahmed Almheiri, Jingyuan Chen, Patrick Hayden, Yuri Lensky, Marcin Nowakowski and Mikhail Lukin for valuable conversations.  JC is supported by the Fannie and John Hertz Foundation and the Stanford Graduate Fellowship program. CMJ's work is partly supported by the Gordon
and Betty Moore Foundations EPiQS Initiative through
Grant GBMF4304. XLQ is supported by the David and Lucile Packard Foundation, and by the National Science Foundation through the grant No. PHY-1720504. FW's work is supported by the U.S. Department of Energy under grant DE-SC0012567, the European Research Council under grant 742104, and the Swedish Research Council under Contract No. 335-2014-7424.

\newpage
\appendix

\section{Channel mapping to initial time density operator}
\label{sec:AppA}

Again suppose that we work on a Hilbert space $\mathcal{H}$ with dimension $d$, and that $\{X_i\}$ is a complete orthonormal basis of operators on $\mathcal{H}$ , satisfying $\tr(X_i^\dagger X_j) = \delta_{ij}$.  Then we have the following theorem: \\

\noindent \textbf{Theorem:} \textit{The map $\mathcal{N}$ which takes}
$$\mathcal{N}[\,|i\rangle \langle j| \,] = X_i^\dagger X_j$$
\textit{is a quantum channel satisfying}
$$\mathcal{N}\left[\frac{1}{d} \sum_{i,j=0}^{d^2 - 1} \tr(X_i \, \rho_0 \, X_j^\dagger) \, |i\rangle \langle j| \right] = \rho_0\,.$$

\noindent \textit{Proof:} We begin by showing that $\mathcal{N}$ is completely positive.  Letting
$$|\Omega\rangle = \sum_{k=0}^{\min(d^2,n)-1}|k\rangle_1 \otimes |k\rangle_2\,,$$
we have
\begin{align*}
(\mathcal{N}\otimes I_n)[|\Omega\rangle \langle \Omega|] &= \sum_{j,k=0}^{\min(d^2,n)-1} |j\rangle \langle k| \otimes X_j^\dagger X_k \,.
\end{align*}
This operator is positive definite because given any state $|\phi\rangle$ in $\mathcal{H} \otimes \mathbb{C}^n$, we have
\begin{align*}
\langle \phi| \left( \sum_{j,k=0}^{\min(d^2,n)-1} |j\rangle \langle k| \otimes X_j X_k^\dagger \right) |\phi\rangle &= \left|\left(\sum_{k=0}^{\min(d^2,n)-1} \langle k|\otimes X_k\right)|\phi\rangle \right|^2 \geq 0
\end{align*}
and thus $\mathcal{N}$ is completely positive.  To show that $\mathcal{N}$ is trace-preserving, it suffices to note that
\begin{equation*}
\tr\left(\mathcal{N}[\,|i\rangle \langle j|\,] \right) = \tr(X_i X_j^\dagger) = \delta_{ij} = \tr(|i\rangle \langle j|)\,.
\end{equation*}
Last, we compute
\begin{align*}
\mathcal{N}\left[\frac{1}{d} \sum_{i,j=0}^{d^2 - 1} \tr(X_i \, \rho_0 \, X_j^\dagger) \, |i\rangle \langle j| \right] &= \frac{1}{d} \sum_{i,j=0}^{d^2 - 1} \tr(X_i \, \rho_0 \, X_j^\dagger) \, X_i^\dagger X_j \\
&= \frac{1}{d} \sum_{i,j=0}^{d^2 - 1} \rho_0 \, X_j^\dagger X_j \\
&= \rho_0
\end{align*}
which gives us the desired result. $\square$ \\ \\
\indent We have the following corollary:

\noindent \textbf{Corollary:} \textit{The map} $\mathcal{N} \otimes \text{tr}(\,\boldsymbol{\cdot}\,) \otimes \cdots \otimes \text{tr}(\,\boldsymbol{\cdot}\,)$ \textit{is a quantum channel satisfying}
$$\bigg( \mathcal{N} \otimes \text{tr}(\,\boldsymbol{\cdot}\,) \otimes \cdots \otimes \text{tr}(\,\boldsymbol{\cdot}\,) \bigg)[\varrho] = \rho_0$$
where
\begin{align*}
&\varrho = \frac{1}{d^{n}} \sum_{\substack{i_1,...,i_n = 0 \\ j_1,...,j_n = 0}}^{d^2 - 1} \tr\left(X_{i_n} U(t_n,t_{n-1}) \cdots U(t_2, t_1) X_{i_1} \, \rho_0 \, X_{j_1}^\dagger \, U^\dagger(t_2,t_1) \cdots U(t_{n},t_{n-1})^\dagger X_{j_n}^\dagger\right) \nonumber \\
&\qquad \qquad \qquad \qquad \qquad \qquad \qquad \qquad \qquad \qquad \qquad \qquad \times\,|i_1\rangle \langle j_1| \otimes |i_2\rangle \langle j_2| \otimes \cdots \otimes |i_n\rangle \langle j_n|\,.
\end{align*}

\section{Relation to other formalisms}

Several formalisms have been proposed over the last few decades which capture, in different ways, temporal correlations.  In this section, we focus on the consistent histories and entangled histories formalisms, and the multi-state vector formalism.  We show that the superdensity formalism encompasses these other formalisms.

\subsection{Consistent histories and entangled histories}

Consistent histories were introduced by Griffiths in \cite{Griffiths1, Griffiths2, Griffiths3}, and were further developed by numerous authors \cite{DF1}--\cite{Isham1}.  We first give a brief overview before explaining how consistent histories fits in to the superdensity formalism.

Consider a system with density operator $\rho_0$ at a fixed initial time.  If we want to know how much of the state lies in some subspace, we can construct the projector $P$ onto that subspace and compute $\tr(P \rho_0 P)$.  The subspace in question can correspond to a physical property.  For example, in a spin system, the subspace could be those states which have an expected value of zero net spin.  As another example, for a single point particle, the subspace could be those states for which the particle lies within some spatial region.  In any case, our system has a specified property if it lies completely within the specified subspace, so that $P \rho_0 P = \rho_0$ and hence $\tr(P \rho_0 P) = 1$.

If we have two subspaces corresponding to two ``properties,''  with projectors $P_1$ and $P_2$ respectively, we might ask:  Is it possible for a system to have the first property but not the second property?  If a system has the first property, then by definition it lies within the first subspace.  But if the system also does not have the second property, it must be the case that the second subspace is orthogonal to the first.  

More generally, we say that some list of properties, corresponding to the projectors $P_1,...,P_n$, are \textit{independent} if they correspond to mutually orthonormal subspaces.  This is intuitive: two observables are independent if they are commuting.  Indeed, if the properties correspond to  mutually orthonormal subspaces, then $[P_i, P_j] = 0$ for $i\not = j$.  Conversely, if two projectors $P_i, P_j$ commute, then they correspond to orthonormal subspaces and thus $P_i P_j = P_j P_i = 0$.

Given two properties corresponding to projectors $P_1$ and $P_2$, we have addressed if it is possible for a system to have these properties independently.  The condition is $[P_i, P_j] = 0$.  But if we have a particular system described by the density operator $\rho_0$, then we can still ask if two properties corresponding to $P_1$ and $P_2$ are independent with respect to $\rho_0$.  The condition is
\begin{equation}
\label{projectioncondition1}
\tr\bigg((P_1 + P_2) \, \rho_0 \, (P_1 + P_2)\bigg) = \tr(P_1 \, \rho_0 \, P_1) + \tr(P_2 \, \rho_0 \, P_2)
\end{equation}
which in words, tells us that the ``amount'' of $\rho_0$ that has the first property does not overlap with the ``amount'' of $\rho_0$ that has the second property.  Note that Eqn.~\eqref{projectioncondition1} holds if\,\footnote{This condition is sufficient but not necessary.  The necessary \textit{and} sufficient condition is that $\text{Re}[\tr(P_i \rho_0 P_j)] = 0$, but in well-studied physical examples the stronger condition $\tr(P_i \rho_0 P_j)$ is used to check the independence of $P_i$ and $P_j$ \,\cite{Griffiths3}.  Thus we will use the stronger condition here for simplicity.}
\begin{equation}
\label{projectioncondition2}
\tr(P_1 \, \rho_0 \, P_2) = 0\,.
\end{equation}
This does not imply $P_2 P_1 = 0$, but rather that $P_2 P_1$ projected onto the support of $\rho_0$ is zero.  If instead we have a list of properties corresponding to $P_1,...,P_n$, then these properties are independent with respect to $\rho_0$ if
\begin{equation}
\label{ConsistentProj1}
\tr(P_i \, \rho_0 \, P_j) = 0
\end{equation}
for all $i \not = j$.

Now we move on to consider a more general case, in which we can ask if a system has a \textit{sequence} of properties in time.  Suppose we have $n$ times $t_1,...,t_n$ and unitary evolution given by $U(t_{i+1},t_i)$.  Further suppose we have a sequence of properties corresponding to the projectors $P_{t_1},...,P_{t_n}$.  To understand to what degree an initial state $\rho_0$ follows this sequence of properties, we can compute
\begin{equation}
\label{historiescorr1}
\tr\bigg(P_{t_n} \, U(t_n,t_{n-1}) \, P_{t_{n-1}} \,\cdots\, P_{t_2} \, U(t_2, t_1) \, P_{t_1} \, \rho_0 \, P_{t_1} \, U(t_2,t_1)^\dagger \, P_{t_2} \, \cdots \, P_{t_{n-1}} \, U(t_n,t_{n-1})^\dagger \, P_{t_n}\bigg)\,.
\end{equation}
The form of this correlation function is familiar to us, and letting
\begin{equation}
P = P_{t_n} \otimes \cdots \otimes P_{t_1}
\end{equation}
and hence
\begin{equation}
|P) = |P_{t_n})\otimes \cdots \otimes |P_{t_1})
\end{equation}
we can write Eqn.~\eqref{historiescorr1} compactly as
\begin{equation}
(P| \, \varrho \, |P)
\end{equation}
where $\varrho$ is the $n$-time superdensity operator with initial state $\rho_0$ and unitaries given by $U(t_{i+1},t_i)$.    Indeed, if $(P| \, \varrho \, |P) = 1$, then $\rho_0$ satisfies the sequence of properties corresponding to $P$.  Using Griffiths' termninology $P$ (or $|P)$) is called a \textit{history}.

Now suppose we have another sequence of properties corresponding to the projectors $Q_{t_1},...,Q_{t_n}$ which we package as the history $Q = Q_{t_n} \otimes \cdots \otimes Q_{t_1}$.  Directly analogous to Eqn.'s~\eqref{projectioncondition1} and~\eqref{projectioncondition2} above, the histories $P$ and $Q$ are independent with respect to $\rho_0$ if
\begin{equation}
\big((P| + (Q|\big) \, \varrho \, \big(|P) + |Q)\big) = (P| \, \varrho \, |P) + (Q| \, \varrho \, |Q)
\end{equation}
which holds if\,\footnote{Similarly, this condition is sufficient but not necessary -- the necessary and sufficient condition is $\text{Re}[(P| \rho_0 |Q))] = 0$, but we will use the stronger condition here as is standard in the consistent histories literature \cite{Griffiths3}.}
\begin{equation}
(P| \, \varrho \, |Q) = 0\,.
\end{equation}
If we have multiple sequences of properties corresponding to
\begin{equation}
|P^{i}) = |P_{t_n}^i) \otimes \cdots \otimes |P_{t_1}^i)\,,
\end{equation}
such that $\sum_i |P^{i}) = |\textbf{1}_{\mathcal{H}_{\text{hist.}}})$, i.e.\! the $|P^i)$'s are a decomposition of the identity on the history Hilbert space, then we call the set $\{|P^{i})\}$ a \textit{family}.  We further say that the set $\{|P^{i})\}$ is a \textit{consistent} family if it additionally satisfies the \textit{consistent histories condition}
\begin{equation}
(P^i| \, \varrho \, |P^j) = 0
\end{equation}
for all $i \not = j$.  The physical content of this condition is that the sequences of properties corresponding to the $|P^{i})$'s do not interfere with each other.  In other words, we can think of the time evolution system as decomposing into parallel sequences of events which do not affect one another.

In the language of superdensity operators, we restrict the superdensity operator to the bilinear map
\begin{equation}
\widetilde{\varrho} : \text{Proj}(\mathcal{H}_{\text{hist.}}) \otimes \text{Proj}(\mathcal{H}_{\text{hist.}})^* \longrightarrow \mathbb{C}
\end{equation}
where $\text{Proj}(\mathcal{H}_{\text{hist.}})$ is the space of projectors on $\mathcal{H}_{\text{hist.}}$ and $\text{Proj}(\mathcal{H}_{\text{hist.}})^*$ is defined similarly.  To find consistent families, we need to partially diagonalize $\widetilde{\varrho}$ with respect to $\text{Proj}(\mathcal{H}_{\text{hist.}})$, namely find a set of $\{|P_i)\}$ satisfying $\sum_i |P_i) = |\textbf{1}_{\mathcal{H}_\text{hist.}})$ such that
\begin{equation}
\label{SuperdensityConsistency}
(P_i | \, \widetilde{\varrho}\, |P_j) = \delta_{ij}\,(P_i |\, \widetilde{\varrho}\, |P_i)\,.
\end{equation}
Note that partially diagonalizing $\widetilde{\varrho}$ with respect to $\text{Proj}(\mathcal{H}_{\text{hist.}})$ such that $\sum_i |P_i) = |\textbf{1}_{\mathcal{H}_\text{hist.}})$ is a constrained problem in two ways: firstly $\text{Proj}(\mathcal{H}_{\text{hist.}})$ is not a linear subspace of $\mathcal{B}(\mathcal{H}_{\text{hist.}})$ (i.e., the sum of two projectors need not be a projector), and secondly we want the solutions $\{|P^i)\}$ to satisfy $\sum_i |P_i) = |\textbf{1}_{\mathcal{H}_\text{hist.}})$.  Furthermore, this constrained, partial diagonalization of $\widetilde{\varrho}$ is not unique, and there are many possible consistent families.

The rephrasing of the consistent histories condition in terms of the superdensity operator in Eqn.~\eqref{SuperdensityConsistency} makes it clear that entangled histories \cite{CW1, CW2, CW3, CW4}, namely higher rank projectors in $\text{Proj}(\mathcal{H}_{\text{hist.}})$ which are not separable across temporal tensor factors, occur naturally when diagonalizing $\widetilde{\varrho}$.  Such entangled histories correspond to a system having multiple properties which may not be independent at any given time.  Histories and more generally entangled histories are in fact observable by coupling to superdensity operator to auxiliary apparatus.  An earlier realization of this idea can be found in \cite{CW4}.

Furthermore, Eqn.~\eqref{SuperdensityConsistency} immediately suggests how to do consistent histories for more elaborate superdensity operators, which may include quantum channels or novel tensor factorizations of Hilbert space into local subsystems.  Conceptually, the superdensity operator is ``dual'' to the consistent histories approach -- the superdensity operator maps histories to numbers.  This means that one can use the quantum information theoretic primitives of the superdensity formalism (i.e., the partial trace, entropies, etc.) to define corresponding meaningful notions for histories, which have remained elusive.  We emphasize that the superdensity formalism is more general than the consistent histories formalism because the superdensity operator can take as input any operators and not just projectors.

\subsection{Multi-time state formalism}

The multi-time state formalism developed in \cite{Aharonov3} generalizes the notion of two-state vectors \cite{Aharonov1, Aharonov2}.  The formalism is a way of organizing measurement procedures in which there are sequential pre- and post-selections.  Furthermore, the combination of pre- and post-selections can be entangled, in the sense that various pre- and post-selection schemes can be applied in superposition.  Let us make this concrete with equations.

As a simple example, suppose we pre-select for the system to be in an initial state $|\phi_0\rangle$ at time $t_1$, and post-select for the system to be in the final state $|\psi_f\rangle$ at time $t_2$.  Then we can write this as the ``two-state vector''
\begin{equation}
\,_{t_2}\langle \phi_f| \,\, |\psi_0\rangle_{t_1}\,.
\end{equation}
Suppose we have some positive-operator valued measure (POVM), which we specify by a collection of Kraus operators $A_k$'s satisfying $\sum_{k} A_k^\dagger A_k = \textbf{1}$.  We will measure the POVM at a time $t$, where $t_1 < t < t_2$.  Then the probability of obtaining the outcome ``$j$'' of the POVM for our proposed pre- and post-selection scheme is
\begin{equation}
\text{Prob}(j) = \frac{\left|\,_{t_2}\langle \phi_f|U(t_2,t) \, A_j \, U(t,t_1) |\psi_0\rangle_{t_1}\right|^2}{\sum_k \left|\,_{t_2}\langle \phi_f|U(t_2,t) \, A_k \, U(t,t_1) |\psi_0\rangle_{t_1}\right|^2}
\end{equation}
where the denominator normalizes the probabilities so that $\sum_k \text{Prob}(k) = 1$.  More generally, we could consider the two-state vector
\begin{equation}
\label{multistatesuperposition1}
\sum_{\alpha, \beta} C_{\alpha \beta} \,_{t_2}\langle \psi_\beta| \,\, |\psi_\alpha\rangle_{t_1}\,.
\end{equation}
where the $C_{\alpha \beta}$'s are complex numbers.  This two-state vector represents a superposition of pre- and post-selection schemes, in which we pre-select on $|\psi_\alpha\rangle$ and post-select on $|\psi_\beta\rangle$ with amplitude $C_{\alpha \beta}$.  Then the probability of obtaining the outcome ``$j$'' of the POVM for this two-state vector is
\begin{equation}
\label{twostateprob1}
\text{Prob}(j) = \frac{\left|\sum_{\alpha, \beta} C_{\alpha \beta} \,_{t_2}\langle \psi_\beta|U(t_2, t) \, A_j \, U(t,t_1) |\psi_\alpha\rangle_{t_1}\right|^2}{\sum_k\left|\sum_{\alpha, \beta} C_{\alpha \beta} \,_{t_2}\langle \psi_\beta|U(t_2, t) \, A_k \, U(t,t_1) |\psi_\alpha\rangle_{t_1}\right|^2}
\end{equation}
where the denominator likewise enforces the normalization of the probabilities.

Before generalizing this scheme to multiple times, we note that we can express the two-state vector formalism compactly in terms of a special class of superdensity operators.  Consider the two-state vector in Eqn.~\eqref{multistatesuperposition1}.  We define the unnormalized superdensity operator
\begin{equation}
\varrho[\mathcal{O}_1, \mathcal{O}_2^\dagger] := \sum_{\alpha_1, \alpha_2, \beta_1, \beta_2}C_{\alpha_1 \beta_1} C_{\alpha_2 \beta_2}^*\,\text{tr}\left(\langle \psi_{\beta_1}| U(t_2,t) \, \mathcal{O}_1\,U(t,t_1) \, |\psi_{\alpha_1}\rangle\langle \psi_{\alpha_2}| U(t,t_1)^\dagger \, \mathcal{O}_2^\dagger\,U(t_2,t)^\dagger |\psi_{\beta_2}\rangle\right)\,.
\end{equation}
This superdensity operator is depicted below in Figure \ref{multistate1}.
\begin{center}\customlabel{multistate1}{12}
\includegraphics[scale=.40]{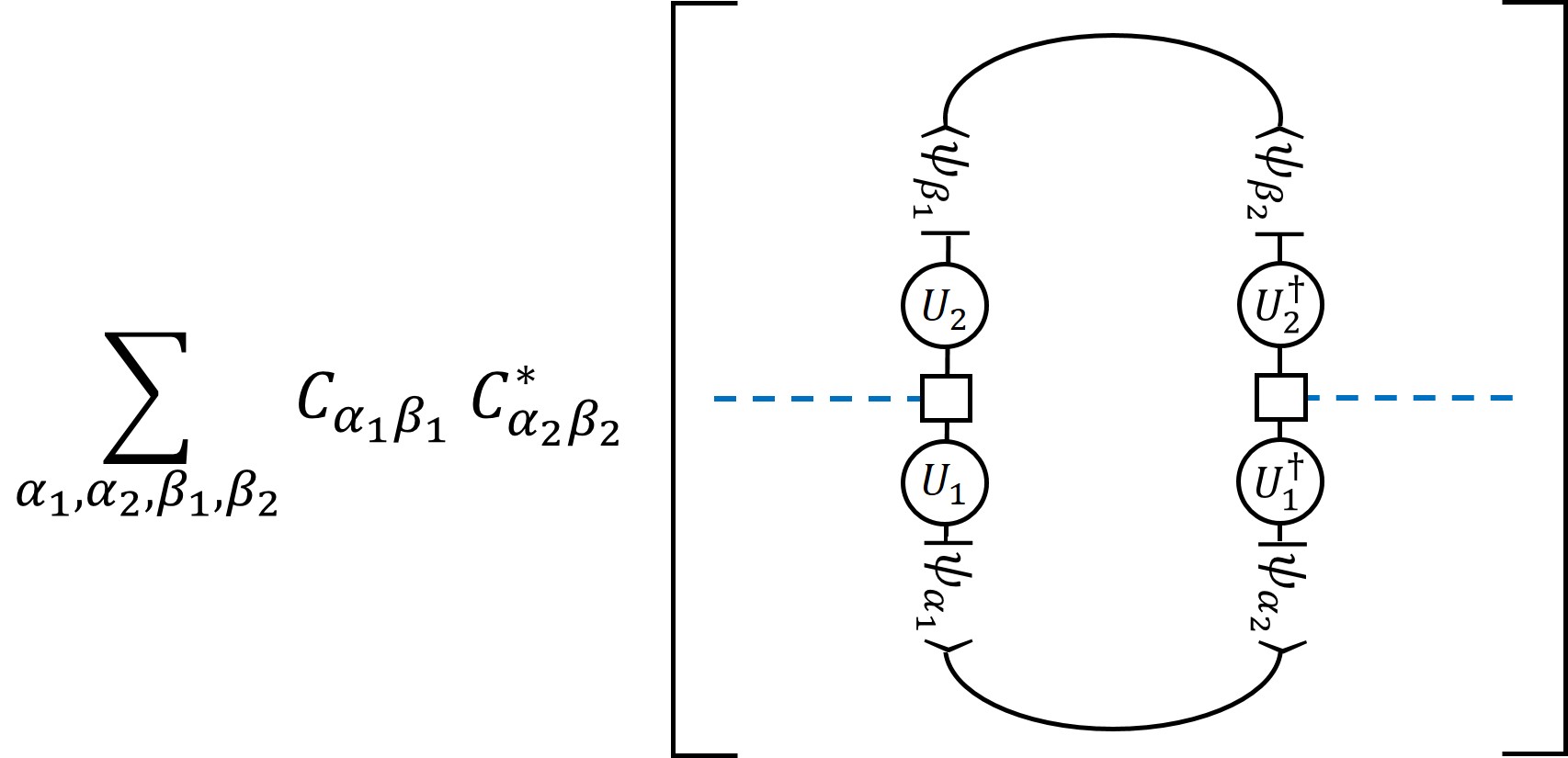}\\
\end{center}
\vskip.1cm
Figure 12: Diagrammatic representation of the unnormalized superdensity operator corresponding to a two-state vector.  In the diagram, $U_1 := U(t_1, t)$ and $U_2 := U(t,t_2)$. \\ \\ \vskip.1cm
\noindent Then we can write the compact equation
\begin{equation}
\text{Prob}(j) = \frac{\varrho[A_j, A_j^\dagger]}{\sum_k \varrho[A_k, A_k^\dagger]}\,,
\end{equation}
corresponding to Eqn.~\eqref{twostateprob1} above.  Thus, we see that expectation values with respect to the two-state vector in Eqn.~\eqref{multistatesuperposition1} can be expressed in terms of a superdensity operator in a superposition of different pre- and post-selection schemes.

Multi-state vectors are multi-time generalizations of the two-state vector in Eqn.~\eqref{multistatesuperposition1}.  An example of a five-time state is
\begin{equation}
\sum_{\alpha \beta \gamma \delta \varepsilon} C_{\alpha \beta \gamma \delta \varepsilon} \,_{t_5}\langle \psi_\varepsilon|\,\,|\psi_{\delta}\rangle_{t_4} \,_{t_3}\langle \psi_\gamma|\,\,|\psi_\beta\rangle_{t_2} \,_{t_1}\langle \psi_\alpha|\,.
\end{equation}
where the $C_{\alpha \beta \gamma \delta \varepsilon}$'s are complex numbers.  Considering three POVM's in term of their Kraus operators $\{A_i^1\}$, $\{A_j^2\}$ and $\{A_k^3\}$ inserted at times $t,t',t''$ with $t<t_1$, $t_2 < t' < t_3$ and $t_4 < t'' < t_5$.  Then utilizing the superposition of pre- and post-selection schemes described by the five-time state above, the probability of obtaining the outcome ``$i$'' for the first POVM, ``$j$'' for the second POVM, and ``$k$'' for the third POVM is
\begin{align}
&\text{Prob}(i,j,k) = \nonumber \\
&\,\,\frac{\left|\sum_{\alpha \beta \gamma \delta \varepsilon} C_{\alpha \beta \gamma \delta \varepsilon} \,_{t_5}\langle \psi_\varepsilon|U(t_5, t'') \, A_k^3 \, U(t'', t_4)|\psi_{\delta}\rangle_{t_4} \,_{t_3}\langle \psi_\gamma|U(t_3,t') \, A_j^2 \, U(t',t_2)|\psi_\beta\rangle_{t_2} \,_{t_1}\langle \psi_\alpha| U(t_1,t) \, A_i^1 \right|^2}{\left|\sum_{\ell,m,n} \sum_{\alpha \beta \gamma \delta \varepsilon} C_{\alpha \beta \gamma \delta \varepsilon} \,_{t_5}\langle \psi_\varepsilon|U(t_5, t'') \, A_n^3 \, U(t'',t_4)|\psi_{\delta}\rangle_{t_4} \,_{t_3}\langle \psi_\gamma|U(t_3,t') \, A_m^2 \, U(t',t_2)|\psi_\beta\rangle_{t_2} \,_{t_1}\langle \psi_\alpha| U(t_1,t) \, A_\ell^1 \right|^2}
\end{align}
Other multi-state vectors have similar form, and are generally coherent superpositions of sequential pre- and post-selection schemes.  Similar to the analysis above, expectation values with respect to any multi-time state in terms of a superdensity operator in a superposition of different pre- and post-selection schemes.

The multi-state vector formalism focuses on the computation of expectation values of POVM measurements with a superposition of sequential pre- and post-selection schemes.  By contrast, the superdensity formalism can capture any spacetime measurement scheme, including those described by multi-state vectors as a proper subset.  
Indeed, the superdensity operator contains them all, in coherent superposition.  

\section{Superdensity operator of free fermion systems}
\label{app:SuperDensity_Free_Fermion}
In Section \ref{sec:free_fermion}, given the operator basis choice in Eqn.~\eqref{Op_Basis_SingleMajorana}, we introduced a fermion operator $f_k$ for the $k$th time instant in the history Hilbert space. Also, we identified $|X_0)$ and $|X_1)$ as states with fermion occupation $f^\dag_k f_k$ being $0$ and $1$, respectively. To show that the superdensity operator $\varrho$ is a Gaussian density operator with respect to the fermion operators $f_k$ and $f_k^\dag$, it is convenient to work with fermionic coherent states in the history Hilbert space:
\begin{align}
| \psi_{1} \psi_{2} \cdots \psi_{n} ) = \left( \prod^n_{k=1} e^{-\overline{\psi}_{k} \psi_{k}/2} e^{-\psi_{k} f^\dag_{k}} \right)  |X_0) \otimes |X_0) \otimes \cdots \otimes |X_0)\,,
\end{align}
where $\psi_{k}$ are complex Grassmann numbers with their complex conjugate denoted by $\overline{\psi}_{k}$. The Grassmann numbers $\psi_k$ anti-commute with physical Majorana fermion operators and the complex fermion operator in the history Hilbert space. Here, the state $|X_0) \otimes |X_0) \otimes .... \otimes |X_0)$ on the right-hand side of this equation is the $n$-fold tensor product of $|X_0)$ corresponding to $n$ time instances. In the coherent state basis, the superdensity operator $\varrho$ has the matrix elements
\begin{align}
& ( \psi_{1}' \psi_{2}' \cdots \psi_{n}' | \varrho | \psi_{1} \psi_{2} \cdots \psi_{n} ) 
\nonumber \\ 
&=
 \tr\left( X(\psi_{n})  U(t_n,t_{n-1}) \cdots U(t_2,t_1)\, X(\psi_{1}) \,\rho_0\, X(\psi_{1}')^\dagger U(t_2,t_1)^\dagger \cdots U(t_n,t_{n-1})^\dagger X(\psi_{n}')^\dagger \right),
\end{align}
where we have introduced the operators
\begin{align}
 X(\psi) = \frac{e^{-\overline{\psi} \psi/2}}{\sqrt{2}} \left( \textbf{1} + \psi \, \chi_\alpha\right),~~~~  X(\psi)^\dag = \frac{e^{-\overline{\psi} \psi/2}}{\sqrt{2}} \left(\textbf{1} +  \chi_\alpha \, \overline{\psi}\right)\,.
\end{align}
Considering Wick's theorem, we can write the superdensity operator matrix elements \newline $( \psi_{1}' \psi_{2}' \cdots \psi_{n}' | \varrho | \psi_{1} \psi_{2} \cdots \psi_{n} )$ in a Gaussian form:
\begin{align}
& ( \psi_{1}' \psi_{2}' \cdots \psi_{n}' | \,\varrho\, | \psi_{1} \psi_{2} \cdots \psi_{n} ) 
\nonumber \\ \nonumber \\
& =
\frac{1}{2^n}\,\exp\left( 
-\frac{1}{2} \sum_{1\leq k \leq n} \overline{\psi}_k' \psi_k'
-\frac{1}{2} \sum_{1\leq k \leq n} \overline{\psi}_k \psi_k
-\frac{1}{2} \sum_{1\leq k,\ell \leq n} \overline{Q}_{k\ell}\, \overline{\psi}_k' \overline{\psi}_\ell'
-\frac{1}{2} \sum_{1\leq k,\ell \leq n} Q_{k\ell}\, \psi_k \psi_\ell
+ \sum_{1\leq k,\ell \leq n} P_{k\ell}\, \overline{\psi}_k' \psi_\ell
 \right), \nonumber \\
\label{SuperDensity_CoherentStates}
\end{align}
where
\begin{align}
& P_{k\ell} = \tr\left( U(t_n,t_{\ell})\, \chi_\alpha \,  U(t_\ell,t_1) \,\rho_0\, U(t_k,t_1)^\dagger \, \chi_\alpha  \, U(t_n,t_{k})^\dagger \right)
\nonumber \\ \nonumber \\
& Q_{k\ell} = 
\begin{cases} 
\tr\left( U(t_n,t_{k}) \, \chi_\alpha \,  U(t_k,t_\ell)\, \chi_\alpha \,  U(t_\ell,t_1) \,\rho_0\, U(t_n,t_{1})^\dagger \right) &  \quad k > \ell \\
- \tr\left( U(t_n,t_{\ell}) \, \chi_\alpha  \, U(t_\ell,t_k)\, \chi_\alpha \, U(t_k,t_1) \,\rho_0\, U(t_n,t_{1})^\dagger \right) & \quad k < \ell \\
      0 & \quad k=\ell 
\end{cases}
\\ \nonumber \\
\label{eq:PQQ_matrix}
& \overline{Q}_{k\ell} = 
\begin{cases} 
-\tr\left( U(t_n,t_{1}) \, \rho_0\, U(t_\ell,t_{1})^\dagger \,\chi_\alpha  \, U(t_k,t_{\ell})^\dagger \, \chi_\alpha \,  U(t_n,t_{k})^\dagger \right) & \,  k > \ell 
\\
\tr\left( U(t_n,t_{1}) \, \rho_0\, U(t_k,t_{1})^\dagger \, \chi_\alpha  \, U(t_\ell,t_{k})^\dagger \, \chi_\alpha \,  U(t_n,t_{\ell})^\dagger \right) & \, k < \ell
\\
0 & \, k= \ell
\end{cases}
\nonumber 
\end{align}
These matrices $P$, $Q$ and $\overline{Q}$ are exactly the ones appear in the definition of $\mathcal{K}$ in Eqn.~\eqref{eq:CalK_Def} and given in Eqn.~\eqref{mind_your_Ps_and_Qs}. The Gaussian form of the superdensity operator matrix elements $( \psi_{1}' \psi_{2}' \cdots \psi_{n}' |\, \varrho \, | \psi_{1} \psi_{2} \cdots \psi_{n} )$ implies that $\varrho$ is a Gaussian (super)density operator when expressed in terms of the fermion operators $f_k$ and $f_k^\dag$. Given the form of the superdensity operator $\varrho$ in Eqn.~\eqref{SuperDensity_CoherentStates} in the coherent state basis, we can calculate the two-point functions  $G_{k\ell} \equiv \tr(f_k^\dag f_\ell \varrho)$, $\widetilde{G}_{k\ell} \equiv \tr(f_k f_\ell^\dag \varrho) = \delta_{k\ell} - G_{\ell k}$ and $\Delta_{k \ell} \equiv \tr(f_k^\dag f_\ell^\dag \varrho)$ in the history Hilbert space.

First of all, in coherent state basis, we have the following useful identities:
\begin{align}
\textbf{1} & = \int D\overline{\psi} D \psi \, |\psi_1 \psi_2 \cdots \psi_n ) ( \psi_1 \psi_2 \cdots \psi_n| \, ,
\\
 \tr( \mathcal{O} ) & = \int D\overline{\psi} D\psi \,( \psi_1 \psi_2 \cdots \psi_n| (-1)^F \mathcal{O} | \psi_1 \psi_2 \cdots \psi_n )
\nonumber \\
& = \int D\overline{\psi} D\psi  D\overline{\psi}' D \psi' \,
e^{-\sum_k \overline{\psi}_k' \psi_k'/2}
e^{-\sum_k \overline{\psi}_k \psi_k/2}
e^{-\sum_k \overline{\psi}_k \psi_k'}
\,( \psi_1' \psi_2' \cdots \psi_n'|\mathcal{O} | \psi_1 \psi_2 \cdots \psi_n )\,.
\end{align}
The first equation provides a resolution of the identity operator in the history Hilbert space. The notation $\int D\overline{\psi} D \psi$ is a shorthand for $\int  d\overline{\psi}_1 \, d \psi_1 \, d\overline{\psi}_2 \, d \psi_2 \cdots d\overline{\psi}_n \, d \psi_n $. The second equation provides an evaluation of the trace of any operator $\mathcal{O}$. The operator $(-1)^F$ is the fermion parity operator. Using these identities, we can obtain 
\begin{align}
G_{k\ell} & = \int D\overline{\psi} D\psi  D\overline{\psi}' D \psi' \,
e^{-\frac{1}{2} \sum_{ k } \left( 
 \overline{\psi}'_k \psi'_k + \overline{\psi}_k \psi_k + 2\overline{\psi}_k \psi'_k \right)} 
 (-\overline{\psi}_k)\psi_\ell'
 \,
 ( \psi_1' \psi_2' \cdots \psi_n'|\,\mathcal{O}\,| \psi_1 \psi_2 \cdots \psi_n )
 \nonumber \\
& =  \int   D\overline{\psi} D\psi  D\overline{\psi}' D \psi'  \, \frac{ -\overline{\psi}_k \psi'_\ell}{2^n}
\exp\left(
\frac{1}{2}
\left(
\begin{array}{cccc}
\psi & \psi' & \overline{\psi} & \overline{\psi}'
\end{array}
\right)
\left(
\begin{array}{cccc}
-Q            &  0 & 1 & -P^T \\
 0            &  0 & 1 & 1 \\
-1            & -1 & 0 & 0 \\
P  & -1 & 0 & -\overline{Q}
\end{array}
\right)
\left(
\begin{array}{c}
\psi \\ \psi' \\ \overline{\psi} \\ \overline{\psi}'
\end{array}
\right)
\right).
\nonumber \\
& =
\left(
\begin{array}{cccc}
-Q            &  0 & 1 & -P^T \\
 0            &  0 & 1 & 1 \\
-1            & -1 & 0 & 0 \\
P  & -1 & 0 & -\overline{Q}
\end{array}
\right)^{-1}_{2n+k,n+\ell}
\label{eq:CorrelMatrix_G}
\end{align}
Here, we have used the matrix notation with $4n$-component vector $(\psi, \psi', \overline{\psi} , \overline{\psi}')^T$ and a $4n\times 4n$ matrix
$$\left(
\begin{array}{cccc}
-Q            &  0 & 1 & -P^T \\
 0            &  0 & 1 & 1 \\
-1            & -1 & 0 & 0 \\
P  & -1 & 0 & -\overline{Q}
\end{array}
\right)$$
with the ``1''s representing $n \times n$ identity matrices. This $4n \times 4n$ is exactly the matrix $\mathcal{K}$ defined in Eqn.~\eqref{eq:CalK_Def} in the main text. For $\Delta_{k\ell}$, we similarly have
\begin{align}
& \Delta_{k\ell}
=  \int   D\overline{\psi} D\psi  D\overline{\psi}' D \psi'\, \frac{ \overline{\psi}_k \overline{\psi}_\ell}{2^n}
\exp\left(
\frac{1}{2}
\left(
\begin{array}{cccc}
\psi & \psi' & \overline{\psi} & \overline{\psi}'
\end{array}
\right)
\left(
\begin{array}{cccc}
-Q            &  0 & 1 & -P^T \\
 0            &  0 & 1 & 1 \\
-1            & -1 & 0 & 0 \\
P  & -1 & 0 & -\overline{Q}
\end{array}
\right)
\left(
\begin{array}{c}
\psi \\ \psi' \\ \overline{\psi} \\ \overline{\psi}'
\end{array}
\right)
\right).
\nonumber \\
& =
-\left(
\begin{array}{cccc}
-Q            &  0 & 1 & -P^T \\
 0            &  0 & 1 & 1 \\
-1            & -1 & 0 & 0 \\
P  & -1 & 0 & -\overline{Q}
\end{array}
\right)^{-1}_{2n+k,2n+\ell}.
\label{eq:CorrelMatrix_D}
\end{align}
With the two-point function in the history Hilbert space obtained, we can calculate the spacetime entropy via the correlation matrix 
$C$ defined in Eqn.~\eqref{eq:Correlation_Matrix_Def}.

The discussion above concentrated on the operator basis choice in Eqn.~\eqref{Op_Basis_SingleMajorana}, which is generated by a single Majorana fermion operator $\chi_\alpha$.  We can straightforwardly generalize the same analysis to the operator basis Eqn.~\eqref{eq:multi_fermion_basis} generated by a collection of Majorana fermion operators $\{ \chi_\alpha \, |\, \alpha \in A \}$.  At each time instant, the history Hilbert space contains $|A|$ fermion modes. The state $| X_{\{n_\alpha\}} )$ that corresponds to the operator $ X_{\{n_\alpha\}} $ in Eqn.~\eqref{eq:multi_fermion_basis} should be identified with the state (in the history Hilbert space) with fermion occupation numbers given by $\{n_\alpha\}$. Similar to the previous discussion, it is convenient to introduce the coherent state basis. At the $k$th time, a coherent state $|\vec{\psi}_k )$ is parametrized by an $|A|$-component Grassmann vector $\vec{\psi}_k$ and its complex conjugate. In the coherent state basis, the superdensity operator can be written as
\begin{align}
& ( \vec{\psi}_{1}' \vec{\psi}_{2}' \cdots \vec{\psi}_{n}' | \,\varrho \, | \vec{\psi}_{1} \vec{\psi}_{2} \cdots \vec{\psi}_{n} ) 
\nonumber \\ 
&=
 \tr\left( X(\vec{\psi}_{n})  U(t_n,t_{n-1}) \cdots U(t_2,t_1)\, X(\vec{\psi}_{1}) \,\rho_0\, X(\vec{\psi}_{1}')^\dagger U(t_2,t_1)^\dagger \cdots U(t_n,t_{n-1})^\dagger X(\vec{\psi}_{n}')^\dagger \right),
\end{align}
where we have introduced the operators
\begin{align}
 X(\vec{\psi}_k) = \frac{1}{2^{|A|/2}} 
\prod_{\alpha \in A} \left( e^{-\overline{\psi}_{k\alpha} \psi_{k\alpha} /2}
 e^{\psi_{k\alpha} \chi_\alpha } \right)
,~~~~  
 X(\vec{\psi}_k)^\dag  = \frac{1}{2^{|A|/2}} 
\prod_{\alpha \in A} \left( e^{-\overline{\psi}_{k\alpha} \psi_{k\alpha} /2}
 e^{ \chi_\alpha \overline{\psi}_{k\alpha} } \right),
\end{align}
where $\psi_{k\alpha}$ denotes the $\alpha$th component of the Grassmann vector $\vec{\psi}_k$. To treat the Fermi statistics in the history Hilbert space properly, we need to specify an ordering 
of the sites in $A$. Different orderings are related to each other by a basis transformation and produce the same physical results. With the ordering of sites specified, we can obtain
\begin{align}
& ( \vec{\psi}_{1}' \vec{\psi}_{2}' \cdots \vec{\psi}_{n}' | \, \varrho \, | \vec{\psi}_{1} \vec{\psi}_{2} \cdots \vec{\psi}_{n} ) 
\nonumber \\
& =
\frac{1}{2^{n|A|}}\exp\left( 
-\frac{1}{2} \sum_{ \substack{1\leq k \leq n \\ \alpha \in A} } 
\left( 
\overline{\psi}_{k\alpha}' \psi_{k\alpha}'
+ \overline{\psi}_{k\alpha} \psi_{k\alpha}
\right)
-\frac{1}{2} \sum_{ \substack{1\leq k,\ell \leq n \\ \alpha,\beta \in A} }
\left(
 \overline{Q}_{k\alpha, \ell\beta} \overline{\psi}_{k\alpha}' \overline{\psi}_{\ell \beta}'
+ Q_{k\alpha, \ell \beta} \psi_{k\alpha} \psi_{\ell \beta}
- P_{k\alpha, \ell\beta} \overline{\psi}_{k\alpha}' \psi_{\ell\beta}
\right)
 \right),
\end{align}
where
\begin{align}
& P_{k\alpha, \ell \beta} = \tr\left( U(t_n,t_{\ell})\, \chi_\beta\, U(t_\ell,t_1) \,\rho_0\, U(t_k,t_1)^\dagger \, \chi_\alpha \, U(t_n,t_{k})^\dagger \right)
\nonumber \\ \nonumber \\
& Q_{k\ell} = 
\begin{cases} 
\tr\left( U(t_n,t_{k}) \,\chi_\alpha\,  U(t_k,t_\ell) \,\chi_\beta \, U(t_\ell,t_1) \,\rho_0\, U(t_n,t_{1})^\dagger \right) &\quad   (k > \ell)~\text{or}~(k=\ell~ \text{and} ~ \alpha >\beta) \\ \\
- \tr\left( U(t_n,t_{\ell}) \,\chi_\beta\,  U(t_\ell,t_k) \, \chi_\alpha  \, U(t_k,t_1) \rho_0\, U(t_n,t_{1})^\dagger \right) &\quad (k < l)~\text{or}~(k=\ell~ \text{and} ~ \alpha <\beta) \\ \\
      0 & \quad k=\ell~ \text{and}~ \alpha = \beta 
\end{cases},
\nonumber \\ \nonumber \\
& \overline{Q}_{k\ell} = 
\begin{cases} 
-\tr\left( U(t_n,t_{1}) \, \rho_0\, U(t_\ell,t_{1})^\dagger \,\chi_\beta \,  U(t_k,t_{\ell})^\dagger \, \chi_\alpha \,  U(t_n,t_{k})^\dagger \right) &   (k > \ell) ~\text{or}~(k=\ell~ \text{and} ~ \alpha >\beta)
\\ \\
\tr\left( U(t_n,t_{1}) \, \rho_0\, U(t_k,t_{1})^\dagger \, \chi_\alpha \, U(t_\ell,t_{k})^\dagger \, \chi_\beta \, U(t_n,t_{\ell})^\dagger \right) & (k < \ell) ~\text{or} ~(k=\ell~ \text{and} ~ \alpha <\beta) 
\\ \\
0 & k=\ell~ \text{and}~ \alpha = \beta  
\end{cases}.
\nonumber
\end{align}
Again, we notice that the matrix elements $( \vec{\psi}_{1}' \vec{\psi}_{2}' \cdots \vec{\psi}_{n}' | \,\varrho\, | \vec{\psi}_{1} \vec{\psi}_{2} \cdots \vec{\psi}_{n} ) $ imply that the superdensity operator is a Gaussian (super)density operator. Therefore, we can use its correlation matrix to capture all of its properties.  Thus we can obtain the spacetime entropy from the correlation matrix as we discussed previously.

\end{document}